\documentclass[11pt, oneside]{article}   	
\usepackage{geometry}                		
\usepackage[parfill]{parskip}    		
\usepackage{graphicx}				
\usepackage{amssymb,amsmath}
\usepackage{latexsym}
\usepackage{url}
\usepackage{hyperref}
\usepackage{color}
\usepackage{multirow}
\usepackage{float}
\usepackage{natbib}

\newcommand{\bA}{\mathbf{A}}
\newcommand{\bB}{\mathbf{B}}
\newcommand{\bC}{\mathbf{C}}
\newcommand{\bI}{\mathbf{I}}

\newcommand{\bP}{\mathbf{P}}
\newcommand{\bQ}{\mathbf{Q}}
\newcommand{\bR}{\mathbf{R}}

\newcommand{\bX}{\mathbf{X}}
\newcommand{\bZ}{\mathbf{Z}}
\newcommand{\bv}{\mathbf{v}}
\newcommand{\bw}{\mathbf{w}}
\newcommand{\bx}{\mathbf{x}}
\newcommand{\by}{\mathbf{y}}
\newcommand{\bzero}{\mathbf{0}}
\newcommand{\bmu}{\boldsymbol{\mu}}
\newcommand{\btheta}{\boldsymbol{\theta}}
\newcommand{\bSigma}{\boldsymbol{\Sigma}}

\DeclareMathOperator{\tr}{tr}
\DeclareMathOperator{\vecx}{vec}
\DeclareMathOperator* {\argmin}{ arg\,min}
\DeclareMathOperator* {\argmax}{ arg\,max}

\newtheorem{remark}{Remark}

\title{Markov-Switching State-Space Models\\ with Applications to Neuroimaging}

 \author{David Degras$^{a}$,  Chee Ming Ting$^{b}$,  Hernando Ombao$^{c}$\\
 $^{a}$\emph{Department of Mathematics, University of Massachusetts Boston} \\
$^{b}$\emph{School of Information Technology, Monash University Malaysia} \\
$^{c}$\emph{Statistics Program, King Abdullah University of Science and Technology}
}

\begin{document}

\maketitle

\begin{abstract}
State-space models (SSM) with Markov switching offer a powerful framework for detecting multiple regimes in time series, analyzing mutual dependence and dynamics within regimes, and asserting transitions between regimes. These models however present considerable computational challenges due to the exponential number of possible regime sequences to account for.  In addition, high dimensionality of time series can hinder likelihood-based inference. This paper proposes novel statistical methods for Markov-switching SSMs using maximum likelihood estimation, Expectation-Maximization (EM), and parametric bootstrap. We develop solutions for initializing the EM algorithm, accelerating convergence, and conducting inference that are ideally suited to massive spatio-temporal data such as brain signals. We evaluate these methods in simulations and present applications to EEG studies of epilepsy and of motor imagery. All proposed methods are implemented in a MATLAB toolbox available at \url{https://github.com/ddegras/switch-ssm}.  
\end{abstract}


\section{Introduction}
\label{intro}


\subsection*{Switching state-space models}

Switching SSMs have an extensive  track record of applications in fields as diverse as control engineering \citep{Chang1978,Melnyk2016}, econometrics \citep{Hamilton1990,Kim1994,Camacho2018,Carstensen2020}, speech recognition \citep{Deng2004,Rosti2004}, computer vision \citep{Bregler1997} and neuroimaging \citep{Prado2013,Samdin2017,Taghia2017,Ombao2018}. These models, also called  switching dynamic factor models, include switching regression \citep{Goldfeld1973,Cosslett1985} and switching vector autoregressive models \citep{Krolzig1997,Yang2000,Lanne2010,Ting2017} as  special cases.

The premise of SSMs is that time series measurements are driven by latent processes representing the true state of the studied system.  Standard SSMs consist of an \emph{observation equation} that encodes the relationship between observations and state vectors, and of a \emph{state equation} that describes the system dynamics. To these elements, switching SSMs add a \emph{switching process} that modulates the observation and state equations over time. This process indicates the \emph{regime} under which the system operates at a given time, for example the  business cycle (growth or recession) in econometrics. While the state process and switching process are both unobserved, the former evolves in a continuous space 
whereas the latter takes discrete values. The switching process may be modeled as a pure innovation process, 
i.e.,  independent innovations,  
or as a Markov or semi-Markov process. 

Several computational methods are available for fitting switching SSMs to data. Frequentist approaches are often based on maximum likelihood estimation and the Expectation-Maximization (EM) algorithm. Indeed, the EM algorithm \citep[e.g.][]{Dempster1977,McLachlan2008} is well suited for model with latent variables (here, the state vectors and switching variables).   
\cite{Chang1978} study nonlinear switching SSMs with time-varying parameters that belong to a known finite set. Assuming Markov dependence in the parameters, they propose to infer the state vectors and regimes with techniques like the extended Kalman filter. 
\cite{Hamilton1990} introduces an EM algorithm for $p$-dependent processes subject to Markov switching. In this situation without latent state vectors, the EM can be implemented exactly with iteration complexity $O(M^{2(p+1)} T^2)$, where $M$ is the number of regimes and $T$ the time series length.  
 \cite{Shumway1991} study dynamic linear systems with independent and nonstationary switching. They propose a pseudo-EM algorithm in which ``smoothed" occupancy probabilities (i.e., probability that the system is in a given regime at  time $t$  
 conditional on all observed data) are approximated by ``filtered" probabilities (i.e., conditioned on all data up to time $t$).
%
\cite{Kim1994} examines general Markov-switching SSMs and pioneers a filtering/smoothing  technique based on  \emph{collapsing} the possible regime sequences. Instead of considering all sequences, which is computationally intractable, this approach recursively approximates required distributions (for the observations, latent state and regimes) at each time point with mixtures of distributions from the recent past. The approximation accuracy is controlled by how far back in time the collapsing goes. 
\cite{Zhou2008} propose a  simple approximate EM algorithm for SSMs with switching in the observation process. Their one-step approximation utilizes the filtering method of \cite{Shumway1991} for independent switching processes. 
Another strategy for approximating the EM is to consider a small number of high probability regime sequences like in multiple object tracking \citep{Cox1996}. 
In recent work, \cite{Pauley2020} develop a grid-based probability 
filter that avoids collapsing but is only suitable for low-dimensional systems. 
Bayesian approaches to switching SSMs include Gibbs sampling \citep{Kim1999}, 
 Markov Chain Monte Carlo with robust nonparametric regression  \citep{Carter1996}, 
 variational Bayes \citep{Ghahramani2000}, and sequential Monte Carlo \citep{Doucet2001}. 
 The monographs of \cite{Kim1999} and \cite{Fruhwirth2006} 
 offer comprehensive accounts of the literature on switching SSMs.

\subsection*{Neuroimaging and dynamic functional connectivity}

In this paper we will employ Markov-switching SSMs to explore the dynamic connectivity structure in brain functional imaging data. 
Functional connectivity (FC), a core theme of recent neuroimaging research, relates to the integration of activity across brain regions \citep[e.g.,][]{Friston2011} and can be defined as statistical associations between recorded brain signals.
Utilizing modalities such as functional magnetic resonance imaging (fMRI) and electroencephalography (EEG), FC studies have advanced the understanding of healthy human brain organization as well as of mental disorders, and cerebral trauma 
 \citep{Jalili2011,Rubia2014,Dennis2014,vanMierlo2014,Siggiridou2014}. 
 They have delivered new insights into brain development and aging over the lifespan. 
In addition, FC studies provide biomarkers for the prediction of clinical and behavioral outcomes in individual subjects 
\citep{Warren2014,Arbabshirani2017,Vergara2017,Heinsfeld2018}.

Early fMRI studies of FC typically assessed global patterns over the course of an entire session.  This has led to the identification of highly reproducible motor, visual, auditory, attentional, and default mode network structures  \citep{Biswal1995,Fox2005}. However, there is now ample evidence that FC varies within and across sessions. In particular FC is known to  correlate with physiological, psychological, and behavioral changes, whether in task or resting condition \citep{Hutchison2013,Calhoun2014,Di2015,Fiecas2016}. These findings have fostered new research on FC as a dynamic phenomenon. \emph{Dynamic functional connectivity} (dFC) has also been investigated using EEG with a larger focus on temporal dynamics, spectral properties, \citep{Sakkalis2011,Chen2013,Karamzadeh2013,Huang2017} and also topographical configurations of sensor measurements known as microstates  \citep{Pascual1995,Lehmann1998}.

A widely used  statistical method for dFC analysis is the \emph{sliding window} method, which consists in calculating FC measures over a time window that slides across the range of the recorded signals \citep{Allen2014,Handwerker2012}. 
FC measures can then be aggregated over time into summary statistics or clustered to detect recurring FC states or regimes. 
Time-frequency analysis based for example on wavelet transforms or short-time Fourier transforms yields fine-grained information on the time-varying coherence and phase of brain signals  \citep{Ombao2005,Chang2010,Sakkalis2011,Park2014,Yang2015,Ombao2021}. 
Time series models offer another path to investigate dFC, representing brain signals by random processes with either smoothly varying dynamics  \citep{Kang2011,Lindquist2014,Molenaar2016,Tronarp2018} or with abrupt regime changes. Examples of the latter include  piecewise constant Gaussian models \citep{Cribben2013,Xu2015}, piecewise multivariate regression \citep{Degras2020}, switching vector autoregressive models \citep{Vidaurre2016,Samdin2017,Ombao2018}, and other switching SSMs \citep{Yang2010,Olier2013,Taghia2017,Ting2017}. 
Switching SSMs are particularly well suited to handle \emph{high-dimensional data} (they can effectively reduce data dimension by mapping observations  to state vectors of much lower dimension) and to conduct \emph{statistical inference}, two central issues in current dFC research. In addition, switching SSMs pool data across their entire time range to jointly identify recurring FC regimes and perform parameter estimation. This may increase statistical accuracy in comparison to two-step approaches that first pool data locally to estimate time-varying parameters and then cluster the parameter estimates to detect  regimes.

\subsection*{Contributions of the paper}

While Markov-switching SSMs (MS-SSMs) have generated substantial research over time, some important issues  remain unaddressed. First, most applications of these models pertain to low-dimensional time series. Usage in a high-dimensional context presents nontrivial challenges  of computation time, memory management, and numerical optimization. In relation to this, efficient software  is needed to implement MS-SSMs with some generality. 
Second, inference in MS-SSMs largely focuses on the unobserved regimes. Indeed occupancy probabilities come for free as by-products of the maximum likelihood procedure. In contrast inference of model parameters is less studied, especially in high dimension, although these parameters provide key insights into the time series dynamics and dependence structure within regimes. 
Lastly, while MS-SSMs have been largely employed in engineering and econometrics, their potential to explore recurrent FC states 
in neuroimaging studies remains underexploited. In response to these needs, we make the following contributions.

\begin{enumerate}

\item \emph{Estimation.} In the context of maximum likelihood estimation and its EM implementation, we propose fast initialization methods based on singular value decomposition, linear regression, $K$-means clustering, and optionally binary segmentation. Finding good starting values for the EM is essential given the tendency of this algorithm to get stuck in local maxima, especially with large datasets. We also propose new  optimization tools for handling parameter constraints (equality across regimes, fixed coefficients, scaling, or eigenvalue constraints) in the M step. These tools  enable flexible model specification based on domain-specific knowledge and analytic requirements;  they also guarantee that the parameter estimates define (asymptotically) stationary processes, which is necessary both for stable computations and rigorous inference.   
Further, we present an acceleration scheme that can speed up convergence of the EM by orders of magnitude. This makes it possible to fit sophisticated MS-SSMs to large datasets, which would otherwise be intractable. The scheme works by alternating regular EM iterations with Kalman filtering/smoothing in a standard (non-switching) SSM where latent regimes are fixed to their most likely values.

\item \emph{Inference.} We develop a simple and efficient parametric bootstrap for inferring identifiable model parameters and functions thereof. This approach circumvents the statistical and computational difficulties of  likelihood-based inference in high dimension. By fully accounting for all sources of variation in the MS-SSM (regimes, state process, and observations), the proposed bootstrap also outperforms nonparametric bootstraps. 
It can be harnessed to infer regime-specific covariance and correlation structures in the observed time series (e.g.,   FC networks in dFC analysis) but also dynamics (e.g.,    autocorrelation functions) and  Granger causality. We study the coverage properties of three  bootstrap confidence intervals (percentile, basic, and normal) under different asymptotics ($N\to\infty$ and/or  $T\to\infty$) in a numerical study.

\item \emph{Modeling and application to neuroimaging.} 
We study three MS-SSMs that represent the main flavors  of switching SSMs. 
The first model features switching at the level of dynamics, that is, in the state equation. 
The second model, namely the popular switching vector-autoregressive model, 
is a special instance of the first with  no hidden state process.    
The third model is fairly sophisticated and can be thought of as a collections of independent SSMs 
together with a gating process that determines 
which  of the SSMs generates the time series at a given time. 
The three models are general enough to be applicable to diverse fields. 
At the same time, they bear useful interpretations in dFC analysis 
relating to sources of brain signals, shared components versus regime-specific components of FC, and more. 
We evaluate these models on two EEG datasets relating to epilepsy and to brain-computer interfaces.

\item \emph{Software.} We provide a highly optimized MATLAB implementation of the EM algorithm for MS-SSMs. 
The toolbox also contains functions to simulate the models, extract regime-specific stationary quantities, 
and conduct parametric bootstrap. It is available at \url{https://github.com/ddegras/switch-ssm}.

\end{enumerate}

\subsection*{Organization of the paper} 

Section 2 gives a general account of MS-SSMs and details three specific models.
Section 3 describes the EM algorithm for MS-SSMs and develops novel techniques for initializing the EM, handling parameter constraints, accelerating convergence, and performing model selection.  
Section 4 introduces a new parametric bootstrap for the maximum likelihood estimator (MLE). 
Section 5 presents a simulation study that assesses the statistical accuracy of the MLE and bootstrap under different models and asymptotics. 
Methods like the sliding windows and oracle estimators are used as  benchmarks for comparison. 
Section 6 presents applications to EEG data from an epileptic seizure study and a motor imagery experiment. 
Section 7 contains a discussion and perspectives for future work.


\section{Linear state-space models with regime switching}
\label{sec:1}

Linear SSMs with regime switching can be viewed as a combination of linear SSMs (also known as linear dynamical systems) and of hidden Markov models  \citep[e.g.,][]{Kim1999,Murphy1998}. 
They are represented by an \emph{observation equation} and a \emph{state equation} as follows: 
\begin{equation} \label{switch-ssm}
\begin{split}
\by_t & =  \bC_{S_t} \bx_{t}+ \bw_t ,  \\
\bx_t & =  \bA_{S_t} \bx_{t-1} + \bv_t ,
\end{split}
\end{equation}
where $t$ is a time index, $\by_t$ an observation vector, 
 $\bw_t $ a measurement error vector, 
$\bx_t$ a hidden state vector,
and $\bv_t $ a random innovation vector. The numbers of observation variables, state variables, and time points are  denoted by $N$,  $r$, and $T$ respectively; typically $r\le N$. The switching variable $S_t$ indicates the regime under which system \eqref{switch-ssm} operates at time $t$. The sequence $(S_t)_ { t=1,\ldots, T}$ is a homogeneous Markov chain taking values in a finite set of regimes  $\{ 1, \ldots, M\}$, with initial state probabilities $\pi_j = P(S_1 = j)$ and transition probabilities $ Z_{ij} = P(S_t = j | S_{t-1}=i ) $ for $1\le i,j \le  M$. In dFC analysis, the regimes represent different modes or states of FC.  
Under regime $S_t=j$, the transition matrix $\bA_j$ governs the dynamics of the state vector $\bx_t$ and the observation matrix $\bC_j$  maps the hidden state vector $\bx_t$ to the observed measurements $\by_t$. 
Conditionally on the regime sequence $(S_t)$, the measurement errors $\bw_t$ are independent over time and have normal distribution $N(\bzero,\bR_{S_t})$. Also conditionally on $(S_t)$, the innovations $\bv_t$ are independent over time, mutually independent with the $\bw_t$, and have normal distribution $N(\bzero,\bQ_{S_t})$. At time $t$, if $S_t=j$, the parameters  at play in model \eqref{switch-ssm} are $(\bA_{j},\bC_{j},\bQ_{j},\bR_{j})$. 
The state vector $\bx_t$ can be interpreted as (a usually small number of) 
factors that drive the signal of interest in the observed data. In the context of neuroimaging, $\bx_t$ may represent  the neural sources of the measured  signals. With this perspective, the observation matrices $\bC_j$ could be the mixing matrices used in EEG/MEG source localization.

\begin{remark}\label{rk: identifiability}
In the absence of model constraints, parameters in (standard or switching) SSMs may  only be identifiable up to classes of invertible matrix transformations \citep{VanOverschee1996,Zhang2011}. 
For example in model \eqref{switch-ssm}, given invertible $r \times r$ matrices  $( \bB_j)_{1\le j \le M }$, the reparameterization defined by $\tilde{\bA}_j =\bB_j^{-1} \bA_j \bB_j  $, 
$ \tilde{\bC}_j = \bC_j \bB_j$, 
 $\tilde{\bQ}_j= \bB_j^{-1} \bQ_j (\bB_j^{-1})' $,
 $\tilde{\bR}_j = \bR_j$, 
 $ \tilde{\bmu}_j = \bB_j^{-1} \bmu_j$, 
 and $ \tilde{\bSigma}_j  = \bB_j^{-1} \bSigma_j ( \bB_j^{-1})' $ 
produces an observationally equivalent model. 
Because of this lack of identifiability, inferential procedures typically pertain to 
quantities such as the stationary covariance matrices $\lim_{t\to \infty}V(\by_t | S_{1}=\cdots = S_{t}= j) $, 
the transition probability matrix $\mathbf{Z}= (Z_{ij})_{1\le i,j\le M}$, or 
the projection matrices $\bC_j (\bC_j' \bC_j)^{-1} \bC_j' $  
which are uniquely determined up to permutations of regime labels. 
\end{remark}

Model \eqref{switch-ssm} is very general and needs to be tailored to the application at hand. 
Hereafter we present three instances of model \eqref{switch-ssm} particularly suitable for the dFC analysis of neuroimaging data. The first and third instances enable dimension reduction from the observation space (e.g.,   EEG channels, fMRI voxels or regions interest) to the state space of underlying signal sources. This is particularly useful with high dimensional measurements that are highly correlated such as high-density EEG.  The first instance posits a single underlying state process with regime-dependent dynamics while the third depicts multiple state processes, each  associated with a FC regime, with uncertainty as to which FC regime is active at a given moment. The second instance directly models dynamics in the observation space and is well suited for time series with relatively low dimension ($N \ll T$).

\subsection{Switching dynamics}
\label{sec:2}

A first  specification of the switching SSM \eqref{switch-ssm} is
 \begin{equation}\label{switch-dyn} 
 \begin{split}
\by_t & =  \bC \bx_t + \bw_t , \\
\bx_t & = \sum_{\ell =1}^p  \bA_{\ell S_t} \bx_{t-\ell} + \bv_t .
\end{split}
\end{equation}
This model
posits a common observation matrix $\bC$ and  error covariance  $\bR$ for all regimes $1,\ldots, M$. In other words the observation equation does not depend on the regime $S_t$. 
On the other hand, the dynamics of the state equation switch with $S_t$.  At time $t$, conditional on  $ S_t $, $\bx_t$ is a vector autoregressive process of order $p$ with transition matrices  $\bA_{\ell  S_t}$ ($ 1\le \ell \le p$  denotes the lag) 
and innovation covariance matrix $\bQ_{S_t}$. 
The dependencies between observations, state vectors, and regimes under the switching dynamics model \eqref{switch-dyn} 
are depicted in Figure \ref{fig: switching dynamics} (left panel). 

The assumption of an  observation matrix $\bC$ common to all regimes is not restrictive because the rank of $\bC$ can be made 
arbitrarily large so as to span the range of any collection of regime-specific observation matrices $( \bC_j)_{ 1\le j \le M}$.
The assumption that the distribution $N(\bzero,\bR)$ of the measurement errors $\bw_t$ is independent of the regime $S_t$ is justified by the fact that FC regimes only modulate the properties of brain signals, not of measurement errors, e.g.,   machine noise, artifacts, etc.

\begin{remark}\label{rk: switch-dyn}
It can be seen that model \eqref{switch-dyn} is a special case of model \eqref{switch-ssm} by expanding the state vector $\bx_t$  so as to include lags up to order $p-1$: 
$\widetilde{\bx}_{t} =\vecx(\bx_{t} ,\ldots, \bx_{t-p+1} ) \in\mathbb{R}^{pr}$, padding the innovation vector $\bv_t$ and observation matrix $\bC$ 
 with zeros: 
$\widetilde{\bv}_t =  \vecx(\bv_{t} ,\bzero_{ (p-1)r}) \in\mathbb{R}^{rp}$ and $\widetilde{\bC}= (\bC , \bzero_{N\times (p-1)r}) \in \mathbb{R}^{N \times pr}$, 
 and defining the $pr \times pr$ companion matrix  
\[ \widetilde{\bA}_{j} = 
\left( \begin{array}{c c c c} 
 \bA_{1 j} &  \bA_{2 j} & \cdots &  \bA_{p j} \\
\bI_r & \mathbf{0}_{r\times r} & \cdots & \mathbf{0}_{r\times r}  \\
\mathbf{0}_{r\times r} & \ddots & \ddots& \vdots \\
\mathbf{0}_{r\times r}&\mathbf{0}_{r\times r}&\bI_r & \mathbf{0}_{r\times r} 
\end{array} \right) \]   
for $1\le j\le M$, 
where $\mathbf{0}_{m\times n}$ denotes the $m\times n$ zero matrix 
and $\bI_r$ the  $r\times r$ identity matrix. 
 One recovers model \eqref{switch-ssm} with 
 $\by_{t} = \widetilde{\bC} \widetilde{\bx}_{t} + \bw_{t} $
 and $\widetilde{\bx}_{t} = \widetilde{\bA}_{S_t} \widetilde{\bx}_{t-1}+ \widetilde{\bv}_{t} $. 
\end{remark}

An ubiquitous special case of model  \eqref{switch-dyn} is the switching vector autoregressive (VAR) model 
\begin{equation}\label{switch-var} 
 \by_t = \sum_{\ell =1}^p  \bA_{\ell S_t} \by_{t-\ell} + \bv_t . 
\end{equation}
This model corresponds to the case where 
$\bC = \bI_N$ and $\bw_t = \bzero_N$  in \eqref{switch-dyn}, 
implying  that $\by_t = \bx_t$ (no hidden state vector).


\subsection{Switching observations}

\begin{figure*}[h]
\begin{center}
\includegraphics[width=.45 \textwidth]{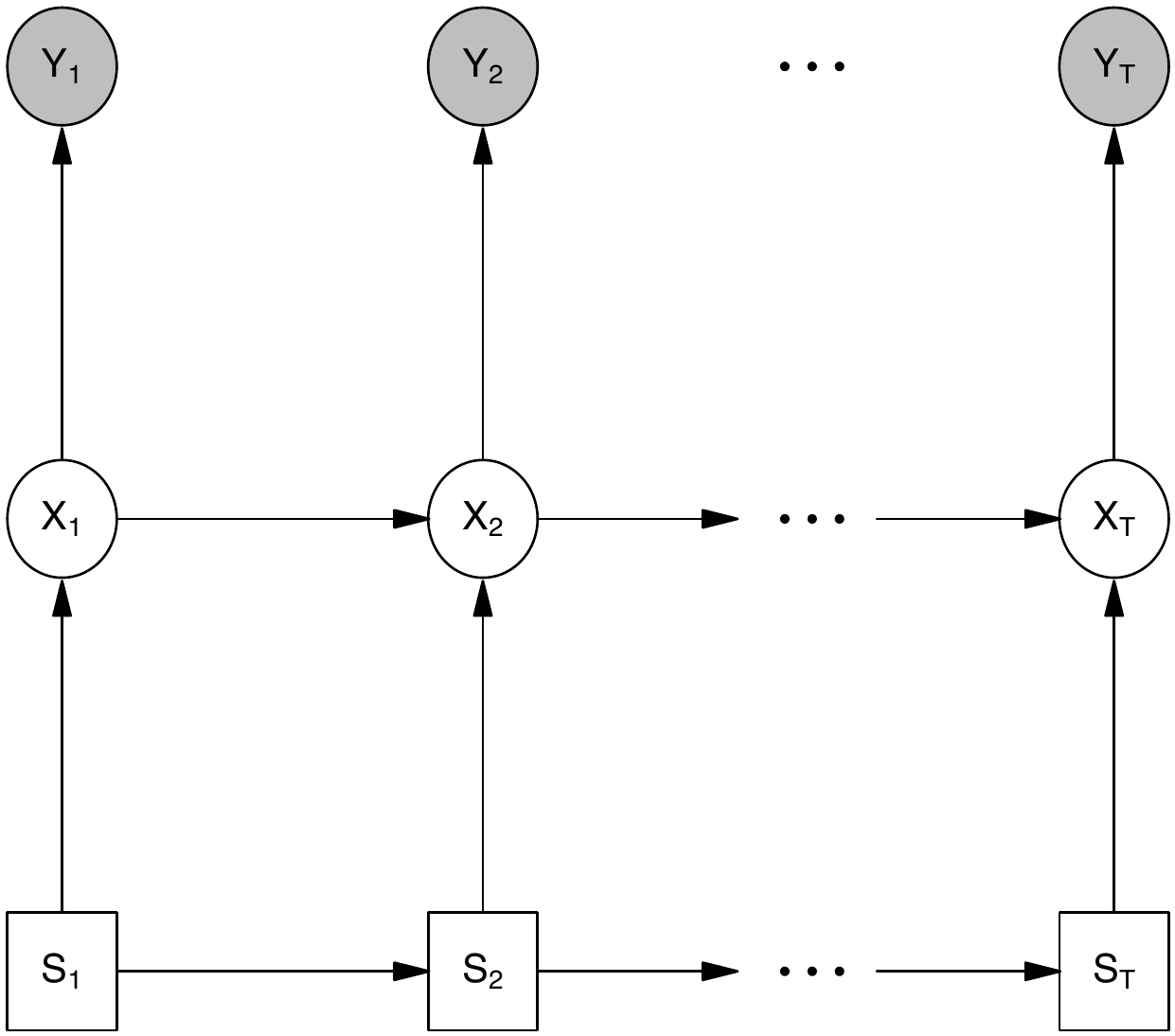} \hfill
\includegraphics[width=.45 \textwidth]{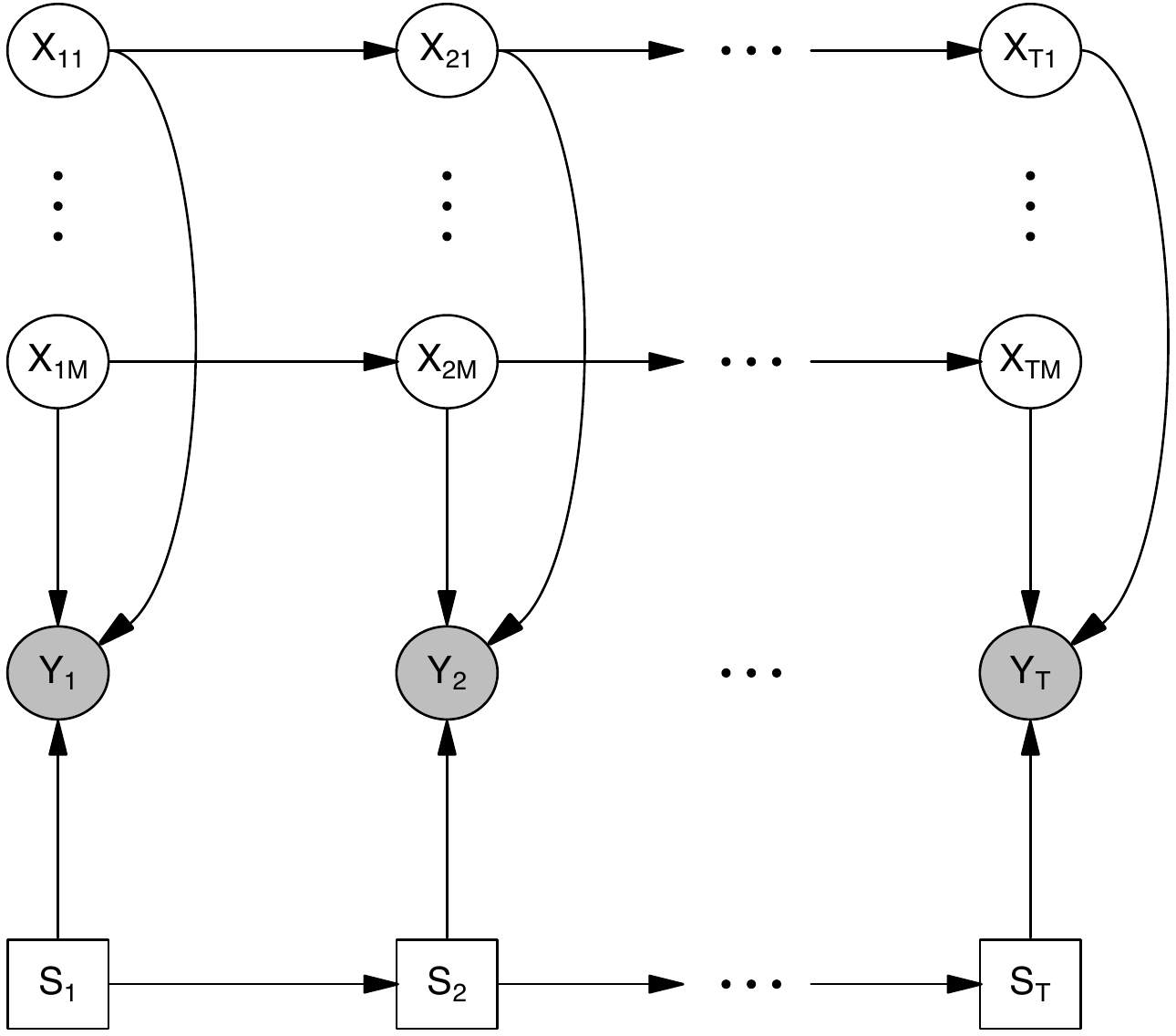}
\vspace*{-2mm}
\caption{Directed acyclic graph representation of the switching dynamics model \eqref{switch-dyn} (left) 
and  the switching observations model \eqref{switch-obs} (right).
Square nodes represent discrete variables and oval ones are Gaussian. 
Shaded nodes are observed and white ones are hidden.}
\label{fig: switching dynamics}
\end{center}
\end{figure*}

In addition to the switching dynamics and switching VAR models, 
the third specification of model \eqref{switch-ssm} we develop in this paper 
is 
\begin{equation}\label{switch-obs}
\begin{split}
\by_t & =  \bC_{S_t} \bx_{t  S_t}+ \bw_t ,\\
\bx_{t j }  & = \sum_{\ell =1}^p  \bA_{\ell j} \bx_{(t-\ell ) j} + \bv_{t j} ,  \quad 1 \le j \le M.
\end{split}
\end{equation}
This model is viewed as a \emph{mixture of experts} by \cite{Ghahramani2000} and as a  \emph{switching observations model with factored states} by \cite{Murphy1998}. 
Both the observation matrix and observed state vector depend on the regime $S_t$. There are in fact $M$ different state vectors $\bx_{tj}, \, 1\le j \le M,$ each evolving independently according to a VAR($p$) model determined by the $\bA_{\ell j}$ and $\bQ_j$. At time $t$, only one of these state vectors generates the observations through $\bC_{S_t}$. 
In neuroimaging this situation can be interpreted as follows: for each $j$ the process $(\bx_{tj})_{1\le t \le T}$ represents \emph{potential} neural activity or source signal associated with the $j$th mode of functional connectivity. At each time $t$, the brain is in only one \emph{actual} mode of connectivity or regime $S_t$ so that only $\bx_{t S_t}$ is observed through the regime-specific observation matrix $\bC_{S_t}$. 
To be clear, the notions of potential and actual FC regimes do not relate to intrinsic brain function: they simply reflect the scientist's uncertainty about the FC regime at a given time. In this sense model \eqref{switch-obs} is strongly tied to the concept of  \emph{association  ambiguity} which also arises in areas such as computer vision and radar systems \citep{Fortmann1983,Shumway1991}.   
The dependencies between observations, state vectors, and regimes under model \eqref{switch-dyn} 
are depicted in Figure \ref{fig: switching dynamics} (right panel).

\begin{remark}\label{rk: switch-obs}
Model \eqref{switch-obs} can be expressed in the form \eqref{switch-ssm} by    
expanding the state vectors $\bx_{tj}$ to include lags up to order $p-1$ and 
concatenating them across regimes, as well as suitably expanding/padding other 
quantities with zeros. More precisely, writing  
$\widetilde{\bx}_t = \vecx (  \bx_{t 1},\ldots, \bx_{(t-p+1) 1} , \bx_{t2} , \ldots )  \in \mathbb{R}^{Mpr}$, 
$\widetilde{\bv}_t = \vecx(\bv_{t 1} ,\bzero_{(p-1)r} , \ldots , \bv_{tM},\bzero_{ (p-1)r})  $,  
$ \widetilde{\bA} = \mathrm{diag}(\widetilde{\bA}_{1} , \ldots, \widetilde{\bA}_{M}) \in \mathbb{R}^{Mpr \times Mpr }$ with   $\widetilde{\bA}_{1} , \ldots, \widetilde{\bA}_{M}$ as in Remark \ref{rk: switch-dyn}, and 
$\widetilde{\bC}_{j} = [ \bzero_{N\times (j-1)pr} ,  \bC_j ,  \bzero_{N\times (p-1+(M-j)p) r} ] \in \mathbb{R}^{N \times Mpr}$,
model \eqref{switch-obs} now reads as $\by_t = \widetilde{\bC}_{S_t} \widetilde{\bx}_t  + \bw_{t} $ and  $\widetilde{\bx}_{t}  = \widetilde{\bA} \widetilde{\bx}_{t-1} + \widetilde{\bv}_{t} $. Other model parameters ($\bmu_j, \bSigma_j$) are redefined in a similar way.     
\end{remark}

\section{Model fitting by the EM algorithm}
\label{sec: EM}

This section starts with a review of the EM algorithm in the general switching SSM \eqref{switch-ssm}. 
A full treatment can be found in \cite{Kim1994} and \cite{Murphy1998}. 
We then propose new methods for initializing the EM in 
models \eqref{switch-dyn}-\eqref{switch-var}-\eqref{switch-obs}. 
We also examine the question of model selection 
and offer practical insights into the EM, 
pointing out common convergence issues  as well as solutions,  
among which an acceleration scheme.

Let $\btheta = \{ (\bA_j, \bC_j, \bQ_j ,\bR_j , \bmu_j, \bSigma_j): 1\le j \le M ; 
\boldsymbol{\pi} ; \mathbf{Z} \}$  be the collection of all  parameters in model \eqref{switch-ssm}, 
with $\boldsymbol{\pi} = (\pi_1,\ldots,\pi_M)'$. 
For brevity we denote the measurements $(\by_t)_{1\le t\le T}$ by $\by_{1:T}$, 
the state vectors $(\bx_{t})_{1\le t\le T}$ by $\bx_{1:T}$, etc. 
We recall that   $\by_{1:T}$ are observed 
whereas   $\bx_{1:T}$ and  $S_{1:T}$ are unobserved. 
We denote the (incomplete) likelihood function by $L(\btheta)$ and the complete likelihood function 
(i.e., if $\by_{1:T} ,  \bx_{1:T}, S_{1:T}$ were all observed) by $L_c(\btheta)$. 
We also denote the probability measure associated to model \eqref{switch-ssm} by $P_{\btheta}$
and expectation under $P_{\btheta}$ by $E_{\btheta}$. 
In the absence of ambiguity, we use the same notation $\btheta$ for 
the true model parameters and for the generic argument/subscript of $L,L_c,P,E$.

\subsection{E-step}

 The complete data log-likelihood  expresses as
 \begin{equation}\label{complete LL}
\begin{split}
 \log  L_c(\btheta )& = \log P_{\btheta}( \by_{1:T} ,  \bx_{1:T}, S_{1:T})  \\
& = \log P_{\btheta}( \by_{1:T} |  \bx_{1:T}, S_{1:T}) + \log P_{\btheta}(\bx_{1:T} | S_{1:T}) + \log P_{\btheta}(S_{1:T})  \\
& = \sum_{t=1}^T \log \phi( \by_t ; \bC_{S_t} \bx_t , \bR_{S_t}) + 
\sum_{t=2}^T \log \phi ( \bx_t ; \bA_{S_t} \bx_{t-1} , \bQ_{S_t})  \\ 
& \quad + \log \phi(\bx_1 ; \bmu_{S_1} , \bSigma_{S_1}) + \log \pi_{S_1}  + \sum_{t=2}^{T} \log P_{\btheta}(S_t | S_{t-1}) 
\end{split}
\end{equation}
where $\phi (\bx;\mathbf{m},\mathbf{V} ) = (2\pi |\mathbf{V}|)^{-d/2} \exp( - (\bx-\mathbf{m})'\mathbf{V}^{-1}(\bx-\mathbf{m}) / 2)$ is the density function of a multivariate normal distribution in $\mathbb{R}^d$
with mean vector $\mathbf{m}$ and covariance matrix $\mathbf{V}$. 

Given a current estimate $\widehat{\btheta}$ of $\btheta$, 
the E step consists in taking the conditional expectation of the complete data log-likelihood 
\eqref{complete LL} with respect to the observed data $\by_{1:T}$ while assuming 
that $\widehat{\btheta}$ is the true model parameter. This produces the $Q$-function 
\begin{equation*}
Q(\btheta ; \widehat{\btheta}) = E_{\widehat{\btheta}} \left(\log L_c(\btheta ) | \by_{1:T} \right)
\end{equation*} 
whose calculation requires the elements 
{\small 
\begin{equation}\label{skfs}
\hspace*{-3mm}
\begin{array}{r l r l }
W_{t|\tau}^j  \hspace*{-2mm} & = P_{\widehat{\btheta}}(S_t = j \, | \, \by_{1:\tau})  ,&
 W_{t-1,t|\tau}^{ij} \hspace*{-2mm} &= P_{\widehat{\btheta}}(S_{t-1} = i , S_{t} = j \, | \, \by_{1:\tau})  , \\
\bx_{t | \tau}^j \hspace*{-2mm} & = E_{\widehat{\btheta} } \big(  \bx_{t} \,  | \,S_t =j,   \by_{1:\tau}  \big) , &
\bP_{t  | \tau }^j \hspace*{-2mm} &=  E_{\widehat{\btheta} } \big(  \bx_{t} \bx_{t}' \,  | \, S_t =j,  \by_{1:\tau}  \big) , \\
\bP_{t -1 | \tau }^{()j} \hspace*{-2mm} &=  E_{\widehat{\btheta} } \big(  \bx_{t-1} \bx_{t-1}' \,  | \, S_t =j,  \by_{1:\tau}  \big)  ,&
\bP_{t,t -1 | \tau }^j \hspace*{-2mm} &=  E_{\widehat{\btheta} } \big(  \bx_{t} \bx_{t-1}' \,  | \, S_t =j,  \by_{1:\tau}  \big),
\end{array}
\end{equation}}
 where $\tau = t-1$ for prediction, $\tau = t$ for filtering, and $\tau = T$ for smoothing.  
The quantities in \eqref{skfs} can be computed approximately 
with the Kim filtering algorithm, also known as Hamilton filtering. 
Details of this procedure can be found e.g.,   in 
\cite{Kim1994} and are omitted here.   
In essence, the calculation of \eqref{skfs} relies on a forward-backward algorithm similar to the Kalman filter/smoother for SSMs. 
Unlike the Kalman filter however, exact calculations in the filtering step would require considering 
at each time $t$  all $M^t$ possible values of the switching variables $S_1,\ldots,S_t$, 
which is not tractable (the computational cost grows exponentially with time). 
Instead, at each  $t$ the Kim filter only considers the recent history 
of the switching variables (say, the $M^2$ possible values of $(S_{t-1},S_t)$) 
to calculate relevant probabilities and expected values, which it then collapses  
to (approximately) recover \eqref{skfs}  (for $\tau=t$). 
Further approximations are involved in the smoothing step 
to make calculations tractable. 
In practice the Kim filter  saves considerable computational time 
and provides accurate approximations to the $Q$-function and log-likelihood, 
at least in low dimensional settings.  
Higher accuracy can be obtained 
at a computational cost by utilizing more recent past regimes, 
say $S_{t-1},\ldots,S_{t-m}$ for some $m \ge 2$.

With the notations of \eqref{skfs}, the $Q$-function expresses as
{\small
  \begin{equation}\label{Q-function}
  \begin{split}
& Q(\btheta ; \widehat{\btheta}) =  -\frac{T(N+r)}{2} \log(2\pi) \\
 & +  \frac{1}{2} \sum_{t=1}^T 
 \sum_{j=1}^M W_{t | T}^j  \left(  \log \big| \bR_j^{-1}  \big| -  \tr  \bR_j^{-1} 
 \big( \by_t \by_t ' -2  \bC_j \bx_{t|T}^j \by_t' + \bC_j\bP_{t | T}^j \bC_j'\big) \right) \\
&  + \frac{1}{2} \sum_{t=2}^T  \sum_{j=1}^M W_{t|T}^{j} \left( \log \big|   \bQ_j^{-1} \big|  -
\tr \bQ_j^{-1}  \big( \bP_{t | T }^j - 2 \bP_{t,t-1|T}^{j}\bA_j'+ \bA_{j} \bP_{t-1 | T }^{()j} \bA_{j}' \big) \right) \\
&  + \frac{1}{2} \sum_{j=1}^M W_{1|T}^j  \left( \log \big| \bSigma_j^{-1} \big| - \tr \bSigma_j^{-1} \big( 
\bP_{1|T}^j - 2 \bx_{1|T}^j \bmu_j' + \bmu_j \bmu_j' \big) \right) \\
& +  \sum_{j=1}^M W_{1|T}^j  \log \pi_j + 
\sum_{t=2}^T \sum_{i=1}^M\sum_{j=1}^M W_{t-1,t|T}^{ij}  \log \pi_{ij} .
\end{split}
\end{equation}}

\subsection{Incomplete data log-likelihood} 

\begin{subequations}
The Kim filtering algorithm gives the approximate (incomplete data) log-likelihood as a by-product of \eqref{skfs}. 
Writing abusively $P_{\widehat{\btheta}}$ for density functions, we have
\begin{equation}
 \log L(\widehat{\btheta}) = \log P_{\widehat{\btheta}} (\by_1) + \sum_{t=2}^T \log P_{\widehat{\btheta}} \left(\by_t | \by_{1:(t-1)} \right) ,
 \end{equation}
 where 
 \begin{equation}
 \begin{split}
 P_{\widehat{\btheta}} (\by_1)&  = \sum_{j=1}^M  P_{\widehat{\btheta}}(\by_1| S_1 = j)  P_{\widehat{\btheta}}( S_1 = j) \\
 & = \sum_{j=1}^M   \phi(\by_1 ; \widehat{\bC}_j \bmu_j , \widehat{\bC}_j \widehat{\bSigma}_j\widehat{\bC}_j ' +\widehat{\bR}_j ) \, \widehat{\pi}_j
\end{split}
\end{equation}
and 
\begin{equation}
\begin{split}
P_{\widehat{\btheta}} (\by_t | \by_{1:(t-1)} ) &  =   \sum_{j=1}^M 
P_{\widehat{\btheta}} (\by_t | \by_{1:(t-1)},S_t=j ) P_{\widehat{\btheta}} ( S_t=j | \by_{1:(t-1)}) \\
&= \sum_{j=1}^M  \phi(\by_t ; \widehat{\bC}_j \bx_{t|t-1}^j , \widehat{\bC}_j \mathbf{V}_{t|t-1}^j  \widehat{\bC}_j ' +\widehat{\bR}_j ) \,
W_{t | t-1}^{j}
\end{split}
\end{equation}
with $\mathbf{V}_{t|t-1}^j := V(\bx_t | \by_{1:(t-1)},S_t=j) = \bP_{t|t-1}^j - \bx_{t|t-1}^j(\bx_{t|t-1}^j)'$ the predictive variance-covariance matrix of $\bx_t$ associated with the regime $S_t=j$. 
\end{subequations}

\subsection{M-step}

The M-step consists in maximizing the $Q$-function \eqref{Q-function} with respect to $\btheta$. 
In the absence of constraints on $\btheta$, this maximization amounts to a simple least squares problem in linear regression.   
Accordingly the unconstrained solutions express as
\begin{equation}\label{unconstrained estimate}
\begin{array}{ll}
 \widehat{\bA}_j  & = \big( \sum_{t=2}^T W_{t|T}^j \bP_{t,t-1|T}^j \big) \big( \sum_{t=2}^T W_{t|T}^j \bP_{t-1|T}^{()j} \big)^{-1} ,\\
 \widehat{\bC}_j & = \big( \sum_{t=1}^T W_{t|T}^j \by_t \big(\bx_{t|T}^j \big)' \big)\big( \sum_{t=1}^T W_{t|T}^j \bP_{t|T}^j \big)^{-1} ,\\
 \widehat{\bQ}_j & = \big( \sum_{t=2}^T W_{t|T}^j\big)^{-1} \big( \sum_{t=2}^T W_{t|T}^j \bP^j_{t|T} -\widehat{\bA}_j \sum_{t=2}^T W_{t|T}^j \big(\bP_{t-1,t | T}\big)' \big) ,\\
 \widehat{\bR}_j & = \big( \sum_{t=1}^T W_{t|T}^j\big)^{-1}   \big(
 \sum_{t=1}^T W_{t|T}^j  \by_t \by_t'  - \widehat{\bC}_j  \sum_{t=1}^T W_{t|T}^j  \bx_{t|T}^j \by_t'  \big) , \\
 \widehat{\bmu}_j & = \bx_{1|T}^j , \qquad 
  \widehat{\bSigma}_j = \widehat{\bP}_{1|T}^j - \widehat{\bmu}_j \widehat{\bmu}_j' ,\\
\widehat{\pi}_j & = W_{1|T}^j , \qquad  
\widehat{Z}_{ij}  =\big( \sum_{t=2}^T W_{t-1 ,t |T}^{ij} \big)  \big( \sum_{t=2}^T W_{t-1  |T}^{i} \big)^{-1} .
\end{array}
\end{equation}

We now discuss the maximization of  \eqref{Q-function}  under constraints. 
\begin{itemize}
\item 
\emph{Fixed coefficient constraints on $\bA_j,  \bC_j$.} 
 Maximizing  \eqref{Q-function} with respect to $\bA_j$ or $\bC_j$ subject to fixed coefficient constraints 
amounts to minimizing a function of the form $f(\bX)=\tr \{\mathbf{W}(-2\mathbf{B}_1\mathbf{X}' + \mathbf{XB}_2\mathbf{X}')\} $ 
under the same constraints, 
where $\bX = \bA_j$ or $\bX=\bC_j$ and  the matrices
$\bB_1, \bB_2,\mathbf{W}$ do not depend on $\mathbf{X}$. 
 Vectorization yields 
 $f(\vecx (\bX))= -2\vecx(\bX)' \vecx(\mathbf{W}\bB_1) + \vecx(\bX)' (\bB_2 \otimes \mathbf{W})\vecx(\bX)$ 
where $\otimes$ is the Kronecker product. After deleting the entries of $\vecx(\mathbf{W}\bB_1)$, resp. rows and columns of $\bB_2 \otimes \mathbf{W}$, that correspond to the fixed coefficients of 
$\vecx(\bX)$, it suffices to minimize the resulting quadratic form to obtain the free coefficients 
of $\vecx(\bX)$, say, $\vecx(\bX)_{free} = (\bB_2 \otimes \mathbf{W})_{free}^{-1}  \vecx(\mathbf{W}\bB_1)_{free}$. 

\item \emph{Fixed coefficient constraints on $\bQ_j, \bR_j, \bSigma_j$.}   
These constraints are more difficult to handle as they do not readily translate to the inverse covariance matrices $\bQ_j^{-1}, \bR_j^{-1} , \bSigma_j^{-1}$ that appear in \eqref{Q-function}. It seems reasonable to restrict fixed coefficient constraints on these parameters to diagonality constraints (i.e., fix all off-diagonal terms to zero) or to fix coefficients in inverse covariances, in which case the optimization is straightforward.

\item 
\emph{Scaling constraints.} The requirement that  columns of  $\bC_j$ have norms of given length, say one, is conveniently handled with the projected gradient method. This optimization method consists in alternating a gradient step 
(to improve the objective function) and a projection step (to enforce parameter constraints) until convergence. 
Here, the projection step is particularly simple as it simply consists in rescaling the columns of the current solution matrix. 

\item \emph{Eigenvalue constraints on $\bA_j$.} The VAR process associated with $\bA_j$ is stable, and hence stationary, if the (complex) eigenvalues of $ \bA_{j}$ are less than 1 in modulus. However, the exact optimization of \eqref{Q-function}, a quadratic form in $\bA_j$, under eigenvalue constraints is a difficult convex optimization problem (especially if $p>1$ as in this case, $\bA_j$ is subject to additional fixed  coefficients constraints; see Remark \ref{rk: switch-dyn}). Instead we adopt the following approach: (i) estimate $\bA_j$ freely as in \eqref{unconstrained estimate} (this formula applies as is when $p=1$ in models \eqref{switch-dyn} and \eqref{switch-obs} and adapts easily when $p>1$), (ii) calculate the spectral radius $\rho(\bA_j) $, i.e., the largest modulus of the  eigenvalues of $\bA_j$, (iii) if $\rho(\bA_j) \ge 1$, multiply the matrices $\bA_{\ell j}$ that make up $\bA_j$ by $((1-\epsilon)/\rho(\bA_j))^\ell$ for $1\le \ell \le p$ and some small $\epsilon \in(0,1)$, (iv) if the resulting matrix increases \eqref{Q-function} in comparison to the current $\widehat{\bA}_j$, update $\widehat{\bA}_j$; otherwise, do  not update. 
The shrinkage step (iii) guarantees that the resulting matrix has eigenvalues at most  $1-\epsilon$ in modulus. 
Although this matrix does not in general maximize \eqref{Q-function} under the eigenvalue constraint, it typically leads to an increase.  
To our knowledge, this algebraic tool has not been described elsewhere in the literature 
and it is the only available method to enforce stationarity of a VAR process in the maximum likelihood framework.

\item \emph{Equality constraints across regimes.} If  one assumes, for example, that  $\bC_1=\cdots = \bC_M$ in \eqref{switch-obs}, 
their common value $\bC$ is estimated by   $\widehat{\bC} =  \big( \sum_{t=1}^T \by_t (\bx_{t|T} )' \big) \big( \sum_{t=1}^T  \bP_{t|T} \big)^{-1} $. If one assumes equality for the $\bA_j$ but not for the $\bQ_j$ ($ 1\le j\le M$), one can first optimize the $Q$-function with respect to the common value $\bA$ while holding the $\bQ_j$ fixed at their current values $\widehat{\bQ}_j$ (this can be done with the vectorization method described in the first item above) and then update the $\widehat{\bQ}_j$ while holding the new $\widehat{\bA}$ fixed. One could also go the other way around and first update the $\widehat{\bQ}_j$, then $\widehat{\bA}$. Either way, this conditional maximization step is guaranteed to not decrease the $Q$-function \eqref{Q-function}.

\end{itemize}

\subsection{Initialization}
\label{subsec: EM initialization}

Given that the EM algorithm is only guaranteed to  converge to a stationary point of the likelihood function $L$ \citep{McLachlan2008,Wu1983}, 
choosing good starting points is essential to increase the chances of convergence to a global maximum. 
We first propose an initialization method for the switching dynamics model \eqref{switch-dyn} 
which also applies to the  switching VAR model \eqref{switch-var}. 
We then adapt this method to the switching observations model \eqref{switch-obs}. 
The hyperparameters $(M,p,r)$ are assumed to be fixed. 

\paragraph{Initialization for  model \eqref{switch-dyn}.}

\begin{enumerate}
\item After centering the rows of the data matrix $\mathbf{Y} = (\by_1,\ldots,\by_T)$, 
perform the singular value decomposition (SVD) $\mathbf{Y=UDV'} $. 
Denoting by $\mathbf{D}_r$ the diagonal matrix containing the $r$ largest singular values in $\mathbf{D}$ 
and by $\mathbf{U}_r$ and $\mathbf{V}_r$ the associated submatrices of $\mathbf{U}$ and $\mathbf{V}$ (singular vectors),
set $\widehat{\bC} =\mathbf{U}_r$ and $\widehat{\mathbf{X}} = (\widehat{\bx}_1, \ldots, \widehat{\bx}_T) = \mathbf{D}_r\mathbf{V}_r'$.

\item Initialize $ \widehat{\bR} $ as the diagonal matrix containing the row-wise sample variances 
 of the residual matrix $\mathbf{Y} - \widehat{\bC} \widehat{\bX}$. 

\item 
For $j=1,\ldots, M$, take  $\widehat{\bmu}_j $ as  the sample mean of $\widehat{\bx}_1,\ldots , \widehat{\bx}_{p}$. 
If $p>1$, take each $\widehat{\bSigma}_j $ as the diagonal matrix containing the component-wise sample variances of these vectors; 
otherwise, set  $\widehat{\bSigma}_j = \bI_r$.

\item Partition  $\{ 1,\ldots,T\}$ into intervals 
$\mathcal{T}_1,\ldots, \mathcal{T}_{\kappa}$ of approximately equal length 
 for some $\kappa > 0$. 
For $k=1,\ldots, \kappa$, fit a VAR($p$) model to $(\widehat{\bx}_t)_{t\in \mathcal{T}_k}$
by ordinary least squares (OLS). Call $\widehat{\bA}^{(k)}= (\widehat{\bA}^{(k)}_1,\ldots,\widehat{\bA}^{(k)}_p)$ the estimated transition matrices and  $\widehat{\bQ}^{(k)}$  the residual covariance matrix. 

\item Partition  the $\kappa$ pairs $(\widehat{\bA}^{(k)},\widehat{\bQ}^{(k)}) $ into $M$ clusters with the $K$-means algorithm.  
Refit a VAR($p$) model by OLS to each of the $M$ subsets $\{ \widehat{\bx}_t\}$ 
defined by the clusters. Take the resulting $(\widehat{\bA}_j,\widehat{\bQ}_j), \, j=1,\ldots, M$
as initial estimates for the EM.    

\item Let  $\widehat{S}_t$ be the estimated regime at time $t$: $\widehat{S}_t=j$ if $ t \in \mathcal{T}_k$ and $(\widehat{\bA}^{(k)},\widehat{\bQ}^{(k)})$ belongs to cluster $j$. For $1\le i,j \le M$, set  $  \widehat{\pi}_j = 1$ if $\widehat{S}_1=j$,  $ \widehat{\pi}_j = 0 $ otherwise, and set $ \widehat{Z}_{ij} = \frac{ \# \{ t: \widehat{S}_{t-1}=i  ,\widehat{S}_t = j\} }{  \# \{ t: \widehat{S}_{t-1}=i \} }$ if $ \# \{ t: \widehat{S}_{t-1}=i \} > 0$, $Z_{ij} = 1/M$ otherwise. 
 \end{enumerate}
The number $\kappa$ of intervals in step 4 expresses a tradeoff between parameter estimation and change point detection. It should be large enough so that most $\mathcal{T}_k$ contain no change points but small enough to enable reasonably accurate estimation of model parameters. The switching VAR model \eqref{switch-var},  a special case of model \eqref{switch-dyn}, 
can be initialized with the above method by skipping steps 1-2-3 and setting $ \widehat{\bx}_t =\by_t$.

\begin{remark}
An alternative to step 4 above is  binary segmentation: first 
fit a $\mathrm{VAR}(p)$ model to $(\widehat{\bx}_t)$ over the range $\{ 1,\ldots, T\}$, 
yielding a sum of squared errors $\mathrm{SSE}(1,T)$. Then, for $1< t \le T$, 
 fit a VAR($p$) over each subinterval $\{1,\ldots, t-1\}$ and $\{t ,\ldots ,  T\}$, yielding a total sum of squares $\mathrm{SSE}(1,t-1) + \mathrm{SSE}(t,T)$. Select  $\tau  = \argmin_{1< t \le T} \{\mathrm{SSE}(1,t-1) + \mathrm{SSE}(t,T)\}$ as a candidate change point. 
If the reduction in SSE is sufficient, say, $(\mathrm{SSE}(1,\tau-1) + \mathrm{SSE}(\tau,T)) \le (1-\epsilon)\mathrm{SSE}(1,T) $ 
for some small $\epsilon >0$,   accept $\tau$ as a change point and split the initial time range $\{1,\ldots ,T\}$  into  $\{1,\ldots, \tau-1\}$ and $\{ \tau,\ldots , T \}$. 
 Repeat the process for each new subinterval and so on so forth until no new change points are found. 
 In practice, the tolerance $\epsilon$ can be selected by trial and error until a reasonable number $\kappa$ of segments has been obtained or through formal hypothesis testing. One may also impose a minimal distance between successive change points for faster computations and more sensible results.
 \end{remark}

\paragraph{Initialization for model \eqref{switch-obs}.}

\begin{enumerate}
\item Build predictors $(\widehat{S}_t)$ of the regimes $(S_t)$ 
with the initialization method for model \eqref{switch-dyn}. 
 For $1\le j \le M$, concatenate the vectors $(\by_{t})_{t\in \mathcal{T}_j}$ into a $N\times T_j$ matrix $\mathbf{Y}^{(j)}$ 
where $\mathcal{T}_j = \{ t: \widehat{S}_t = j\}$ and $T_j = \# \mathcal{T}_j$;  
compute the SVD decomposition $\mathbf{Y}^{(j)} = \mathbf{U}^{(j)}\mathbf{D}^{(j)}(\mathbf{V}^{(j)})'$;  
and set $\widehat{\bC}_j = \mathbf{U}_r^{(j)}$ and $\widehat{\mathbf{X}}^{(j)} = (\widehat{\bx}_{tj})_{t\in\mathcal{T}_j} = \mathbf{D}_r^{(j)}(\mathbf{V}_r^{(j)})'$  using the $r$ largest singular values of $\mathbf{D}^{(j)}$ and the associated singular vectors in $\mathbf{U}^{(j)}$ and $\mathbf{V}^{(j)}$.

\item 
For $1\le j \le M$, define $ \widehat{\bR}_j $ as the sample covariance of the residuals 
$\by_t - \widehat{\bC}_j \widehat{\bx}_{tj}, \, t\in \mathcal{T}_j$. Initialize $\widehat{\bR}$ as the weighted average 
of the $\widehat{\bR}_j$ with weights $w_j=T_j/T$.  

For $1\le j \le M$, take  $\widehat{\bmu}_j $ as  the sample mean of $\widehat{\bx}_{1j},\ldots , \widehat{\bx}_{pj}$. 
If $p>1$, take each $\widehat{\bSigma}_j $ as the diagonal matrix containing the component-wise sample variances of these vectors; 
otherwise, set  $\widehat{\bSigma}_j = \bI_r$.

\item For $1\le j \le M$, fit a VAR($p$) model to $(\widehat{\bx}_{tj})_{t\in \mathcal{T}_j}$ 
by OLS; call  $\widehat{\bA}_{j} = (\widehat{\bA}_{1j}, \ldots,\widehat{\bA}_{pj})$ the estimated transition matrices 
and take $\widehat{\bQ}_{j}$ as the residual covariance matrix. 

\item Define the initial probability estimates $\widehat{\pi}_j$ and transition probability estimates $\widehat{Z}_{ij}$ 
as in step 5 of the initialization method for model \eqref{switch-dyn}. 

\end{enumerate}

\subsection{Model selection} 

When fitting models \eqref{switch-dyn}-\eqref{switch-obs} to data, 
depending on analytic goals, 
one or several techniques can be utilized to   
select the hyperparameters $(M,p,r)$. 
A common approach based on  model complexity is to
minimize the Akaike Information Criterion (AIC) or the Bayesian Information Criterion (BIC)
\begin{equation}\label{aic-bic}
-2\log L(\widehat{\btheta};M,p,r) + k \left( \textrm{\# free parameters}\right) 
\end{equation}
with respect to $(M,p,r)$, where $\widehat{\btheta}$ is the maximum likelihood estimator, $k=2$ for AIC and $k=\log(T)$ for BIC. 
Assuming no coefficient constraints, the number of free parameters is approximately  
$O(Mpr^2 +M^2+N^2)$ in models \eqref{switch-dyn}-\eqref{switch-var} 
and $O(Mpr^2+M^2+N^2+MNr)$ in model \eqref{switch-obs}. 
As in other statistical contexts, these model complexity approaches seem to work well in ``standard" situations where the number of model parameters and time series dimension $N$ are small compared to the time series length $T$. Adaptation may be required in high-dimensional setups, for example in fMRI data analysis.

Model selection may also be based on predictive ability.  
One may for example seek to 
minimize the average one-step ahead prediction error 
\begin{equation}\label{MAPE} 
\frac{1}{(T-p)r} \sum_{t=p+1}^{T} \left\| \by_t - \by_{t|t-1} \right\|_1 
\end{equation}
where 
$ \by_{t|t-1} = \widehat{\bC} \bx_{t|t-1} $    
in model  \eqref{switch-dyn},  
 $ \by_{t|t-1} =  \sum_{j=1}^{M} \widehat{Z}_{(\widehat{S}_{t-1}) j} \sum_{\ell=1}^{p} \widehat{\bA}_{j \ell } \by_{t-\ell } $ 
in model \eqref{switch-var},   and 
$ \by_{t|t-1} =  \sum_{j=1}^M   \widehat{Z}_{(\widehat{S}_{t-1})j} \widehat{\bC}_j \bx_{tj | t-1} $
in model  \eqref{switch-obs}. This criterion is convenient as 
all necessary elements are already calculated in the original model fit (unlike, say, cross-validation, which requires multiple  fits). 

It may be useful to complement the process of model selection 
 using \eqref{aic-bic} or \eqref{MAPE} with model diagnostics based on visual examination or hypothesis tests.  
For example, inspection of the dwell times $\frac{1}{T}\sum_{t=1}^T P_{\widehat{\btheta}}(S_t = j | \by_{1:T}) , \, 1\le j \le M,$ 
may suggest to reduce $M$ if some dwell times are too small. 
Also, the presence of substantial autocorrelation in the residuals of fitted state vectors 
(e.g.,   $\widehat{\bv}_t = \bx_{t|T} - \sum_{j=1}^M W_{t|T}^{j} \sum_{\ell = 1}^{p} \widehat{\bA}_{\ell j} \bx_{(t-\ell)|T} $ in model \eqref{switch-dyn}) or of fitted observation vectors may lead one to increase $p$ or $r$, respectively.

\subsection{Convergence of the EM} 
\label{sec: EM issues}

We list here 
numerical issues that may  arise in fitting switching SSMs with the EM algorithm and propose solutions.

\begin{enumerate}

\item \emph{Non-monotonicity.} 
To be tractable, the calculation of the likelihood function $L$ and conditional expectation function 
$Q$ relies on several approximations (principally, collapsing information on the regime histories $(S_t)$). As a result, although  successive EM iterations never decrease $L$  in theory, this may happen in practice  especially in the switching observations model \eqref{switch-obs}  which contains a large proportion of latent variables. To handle this issue, we terminate the EM if there is insufficient increase in $L$ in 5 consecutive iterations. 
Also, in the M step, updates of parameter estimates are realized only if they increase $Q$.

\item \emph{Convergence to local maxima of the likelihood function.} 
This well-known problem of the EM can be diagnosed  by randomly perturbing the final parameter estimate, re-running the EM, and checking whether it returns to same solution. 
One way to mitigate the problem is to use multiple start points, for example a ``cold" start point obtained as in Section \ref{subsec: EM initialization} and a ``warm" start point obtained from a previous model fit with lower $M$, $p$, or $r$.
 Another solution is the \emph{deterministic annealing  EM} \citep[DAEM,][]{Ueda1998},  
 which replaces the smoothed regime probabilities $P(S_t = j| \by_{1:T})$ by $P(S_t = j| \by_{1:T})^\beta$ and rescales them, where $0 < \beta \le 1$ is an inverse temperature parameter that increases to 1 over iterations ($\beta=1$ corresponds to the regular EM). The initial flattening of regime probabilities allows the DAEM to explore the huge space of $M^T$ possible regime sequences $(\widehat{S}_t)$ for long enough before converging to a likely sequence.

\item \emph{Slow convergence.} The EM is known to  converge slowly, a problem  compounded by the size of $\btheta$ and the number of latent variables $(\bx_t), (S_t)$.  
Although acceleration schemes exist for the EM, we have not found them to be effective in fitting switching SSMs to long, possibly high-dimensional time series. The estimated information matrices used in these methods often fail to be negative definite as required or they yield update directions that decrease the likelihood instead of increasing it (maybe due to poor approximations, see item 1). 
We propose a simple acceleration scheme that reduces computation time for models \eqref{switch-dyn}-\eqref{switch-obs}   by a factor at least 3 (often much more) in our data analyses. The scheme consists in alternating these two steps until convergence.  
\begin{enumerate}
\item Run 20-50 EM iterations with  stopping criterion
$L(\btheta_{k}) \le (1+\epsilon) L(\btheta_{k-1})  $, 
where $\btheta_0,\btheta_1,\ldots$ are the successive EM iterates, $\epsilon \in [10^{-5},10^{-4}]$, 
and where $\btheta_0 = \btheta^{+}$ after step (b) has been completed at least once. 
Call $\widehat{\btheta}=\argmax_k L(\btheta_k)$ the best solution and $(\hat{S}_t)$ the estimated regime history. 
\item Run 500-1000 EM iterations with the regimes $(S_t)$ fixed to $(\widehat{S}_t)$,  
 starting from $\widehat{\btheta}$   and  stopping with tolerance  $\epsilon \in [10^{-6}, 10^{-5}]$.  
Call $\btheta^{+}$ the best solution.   
\end{enumerate}
The scheme ends when the increase in $L(\widehat{\btheta})$  between two successive steps (a) becomes insufficient. 
The efficiency of this scheme comes from the facts that: (i) the EM tends to converge quickly to a  sequence $(\widehat{S}_t)$ with probabilities $P(S_t = j| \by_{1:T})$ close to 0 or 1 for most $t$, so that fixing these values to exactly 0 or 1 is a good approximation, and (ii) step (b) amounts to a standard (non-switching) SSM whose EM iteration complexity (leading term $O(TN^3)$) is far less than in switching SSMs.

\end{enumerate}


\section{Bootstrap inference of model parameters}
\label{sec: bootstrap}

MLEs are  consistent and asymptotically normal in the context of  linear Gaussian SSMs \citep[][Chap. 50]{Hamilton1994a}. These properties extend to the switching SSM \eqref{switch-ssm} under identifiability conditions and mild assumptions on the Markov chain $(S_t)$  \citep{Cappe2005,Ailliot2013}. 
Thus, in principle, one could infer the parameters $\btheta$ of this model  using the limiting normal distribution of the MLE. 
This  in turn requires knowledge of the asymptotic covariance matrix of the MLE, that is, the inverse of the Fisher information matrix evaluated at $\btheta$.  In the EM framework, the available methods for estimating the information matrix  \cite[][Chap. 4]{McLachlan2008} are impractical due to the high dimension of $\btheta$. 
Furthermore, the fact that statistical inference may not only target  $\btheta$ but also functions thereof would entail additional tedious calculations based on the delta method. We thus propose parametric bootstrap as an alternative to likelihood-based inference \citep[see also][for nonparametric bootstrap]{Stoffer1991,Airlane2013}.
This approach enjoys a simple implementation and lends itself to parallelization, which greatly reduces the computational load. The bootstrap method goes as follows. 

\begin{enumerate}
\item Apply the EM algorithm of Section \ref{sec: EM} to the data $\by_{1:T}$.  
Denote by $\widehat{\btheta}$ the resulting MLE  of $\btheta$.  

\item Generate a bootstrap replicate $S^{\ast}_1$ 
according to the multinomial distribution with probabilities $\widehat{\boldsymbol{\pi}}$. 
For $t=2,\ldots, T$, generate a bootstrap replicate  $S_{t}^{\ast}$ 
according to the  multinomial distribution with probabilities  
$( \widehat{Z}_{S_{t-1}^{\ast}, 1} , \ldots, \widehat{Z}_{S_{t-1}^{\ast} ,M} )$. 

\item  Draw a bootstrap replicate $\bx_{1}^{\ast}$ 
from $N(\widehat{\bmu}_{S_{1}^{\ast}},\widehat{\bSigma}_{S_{1}^{\ast}})$. 
For $t=2,\ldots, T$, draw a bootstrap replicate $\bx_t^{\ast}$ 
from $N( \widehat{\bA}_{  S_{t-1}^{\ast}} \bx_{t-1}^{\ast} , \widehat{\bQ}_{S_t^{\ast}})$.

\item For $t=1,\ldots, T$, draw a bootstrap replicate $\by_t^{\ast}$ 
from $N(\bC_{S_t^{\ast}} \bx_t^\ast , \widehat{\bR}_{S_t^{\ast}} )$.

\item Apply the EM algorithm of section \ref{sec: EM} to the bootstrap sample $\by_{1:T}^{\ast}$. 
Denote by  $\widehat{\btheta}^{\ast}$ the resulting MLE. 

\item Repeat steps 2--7 many  times, say $100\le B \le 1000$, 
to obtain the probability distribution of the bootstrap estimator $\widehat{\btheta}^{\ast}$ conditional on  $\by_{1:T}$.  
\end{enumerate}

The bootstrap distribution of $\widehat{\btheta}$ can be harnessed to build confidence intervals (CI) for $\btheta$ 
that have asymptotically correct coverage and usually good finite-sample properties \citep{Efron1993}.  
For example, given $\widehat{\btheta}$, bootstrap replicates $\widehat{\btheta}^{\ast}_1,\ldots, \widehat{\btheta}^{\ast}_{B}$, 
and  $ \alpha \in(0, 0.5)$, the \emph{percentile} bootstrap CIs  
$\big[ \widehat{\btheta}^{\ast  (\alpha/2) }_B, \widehat{\btheta}^{\ast  (1-\alpha/2) }_B  \big]$
have approximate pointwise coverage $1-\alpha$, 
where   $\widehat{\btheta}^{\ast  (\alpha/2) }_B$ and $ \widehat{\btheta}^{\ast  (1-\alpha/2) }_B$ 
are the empirical quantiles of $\{ \widehat{\btheta}^{\ast}_1,\ldots, \widehat{\btheta}^{\ast}_{B}\}$ 
 of level $\alpha/2  $ and $1-\alpha/2$, respectively. 
Other common bootstrap CIs include the \emph{basic} bootstrap CIs 
$\big[ 2 \widehat{\btheta}- \widehat{\btheta}^{\ast  (1-\alpha/2) }_B, 
2 \widehat{\btheta}- \widehat{\btheta}^{\ast  (\alpha/2) }_B  \big]$ 
and the \emph{normal} bootstrap CIs 
$\big[ 
 \widehat{\btheta}- \widehat{\mathrm{bias}} ( \widehat{\btheta}) 
 \pm z^{(1-\alpha/2)} \widehat{\mathrm{se}} ( \widehat{\btheta}) \big]$, 
 where $ \widehat{\mathrm{bias}} ( \widehat{\btheta})  =
  (1/B)\sum_{b=1}^{B}  \widehat{\btheta}^{\ast}_b - \widehat{\btheta}$
and $ \widehat{\mathrm{se}} ( \widehat{\btheta}) $ are the bootstrap estimates 
of the bias and standard error of $  \widehat{\btheta}$, 
with $z^{(1-\alpha/2)}$ the usual quantile of $N(0,1)$.


\section{Simulation study}
\label{sec: simulations}

In this section we study the statistical performance of the EM algorithm (Section \ref{sec: EM}) and parametric bootstrap (Section \ref{sec: bootstrap}) in Markov-switching SSMs via simulation. Specifically, we evaluate these methods with respect to inferring the regime sequence $(S_{t})$ and estimating identifiable functions of the parameters $\btheta$ with point estimates and confidence intervals.

\subsection{Simulation setup}
\label{sec: simulation setup}

\paragraph{Data generation.} 
Models  \eqref{switch-dyn}-\eqref{switch-var}-\eqref{switch-obs} were simulated to generate time series of dimension 
$N\in\{10,50,100\}$ and length $T\in\{ 400,600,800,1000\}$. The large range of values for $N$ and $T$ enables us to examine   
various asymptotics in which $N\to\infty$ and/or $T\to \infty$. The number of regimes was set to $M=2$ and the autoregressive order to $p=2$. 

We first describe the data generation mechanism for models \eqref{switch-dyn}-\eqref{switch-obs}.
The state vector $\bx_t$ had size $r=2$. For each regime $ j=1,2$, the entries of the $r\times r$ transition matrices $\bA_{1 j}$ (lag 1) and $\bA_{2 j}$ (lag 2) were independently drawn from  uniform distributions on $[0,0.7]$ and $ [0,0.3]$, respectively. It was verified that the $(\bA_{1j},\bA_{2j})$ define an asymptotically stationary process $(\bx_t)$. 
The observation matrices $\bC$ in \eqref{switch-dyn} and $\bC_1,\bC_2$ in \eqref{switch-obs} were generated as the left-singular vectors of a random matrix of size $N \times r$ with independent $N(0,1)$ entries, so that $\bC'\bC=\bI_r$.  
 The initial state mean vectors and covariance matrices were set to $\bmu_j=\bzero_{r}$ and $\bSigma_j=0.1\, \bI_r$ for $j=1,2$. 
The state innovation variance matrices $\bQ_j$ were taken as realizations of a Wishart distribution with $r$ degrees of freedom and variance matrix $0.005\, \bI_r$. The observation noise variance $\bR = ( R_{ij})$ had an exchangeable structure: 
$R_{ii} = \sigma_{R}^2 $ and $R_{ij} = \rho \sigma_{R}^2  $ for $i\ne j$ 
where $1 \le i,j \le N$, $\sigma_R^2 = 0.005/N$, and $\rho=0.1$. 
Model parameters were jointly selected to produce a signal-to-noise ratio $\tr V(\bC \bx_t)/ \tr \bR$ between 5 and 10. 
This is a realistic ratio  in neuroimaging data such as EEG recordings. 
The regimes $(S_t)$ were simulated according to a Markov chain with initial probabilities $\boldsymbol{\pi} = (1,0)$ 
and transition probabilities $Z_{11} = Z_{22} =0.98 $ and $Z_{12}=Z_{21} = 0.02$. 
 For each combination $(N,T)$, the simulations were replicated 500 times for estimation 
and 200 times for bootstrap inference to keep computations manageable.

In the switching VAR model \eqref{switch-var},  the $N\times N$ transition matrices $\bA_{1 j}$  and $\bA_{2 j}$  
had diagonal entries independently drawn from  uniform distributions on $[0.85,0.95]$ and $ [-0.05,0.05]$, respectively, 
and off-diagonal entries set to zero. (Note that accurate estimation of dense matrices $\bA_{\ell j}$ 
in \eqref{switch-var} is in general infeasible unless $MpN \ll T$.) The noise variance matrices $\bQ_j$ were generated 
as for models \eqref{switch-dyn}-\eqref{switch-obs} but with  Wishart variance matrix  $(0.01/N) \bI_N$. 
The other parameters ($\bmu_j, \bSigma_j, \mathbf{Z}$) were obtained in the same way as in  \eqref{switch-dyn}-\eqref{switch-obs}. The simulations were run 500 times for each combination $(N,T)$. 

 \paragraph{Estimation targets and  methods.} 
 \label{sim: estimators}
 
Three types of quantities are of primary concern in estimation: 
the regime sequence $(S_t)$, the model parameters $\btheta$, 
and stationary covariance and correlation matrices derived from $\btheta$. 
Important parameters  in models \eqref{switch-dyn}-\eqref{switch-obs} are 
$\bC$ or $\bC_j$ (to be precise, their associated projection matrices),  
$\bA_{\ell j}, \bQ_{j}, \bR,$ and  $\mathbf{Z}$ ($\ell=1,2,\ j=1,2$). 
In model \eqref{switch-var} we are concerned with the estimation of the 
$\bA_{\ell j}, \bQ_{j}, $ and  $\mathbf{Z}$. 
Recall that model \eqref{switch-var} is a special case of model \eqref{switch-dyn} 
where $\by_t = \bx_t$, $\bC = \bI_N$, and $\bR = \bzero_{N\times N}$. 
We also consider the estimation of the (long run) stationary variance-covariance matrix 
 \begin{equation}\label{eq: stationary covariance}
 \bSigma_{j}^{y}= \lim_{t \to \infty} V(\by_{t}|S_{1}=\cdots = S_t = j)
\end{equation}
of the observation process $(\by_t)$ under regime $ j=1,\ldots, M$.
This matrix is uniquely determined by the relations  
\begin{equation}\label{eq: stationary covariance 2}
\bSigma_{j}^{y} = \bC_j \bSigma_{j}^{x} \bC'_j + \bR  \quad \textrm{and} \quad 
\bSigma_j^{x} = \bA_j  \bSigma_j^x \bA_j' + \bQ_j
\end{equation}
where $\bSigma_{j}^{x} = \lim_{t \to \infty} V(\bx_{t}|S_{1}=\cdots = S_t = j)$  
is the stationary covariance matrix of the state process $(\bx_t)$ under regime $j$. 
(For simplicity we have replaced the $\bA_{\ell j}$ and $\bx_{t- \ell}$ 
in \eqref{eq: stationary covariance 2} by a single $\bA_j$ and $\bx_t$, noting that models 
\eqref{switch-dyn}-\eqref{switch-var}-\eqref{switch-obs} can always be reduced to order $p=1$ 
as in Remarks \ref{rk: switch-dyn}-\ref{rk: switch-obs}).
Related quantities of interest include the correlation matrices $\mathbf{R}_j^y$ 
deduced from  \eqref{eq: stationary covariance} and the autocorrelations
$ \rho_{kj}^{y}(\ell) = \lim_{t \to \infty} \mathrm{Cor}(y_{kt}, y_{k(t-\ell)} | S_{1}=\cdots = S_t = j)$
where $\by_t = (y_{1t} , \ldots, y_{Nt})'$.

We consider the following benchmarks for comparison with the MLE of Section \ref{sec: EM} (SSM-ML): 
the approach based on sliding-window covariances and $K$-means clustering 
mentioned in Section \ref{intro}  (SW-KM); 
the initialization method of  Section \ref{subsec: EM initialization} (SSM-OLS);  
and oracle estimators based on the unobserved regimes  $(S_t)$ (OR-OLS, OR-ML). 
Comparison of the SSM-ML with SW-KM probes the benefits of explicitly modeling time series 
 dynamics and simultaneously estimating regimes and model parameters. (In contrast, the SW-KM  
is only based on instant covariances and estimates regimes in two steps.) 
By comparing  SSM-ML with its initialization SSM-OLS, we question to which extent 
maximum likelihood estimation improve upon a simple OLS approach. 
Finally, the oracle estimators, which are not affected by regime estimation error, 
make it possible to assess how regime estimation error compounds 
parameter estimation error in SSM-ML.

For SSM-ML and SSM-OLS, the regimes   $S_{t}$ are estimated by the most likely regimes  
\begin{equation}\label{regime estimator}
\widehat{S}_{t} = \argmax_{1\le j \le M} P_{\widehat{\btheta}}(S_t = j \, | \, \by_{1:T}).
\end{equation}
The matrix  $\bSigma_j^x$ is estimated by plugging the parameter estimates 
$\widehat{\bA}_j$ and $\widehat{\bQ}_j$ in the second equation of \eqref{eq: stationary covariance 2}, 
premultiplying this equation by $\widehat{\bA}_j^{-1} $, and solving the Sylvester equation 
$ \widehat{\bA}_j^{-1}  \bSigma_j^x =  \bSigma_j^x \bA_j' + \widehat{\bA}_j^{-1} \bQ_j$ 
(see Appendix \ref{sec: stationary covariance} for details).  
The matrix $\bSigma_j^y$ is estimated by plugging $\widehat{\bSigma}_j^x$ and the estimates $ \widehat{\bC}_j$ and $\widehat{ \bR }$ in the first equation of \eqref{eq: stationary covariance 2}.  
The SW-KM approach consists in calculating sample covariance matrices over successive time windows. 
 In a second stage, the (half) covariance matrices are clustered into $M$ regimes 
 with the $K$-means algorithm to estimate the underlying regimes $(S_t)$.  
 After that, each $\bSigma_j^y$  is estimated by the 
sample covariance matrix calculated over the cluster $\{t: \widehat{S}_t = j\}$. 
The stationary correlation matrices $\mathbf{R}_j^y$ 
and autocorrelation functions $\rho_{kj}^y$ 
are estimated by their sample counterparts in the same fashion.   
In the simulations, windows of length 31 time points were used 
with a shift of 1 time point between successive windows. 
This length gave optimal results but overall,  
SW-KM performance was not very sensitive to window length. 
The OR-OLS estimator is simply the SSM-OLS estimator with the $K$-means clustering step skipped and the regime estimates $(\widehat{S}_t)$ set to the true regimes $(S_t)$. 
OR-ML amounts to a standard maximum likelihood approach to SSM with the regimes $(S_t)$  known.

\begin{center}
\begin{tabular}{ r l }
\hline
Notation & Estimator \\ \hline
SW-KM & Sliding-window covariance/$K$-means clustering \\ 
SSM-OLS & EM initialization (Section \ref{subsec: EM initialization}) \\
SSM-ML & EM-based maximum likelihood estimator (Section \ref{sec: EM}) \\ 
OR-OLS & EM initialization with regimes $(S_t)$ known \\ 
OR-ML &  EM-based maximum likelihood estimator with regimes $(S_t)$ known\\ \hline
\end{tabular}
\end{center}

The parameter  $\btheta $ of the general switching SSM \eqref{switch-ssm}  
has $M$ regime-specific components $\btheta_j =(\bA_j,\bC_j,\bQ_j,\bR_j,\ldots) $ for $ j=1,\ldots,M$. 
 Accordingly,  association ambiguity arises when estimating  $\btheta$. 
 To properly assess the accuracy of an estimator $\widehat{\btheta}$, 
 we must resolve association ambiguity 
 between $\widehat{\btheta}_1, \ldots, \widehat{\btheta}_M$ and $  \btheta_1,\ldots, \btheta_M$. 
Failure to do so would lead to 
 mismatches between parameters and estimates and would artificially lower coverage levels. 
We match components between $\widehat{\btheta}$ and $\btheta$ 
by maximizing the classification rate in regime estimation:  
$\widehat{\sigma} = \argmax_{\sigma} \# \{ t: \sigma(\widehat{S}_t) = S_t \}$ where 
$(S_t)$ is known from the simulation, $(\widehat{S}_t)$ is obtained through \eqref{regime estimator}, and 
the search is over all permutations $\sigma$ of $\{1,\ldots, M\}$. Each $\btheta_j $ is then estimated by $\widehat{\btheta}_{\widehat{\sigma}(j)}$. This matching procedure is more accurate than directly matching the $\widehat{\btheta}_j$ to the 
$\btheta_j $ because: (i) the elements $\bA_j,\bC_j,\ldots$ of   
 $\btheta_j$ are for the most part non-identifiable, and (ii) estimation error about these elements is   
amplified  by estimation error about the regimes $(S_t)$.

 \paragraph{Bootstrap inference.}
  
This part of the numerical study evaluates the statistical accuracy of the
bootstrap confidence intervals described in Section \ref{sec: bootstrap}. 
In models \eqref{switch-dyn}-\eqref{switch-var}-\eqref{switch-obs}, 
we construct percentile, basic, and normal bootstrap CIs 
 for identifiable model quantities and report their attained coverage.  
 The CIs are built at the pointwise nominal  level $ 1-\alpha = 0.9$ using $B=100$ bootstrap replicates. This number is sufficient to estimate bootstrap means, standard deviations, and quantiles of level 0.05 and 0.95 with reasonable accuracy while keeping the high computational load of bootstrap manageable. 
We will also report on attempts at nonparametric bootstrap \citep{Stoffer1991,Airlane2013}.

Our main inferential targets are the stationary covariance matrices $\bSigma_y^{j}$, correlation matrices $\mathbf{R}_j^y $, 
and autocorrelations functions $\rho_{kj}^y $ for $j =1,\ldots,  M$ and $k=1,\ldots, N$.  
Up to label permutations, these quantities are identifiable in all three models. 
We also build bootstrap CIs for the probability transition matrix $\bZ$ (identifiable in all three models). 
Finally, we infer the identifiable quantities 
$\bC \bC'$ or $\bC_j \bC_j'$ (projection matrix on the subspace spanned by the state vectors $\bx_t$ or $\bx_{tj}$ in the observation space $\mathbb{R}^N$)  
and $\bR$ in models  \eqref{switch-dyn}-\eqref{switch-obs} and the parameters $\bA_{\ell j}$ and $ \bQ_j$
in model \eqref{switch-var}.  

In addition to the association ambiguity between the components of the MLE $\widehat{\btheta}$ and  of $ \btheta$, 
there is also association ambiguity between the components of the bootstrap replicates $\widehat{\btheta}^{\ast}_{b}$ and of $\widehat{\btheta}$ for $b=1,\ldots,B$.  The matching procedure based on estimated regimes is no longer available here because bootstrap regime sequences are generated independently of $(\widehat{S}_t)$. 
However, $\widehat{\btheta}^{\ast}_{b}$  and  $\widehat{\btheta}$ can be matched with quasi certainty 
using the estimates $\widehat{\boldsymbol{\pi}}^{\ast}_b$ and $\widehat{\boldsymbol{\pi}}$ 
of initial regime probabilities   (at least for $M=2$). Indeed the fact that both vectors invariably 
have one component almost equal to 1 and the other to 0 enables immediate regime identification.


\subsection{Results}

\paragraph{Regime estimation.}

\begin{figure}[ht!]
\begin{center}
\hspace*{-10mm}
\begin{tabular}{ccc}
\includegraphics[width = .34 \textwidth]{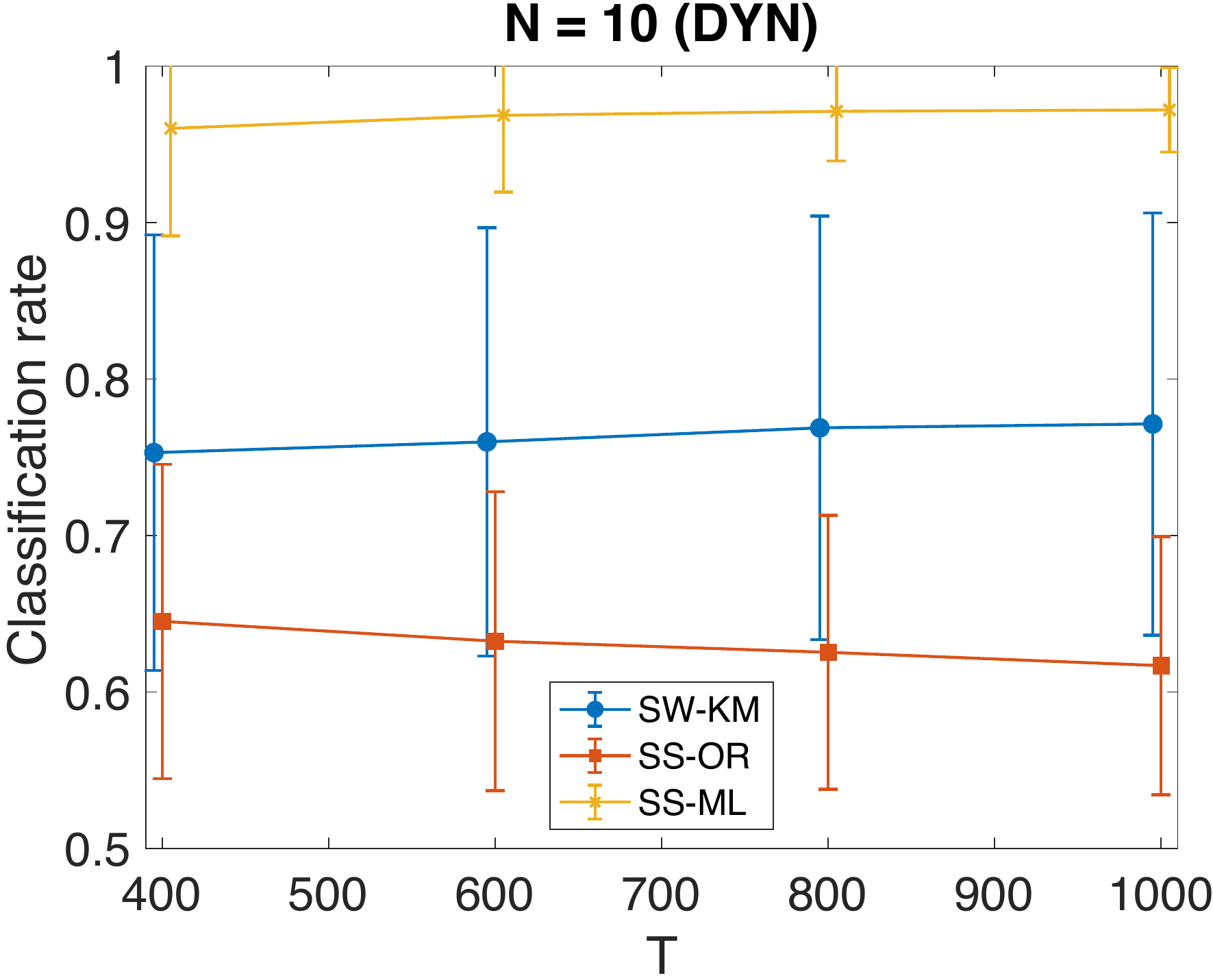} & 
\includegraphics[width = .34 \textwidth]{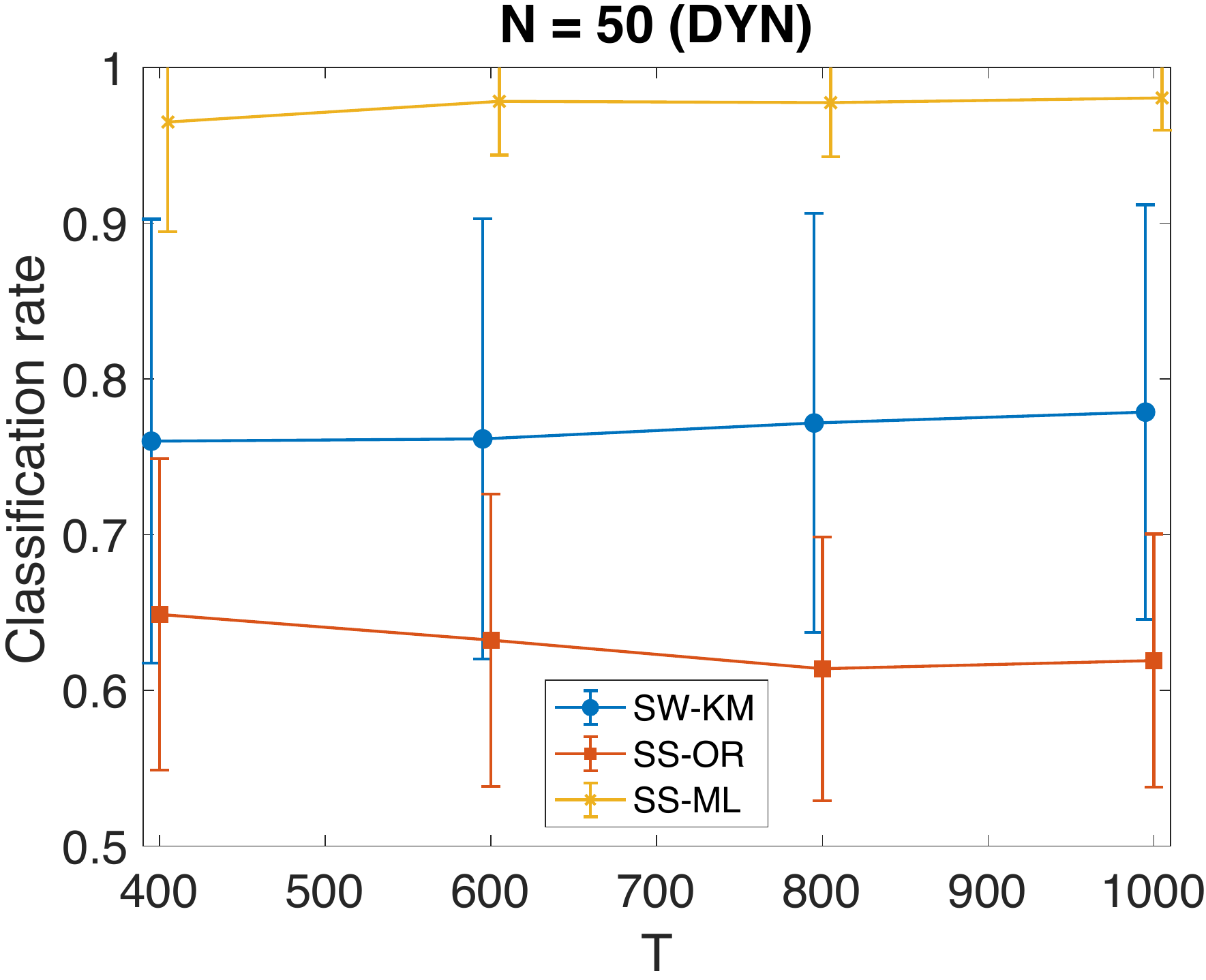} & 
\includegraphics[width = .34 \textwidth]{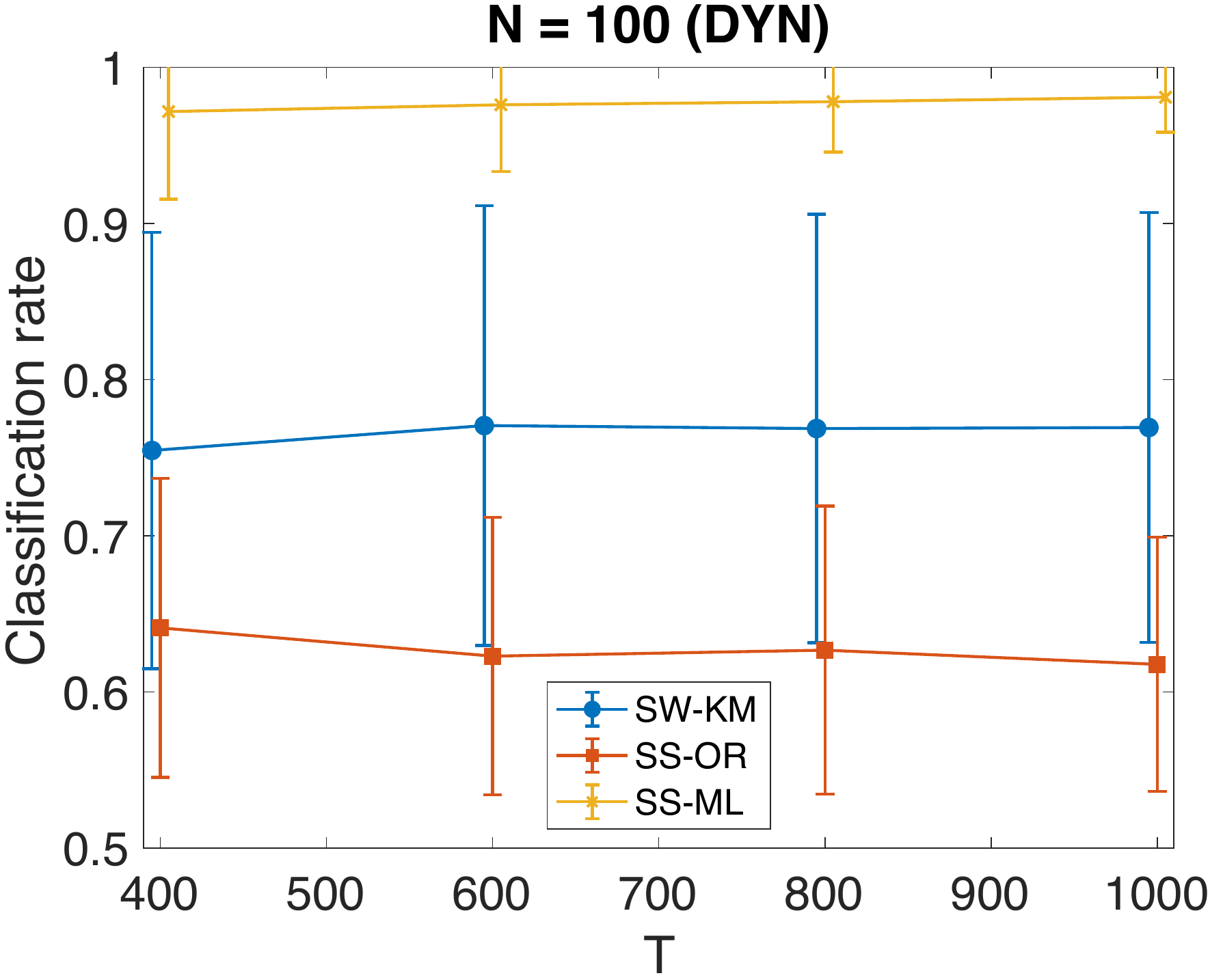} \\
\includegraphics[width = .34 \textwidth]{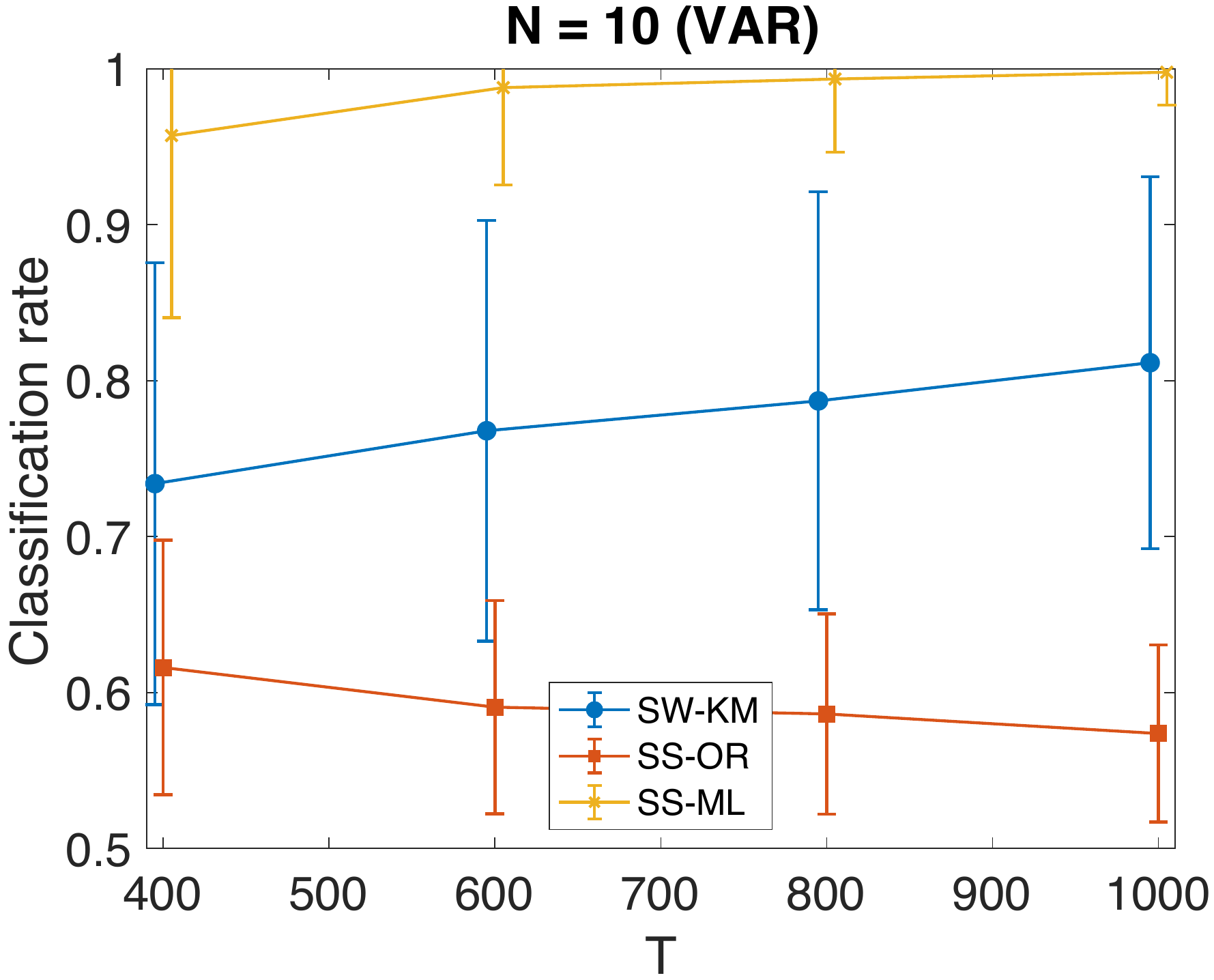} & 
\includegraphics[width = .34 \textwidth]{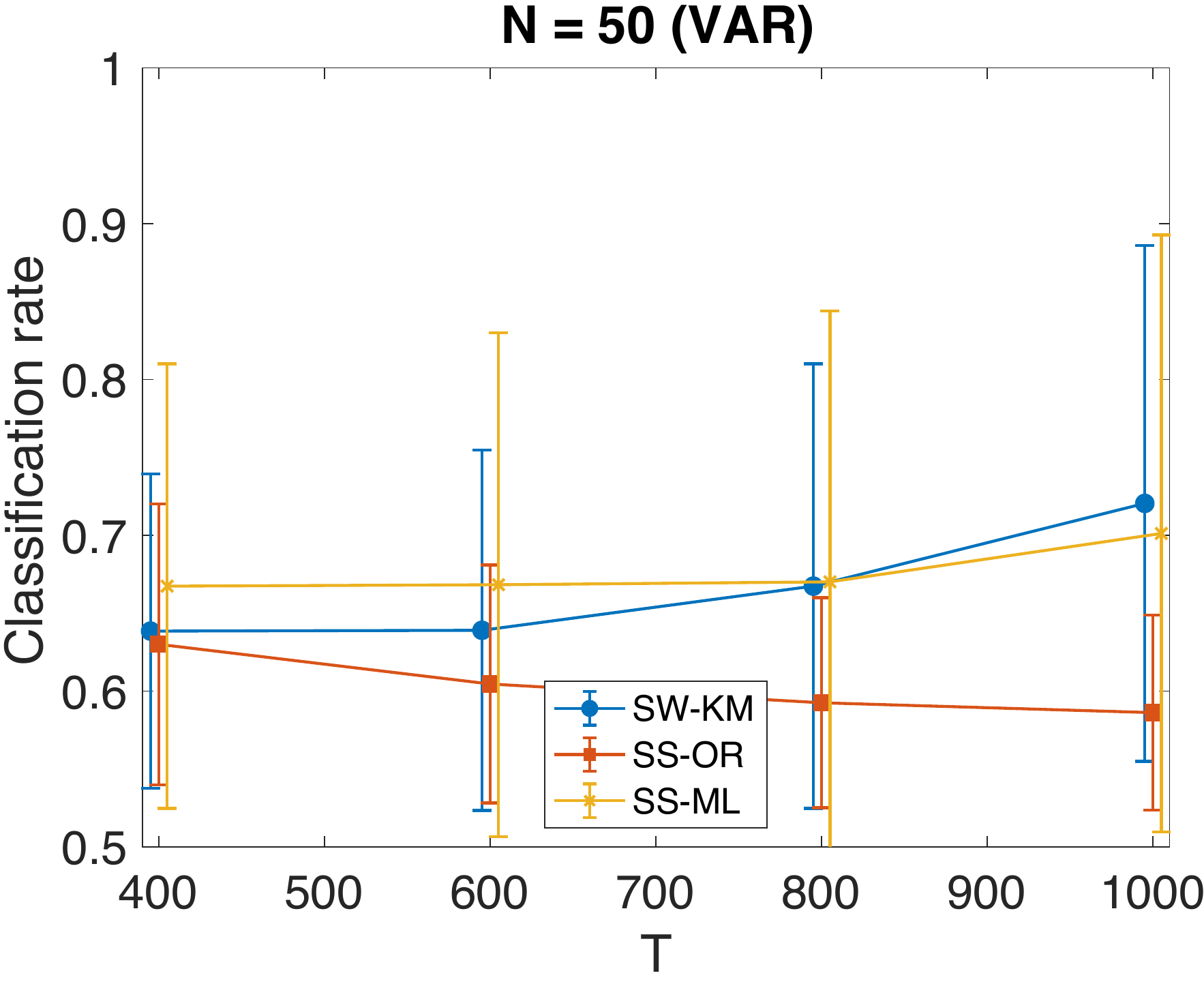} & 
\includegraphics[width = .34 \textwidth]{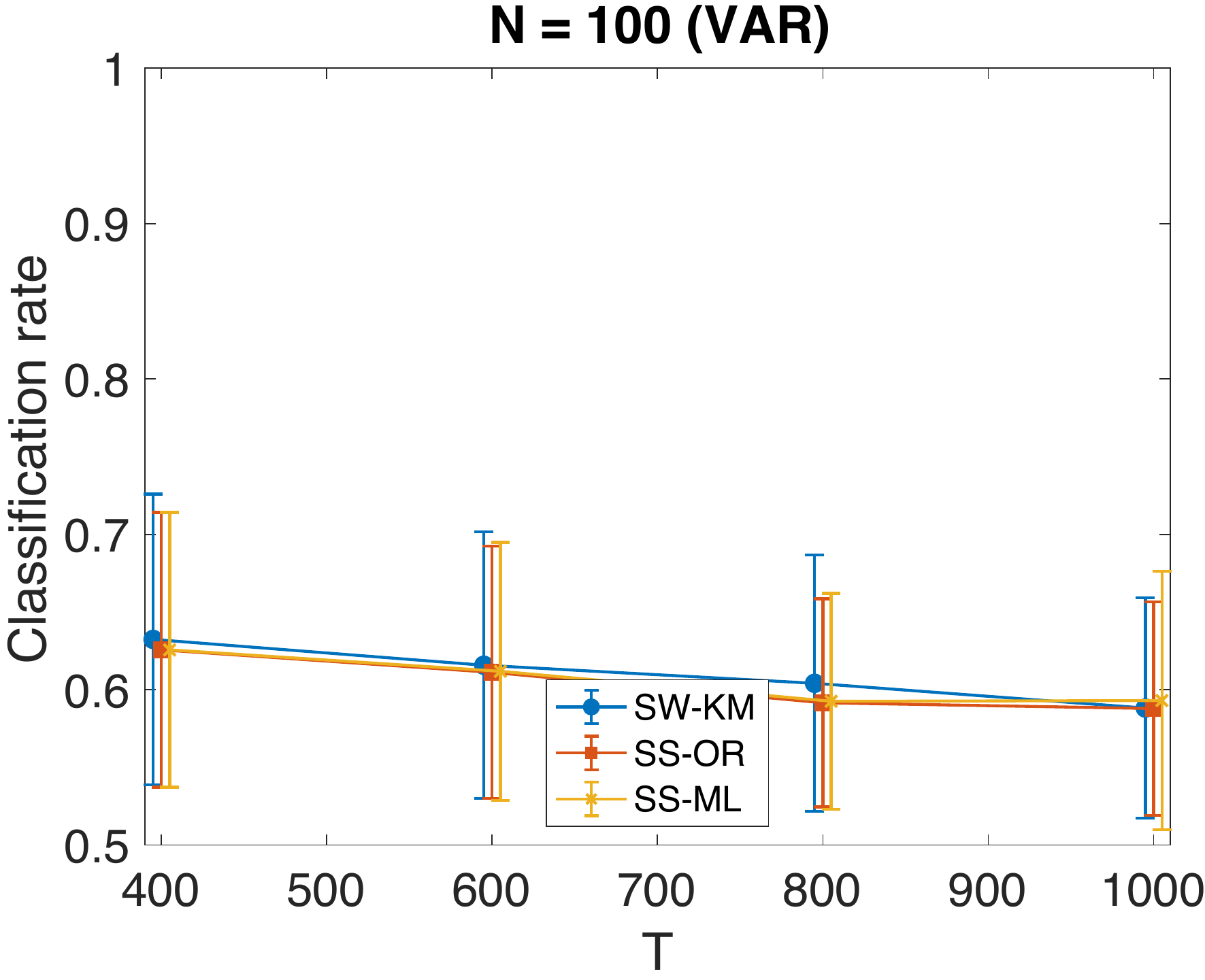} \\
\includegraphics[width = .34 \textwidth]{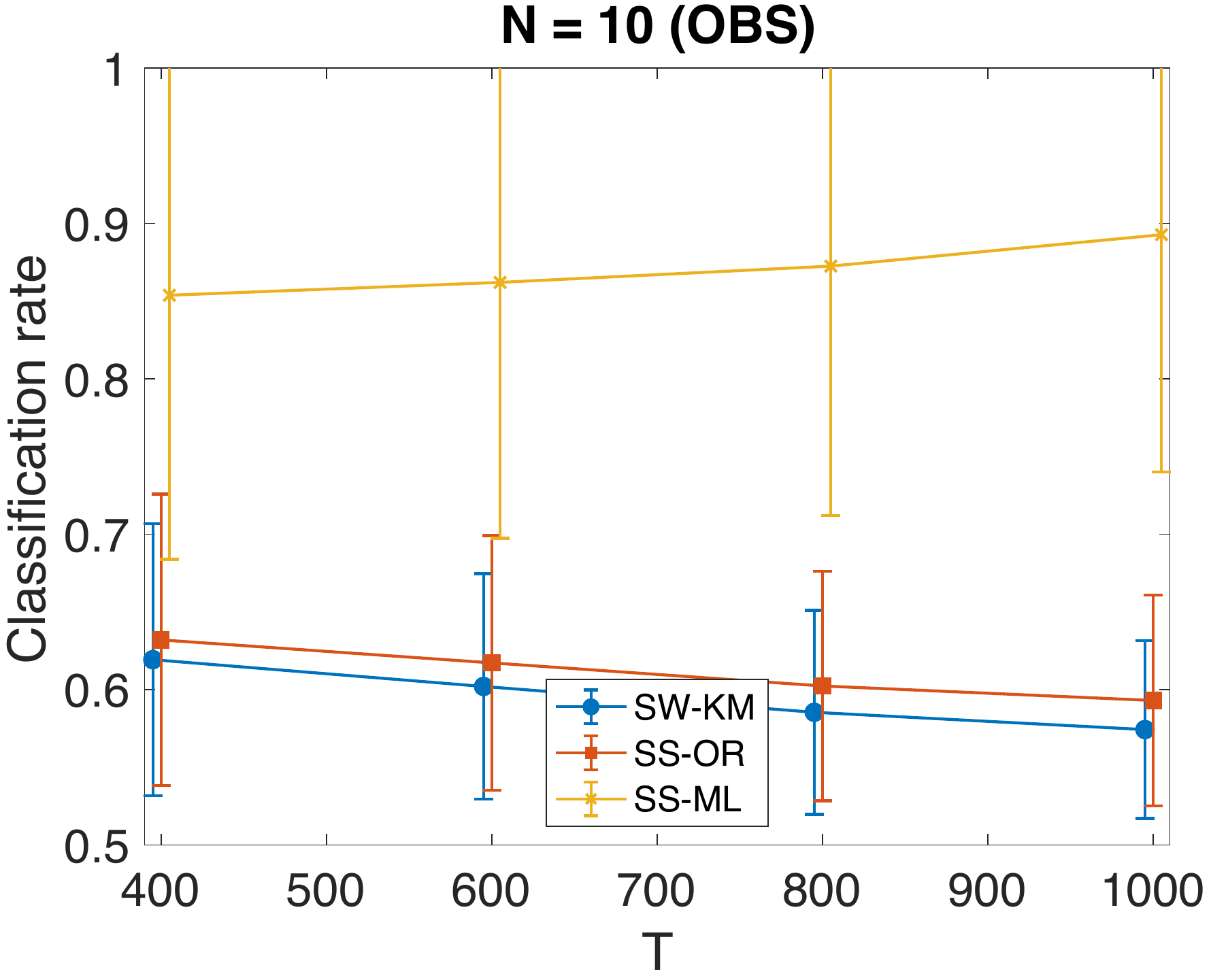} & 
\includegraphics[width = .34 \textwidth]{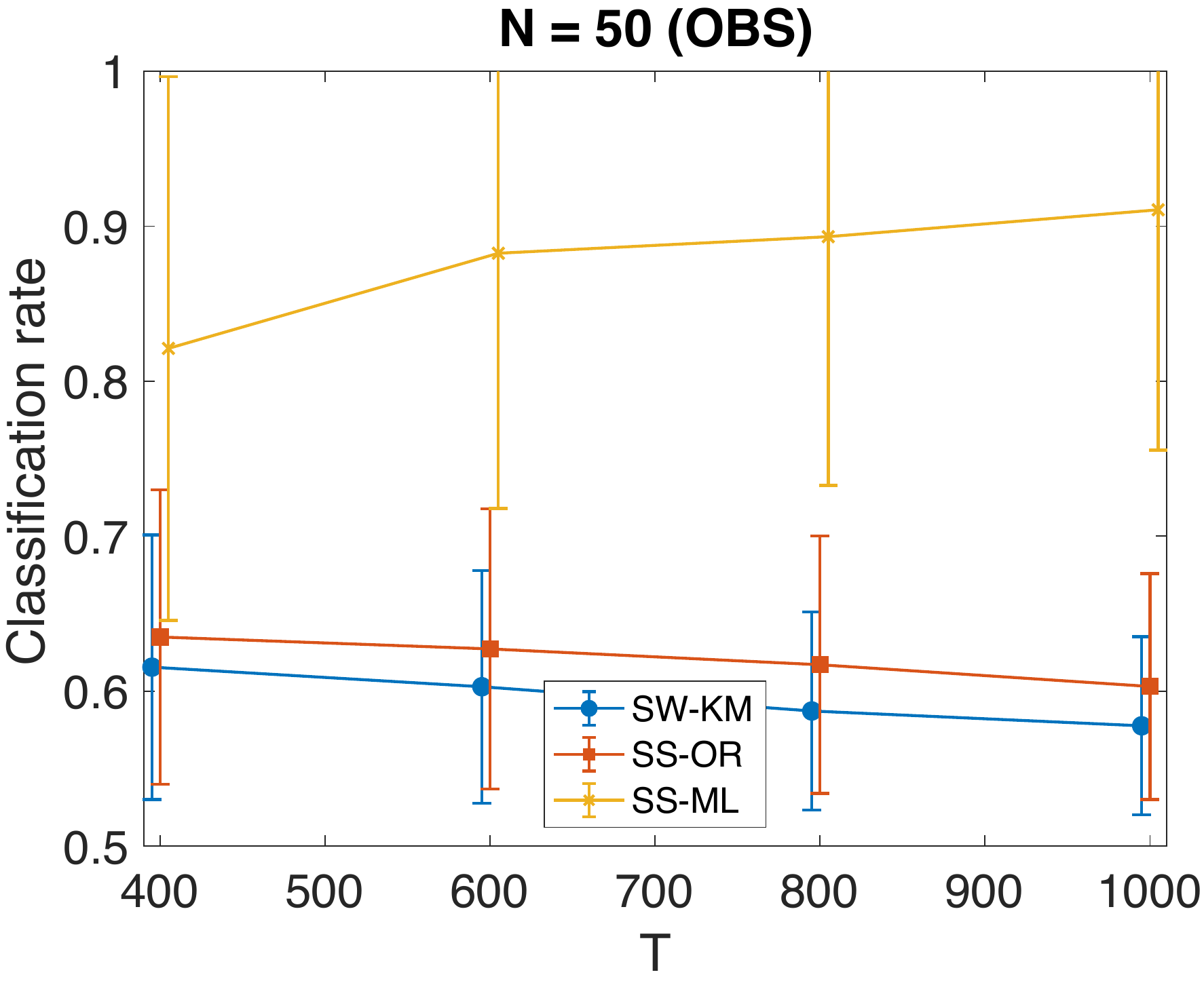} & 
\includegraphics[width = .34 \textwidth]{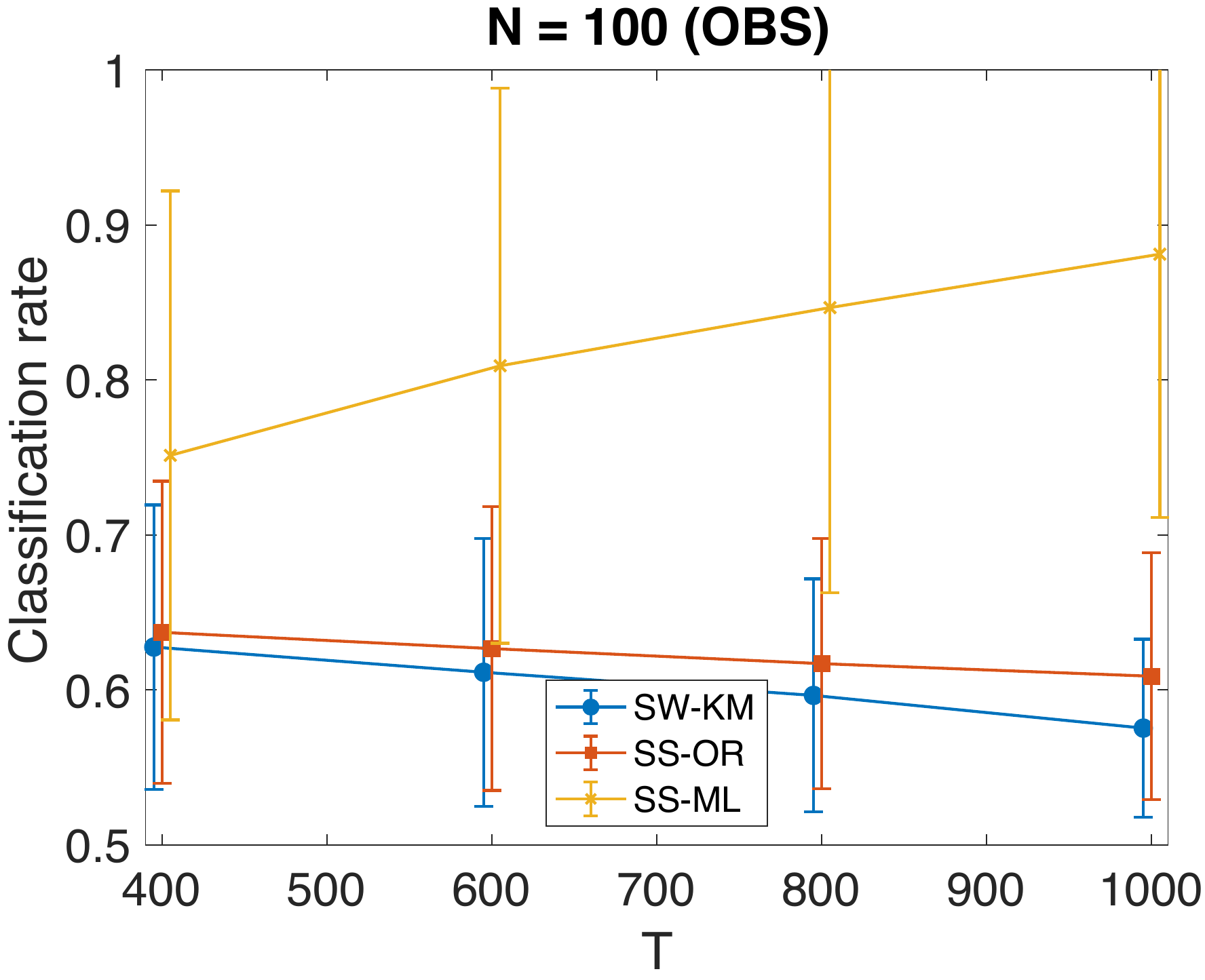} 
\end{tabular}
\caption{Accuracy of regime estimation in models \eqref{switch-dyn}-\eqref{switch-var}-\eqref{switch-obs} (DYN, VAR, OBS). 
For each model and dimension $(N,T)$, 500 simulations are performed. 
For each estimation method, the average and standard deviation of the classification rate are displayed 
as solid line and error bars. A classification rate 0.5 corresponds to random guessing.}
\label{fig: classification rate}
\end{center}
\end{figure}

We measure accuracy in the estimation 
of  $(S_t)$ by $(\widehat{S}_t)$  with the classification rate 
$ \max_{\sigma} \# \{ t: \sigma(\widehat{S}_t) = S_t \} / T$, 
where the maximum is taken over all permutations $\sigma$ of $\{1,\ldots ,M\}$. 
(Recall that switching SSMs are invariant under permutations of regime labels.)
As a baseline for comparison, a classification rate $0.5$ amounts to 
random guessing.    
Figure \ref{fig: classification rate} displays the classification rate 
of SW-KM, SSM-OLS, and SSM-ML 
in models \eqref{switch-dyn}-\eqref{switch-var}-\eqref{switch-obs}.
OR-OLS and OR-ML, which by definition achieve perfect classification, 
are not discussed here. 
SSM-ML  largely dominates  SW-KM and SSM-OLS with respect to classification  
and shows good to excellent  performance in all simulation setups 
except for  $N=50,100$ in model \eqref{switch-var}.  
In these cases,   the poor performance of all three methods is due to the fact that 
the number of model parameters is comparable to or larger than the number of measurements. 
Breaking down results by model,  
SSM-ML has average classification rates 0.973, 0.756, 0.856 across all combinations of $(N,T)$ 
for model \eqref{switch-dyn}-\eqref{switch-var}-\eqref{switch-obs} respectively  
versus   0.766, 0.684, 0.602 for SW-KM and  0.629, 0.600, 0.618 for SSM-OLS.

Turning to change point detection,  for each model, 
the average number of true change points is 14.0  across all combinations of $(N,T)$. 
In comparison, SSM-ML   detects on average 12.8, 6.9, 16.3 change points  in models 
 \eqref{switch-dyn}-\eqref{switch-var}-\eqref{switch-obs} respectively 
versus 17.7, 4.2, 15.2 for SW-KM and 5.1, 3.3, 5.3 for SSM-OLS. 
That is, SSM-ML and SW-KM estimate 
the number of change points relatively well in the switching dynamics and switching 
observations models
but strongly underestimate it  in the switching VAR model for large $N$. 
SSM-OLS  strongly underestimates the number of change points in all setups, 
which is not surprising for a crude initialization method, 
but it is good enough a starting point for SSM-ML to estimate regimes accurately.

\paragraph{Parameter estimation.}  
As stated in Remark \ref{rk: identifiability}, model \eqref{switch-ssm} admits an infinity of equivalent parameterizations. Accordingly, precautions must be taken when comparing estimates to their target parameters (in addition to resolving association ambiguity between regimes). For example $\bC$ is not identifiable in model \eqref{switch-dyn}. Instead of comparing $\hat{\bC}$ to $\bC$ directly, 
we compare the associated projection matrices $\widehat{\bC} (\widehat{\bC}'\widehat{\bC})^{-1} \widehat{\bC}'$ and $\bC (\bC'\bC)^{-1} \bC' =\bC\bC' $ that only depend on the linear spaces spanned by 
$\widehat{\bC}$ and $\bC$ and not on the choice of basis for these spaces. 
For the estimators $\widehat{\bA}_{\ell j}$ and $\widehat{\bQ}_j$ ($ j=1,\ldots, M, \ \ell=1,\ldots,p$), we first find the matrix $\bB$ that best maps $\widehat{\bC}$ to $\bC$, 
i.e., $\bB = \argmin_{\bX} \| \bC - \widehat{\bC} \bX \|_F^2 = (\widehat{\bC}' \widehat{\bC})^{-1} \widehat{\bC}'\bC
$ with $\| \cdot \|_F$  the Frobenius norm 
and then replace $\widehat{\bA}_{ \ell j}$ by $\bB^{-1}\widehat{\bA}_{ \ell j} \bB $ and 
 $\widehat{\bQ}_j$ by  $ \bB^{-1} \widehat{\bQ}_j (\bB^{-1})'$ as in Remark \ref{rk: identifiability}. 
 While this procedure seems a reasonable way to compare $\widehat{\bA}_{\ell j}$ with $\bA_{ \ell j}$ 
 and $\widehat{\bQ}_j$ with $\bQ_j$, the resulting error reflects not only error in the estimation of $\bA_{\ell j}$ or $\bQ_j$ but also error in the estimation of $\bC$. 
 The parameters $\bR$ is invariant under model reparameterization
  and can be directly compared to $\widehat{\bR}$. 
For each type of parameter ($\bA,\bC,\bQ,\bR,\bZ$), we measure the estimation accuracy by the relative error across all regimes, 
e.g.,   $\big( \sum_{j=1}^M \sum_{\ell=1}^{p}  \| \bB^{-1}\widehat{\bA}_{\ell j} \bB - \bA_{\ell j} \|_{1,1} \big/  \sum_{j=1}^M 
\sum_{\ell=1}^p  \| \bA_{ \ell j} \|_{1,1} \big)$ for the matrices $\bA_{ \ell j}$, $\big( \| 
\widehat{\bC} (\widehat{\bC}'\widehat{\bC})^{-1} \widehat{\bC}' - 
 \bC  \bC' \|_{1,1}  \big/  \|\bC  \bC' \|_{1,1} $ for $\bC$, etc., 
 where $\| \cdot \|_{1,1}$ is the $L_{1,1}$ matrix norm (sum of absolute values). 
It is reasonable to aggregate the estimation errors across regimes because by design, 
the model parameters have similar magnitude and signal-to-noise ratios across regimes. 
The use of the $L_{1,1}$ norm instead of, say, the Frobenius norm 
adds robustness against extreme but infrequent estimation errors. 

\begin{figure*}[htp]
\begin{center}
\small{
\begin{tabular}{cccc}
\hline
 & $N=10$ & $N=50$ & $N=100$ \\
\hline
\noalign{\medskip} 
$\bA$ \hspace*{3mm}   & 
\parbox[c]{11em}{\includegraphics[width = .28 \textwidth, keepaspectratio]{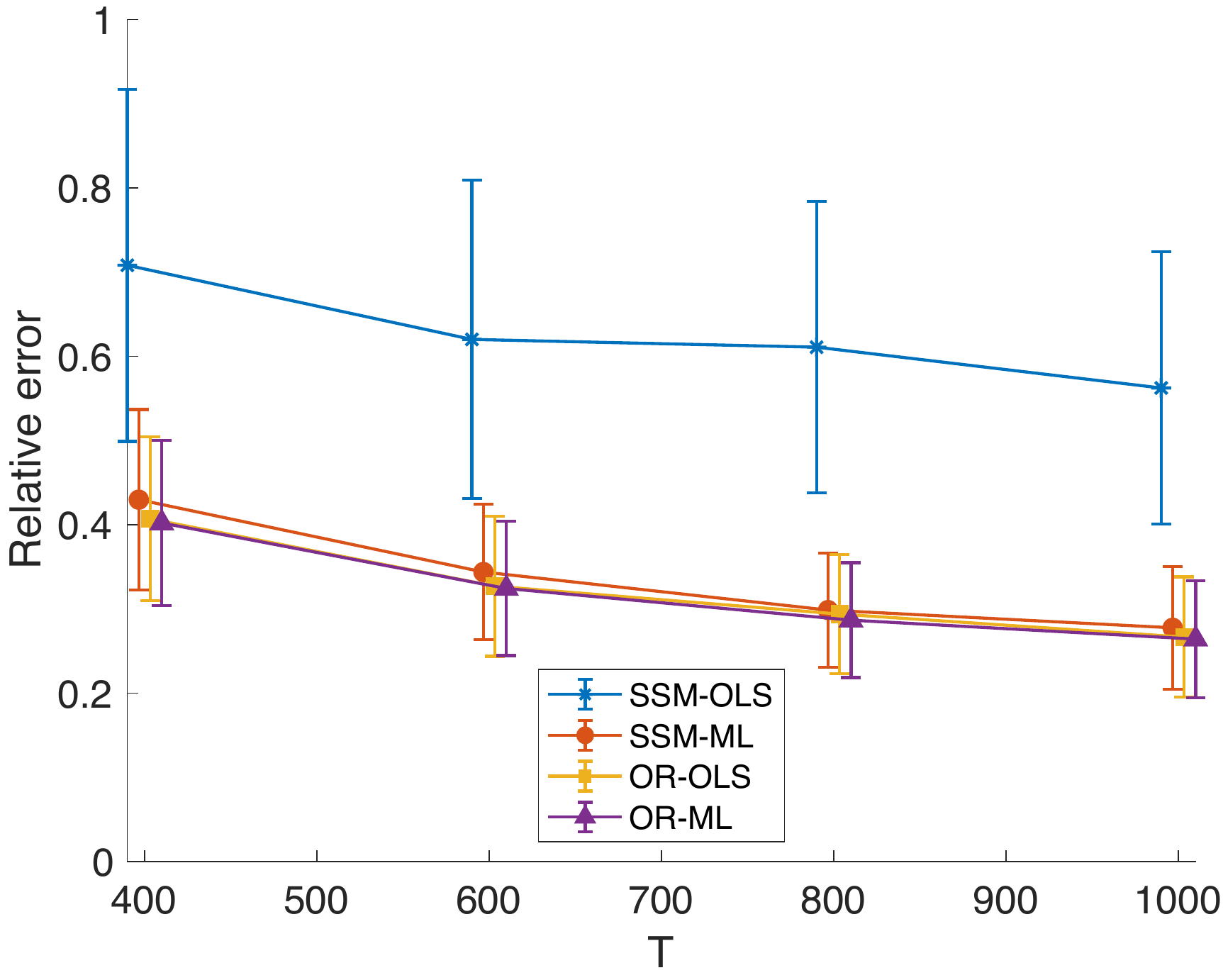}} &
\hspace*{-3mm}
\parbox[c]{11em}{\includegraphics[width = .28 \textwidth, keepaspectratio]{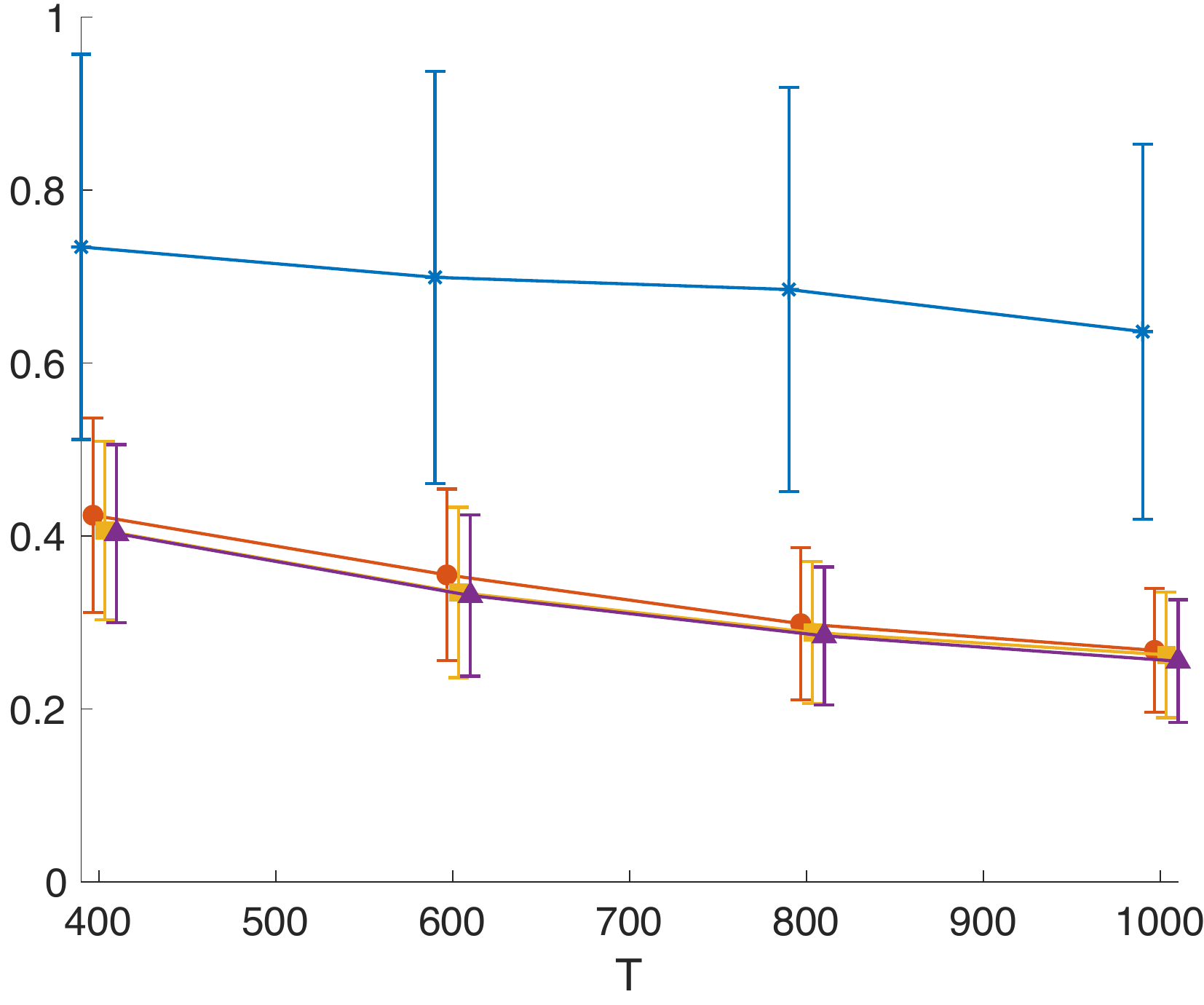} }&
\hspace*{-3mm}
\parbox[c]{11em}{\includegraphics[width = .28 \textwidth, keepaspectratio]{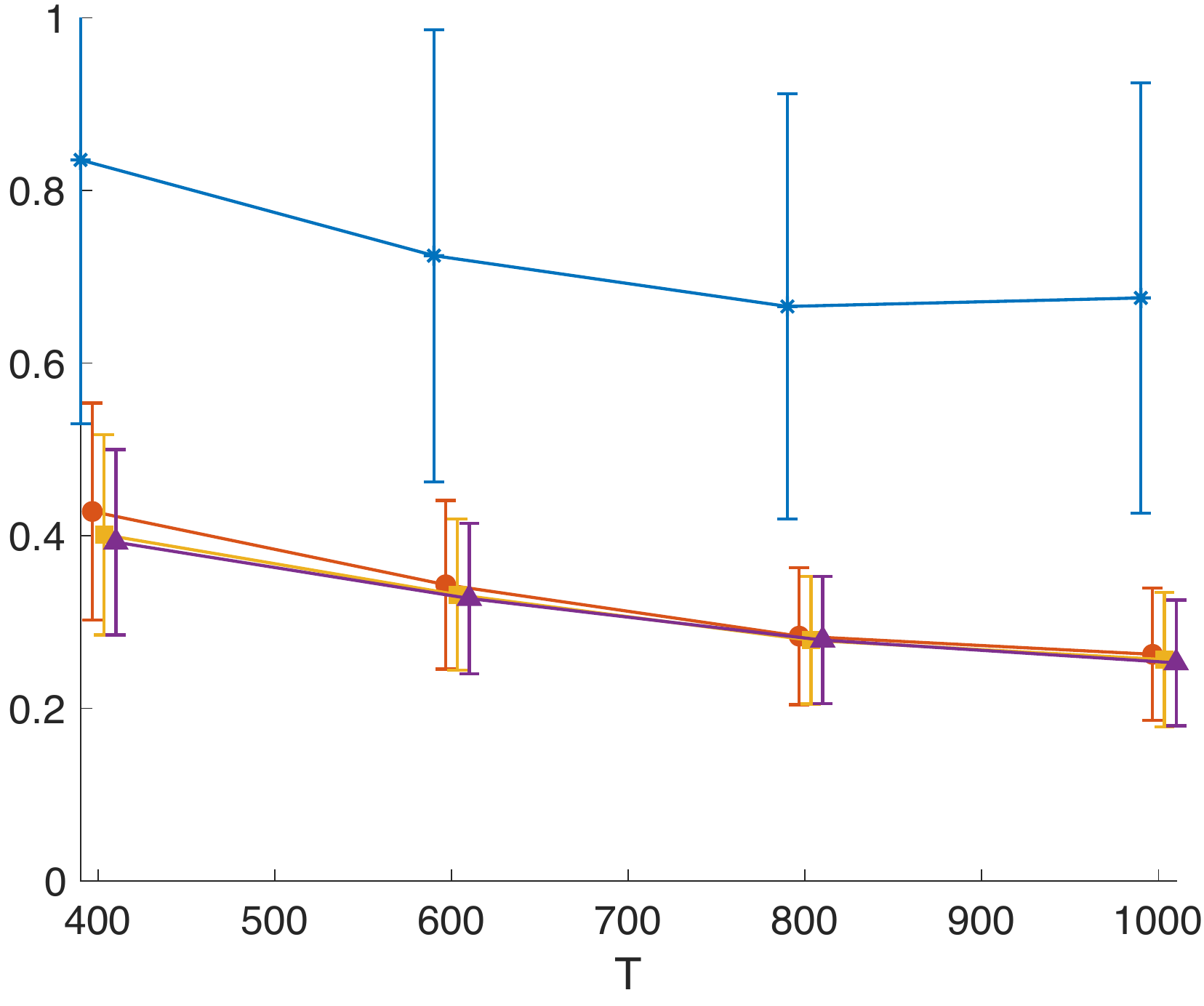}} \\
$\bC$  \hspace*{3mm} & 
 \parbox[c]{11em}{\includegraphics[width = .28 \textwidth, keepaspectratio]{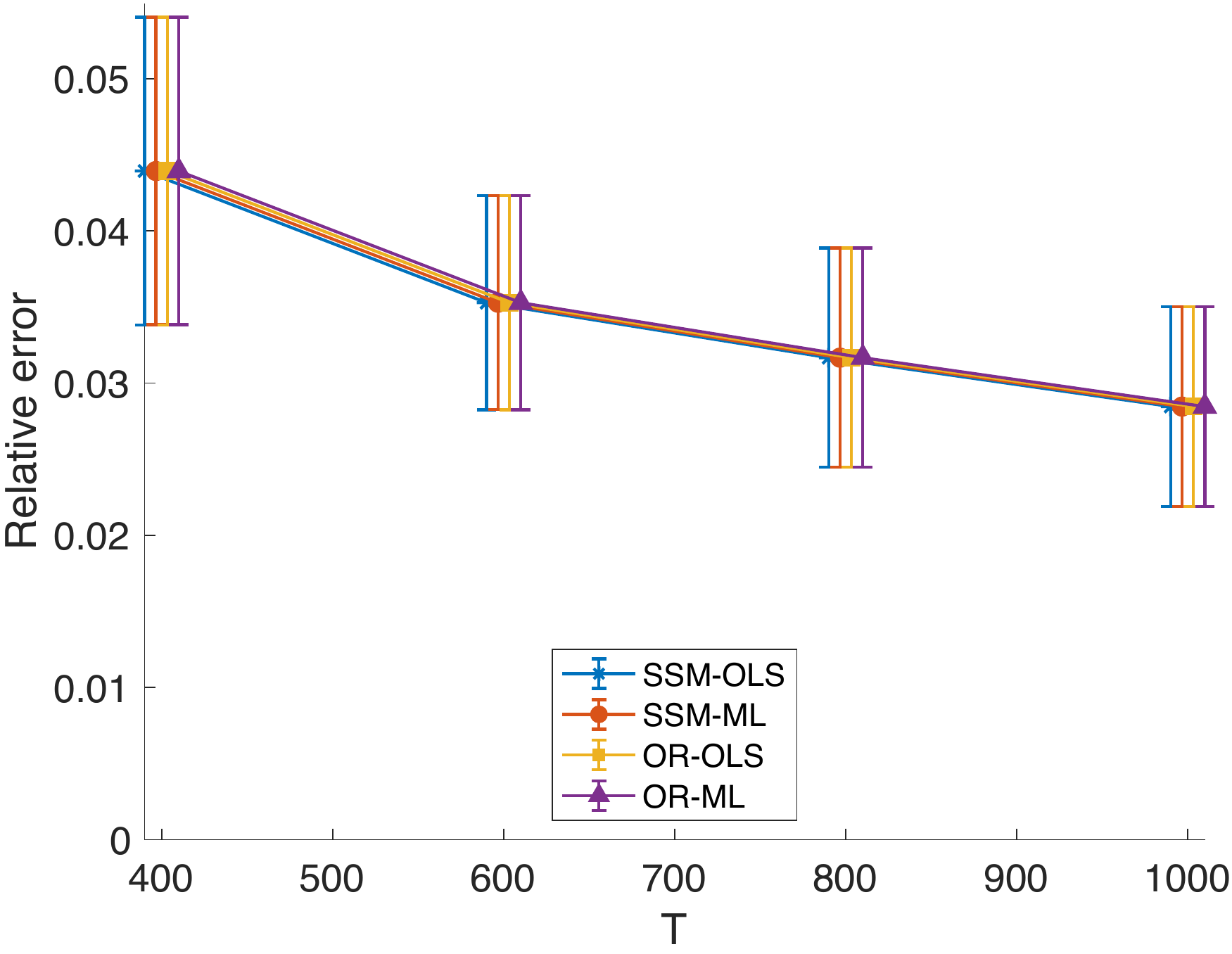}} &
 \hspace*{-3mm}
\parbox[c]{11em}{\includegraphics[width = .28 \textwidth, keepaspectratio]{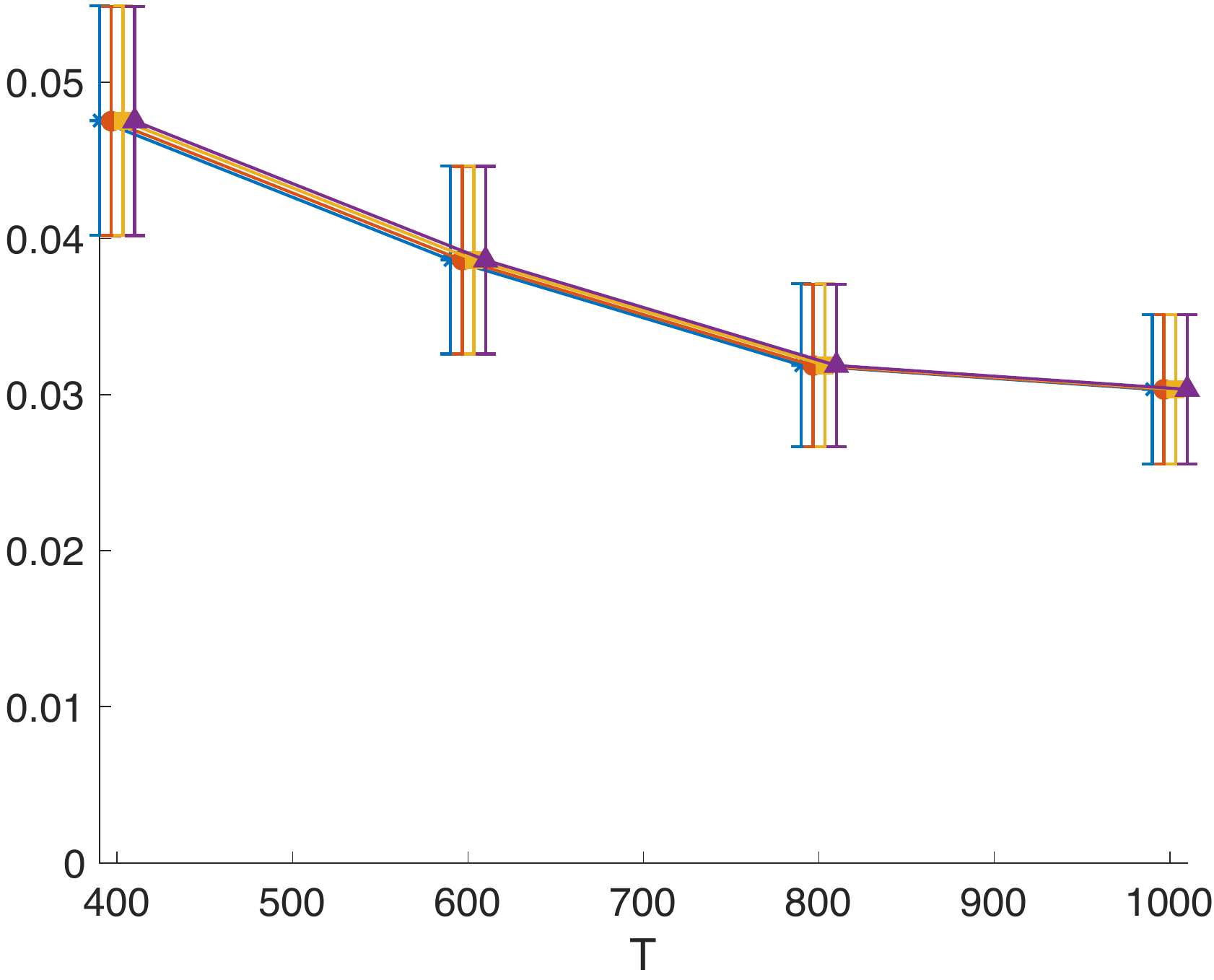}} &
\hspace*{-3mm}
\parbox[c]{11em}{\includegraphics[width = .28 \textwidth, keepaspectratio]{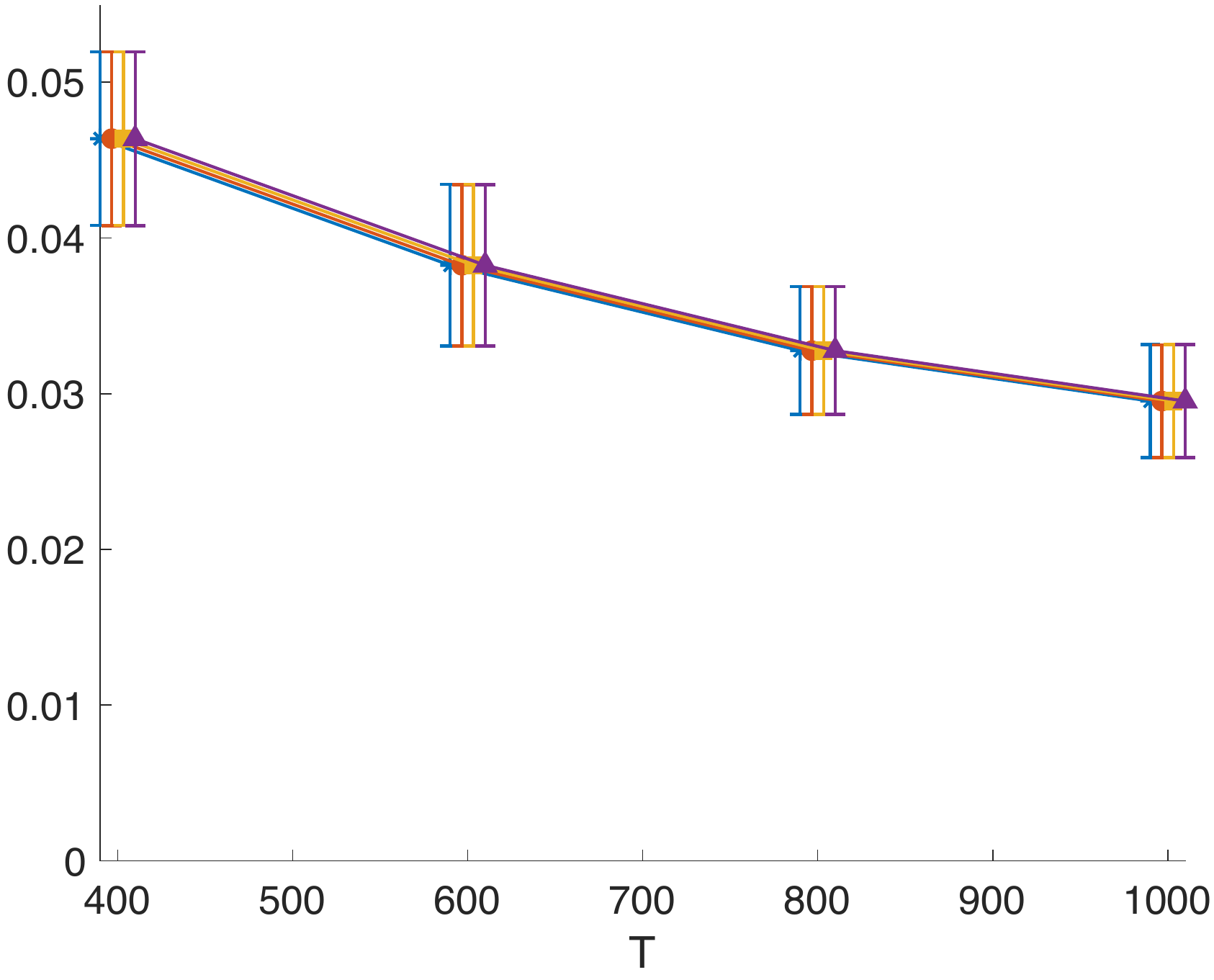}} \\
$\bQ$  \hspace*{3mm} & 
 \parbox[c]{11em}{\includegraphics[width = .28 \textwidth, keepaspectratio]{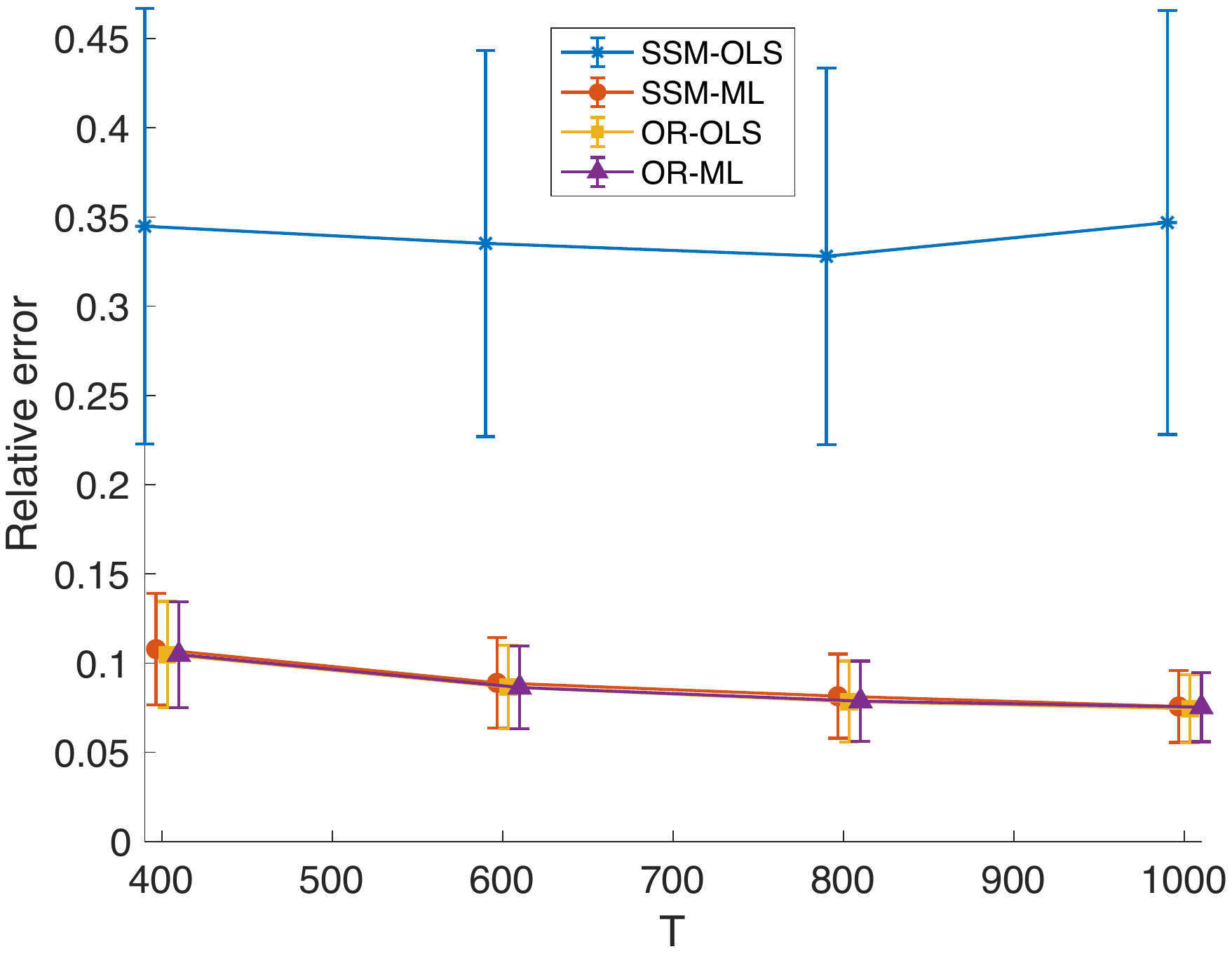}} &
 \hspace*{-3mm}
\parbox[c]{11em}{\includegraphics[width = .28 \textwidth, keepaspectratio]{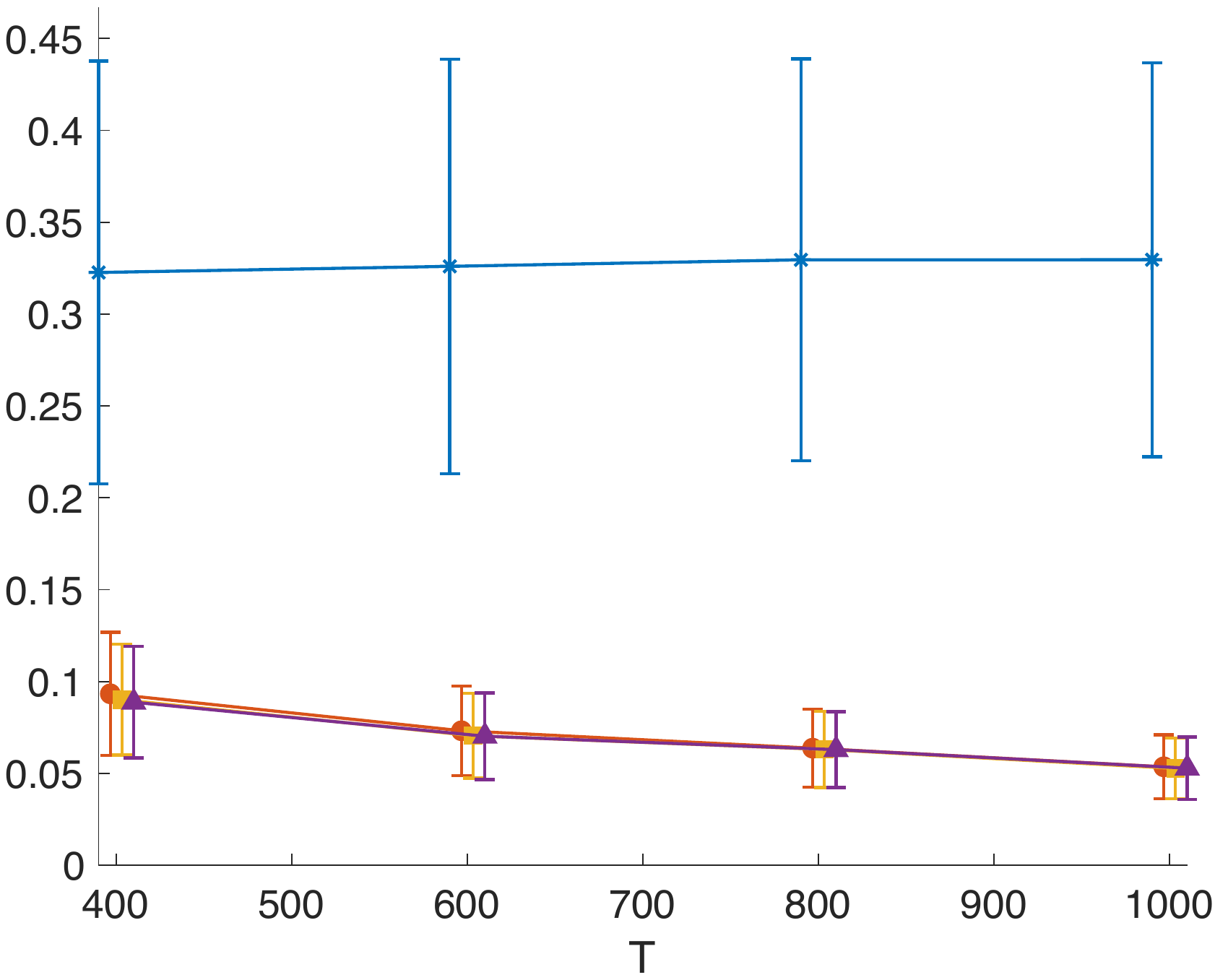}} &
\hspace*{-3mm}
\parbox[c]{11em}{\includegraphics[width = .28 \textwidth, keepaspectratio]{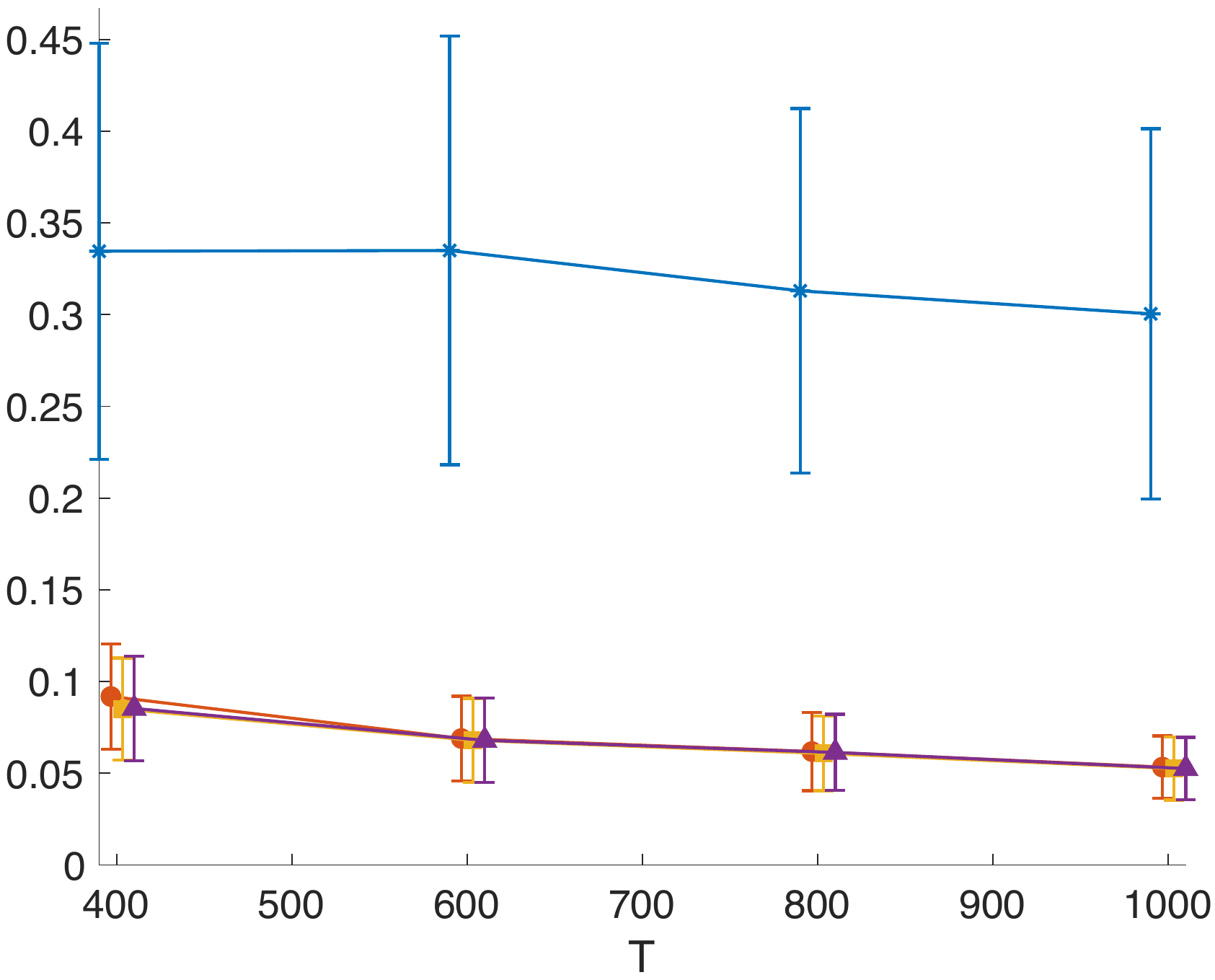}} \\
$\bR$  \hspace*{3mm} & 
 \parbox[c]{11em}{\includegraphics[width = .28 \textwidth, keepaspectratio]{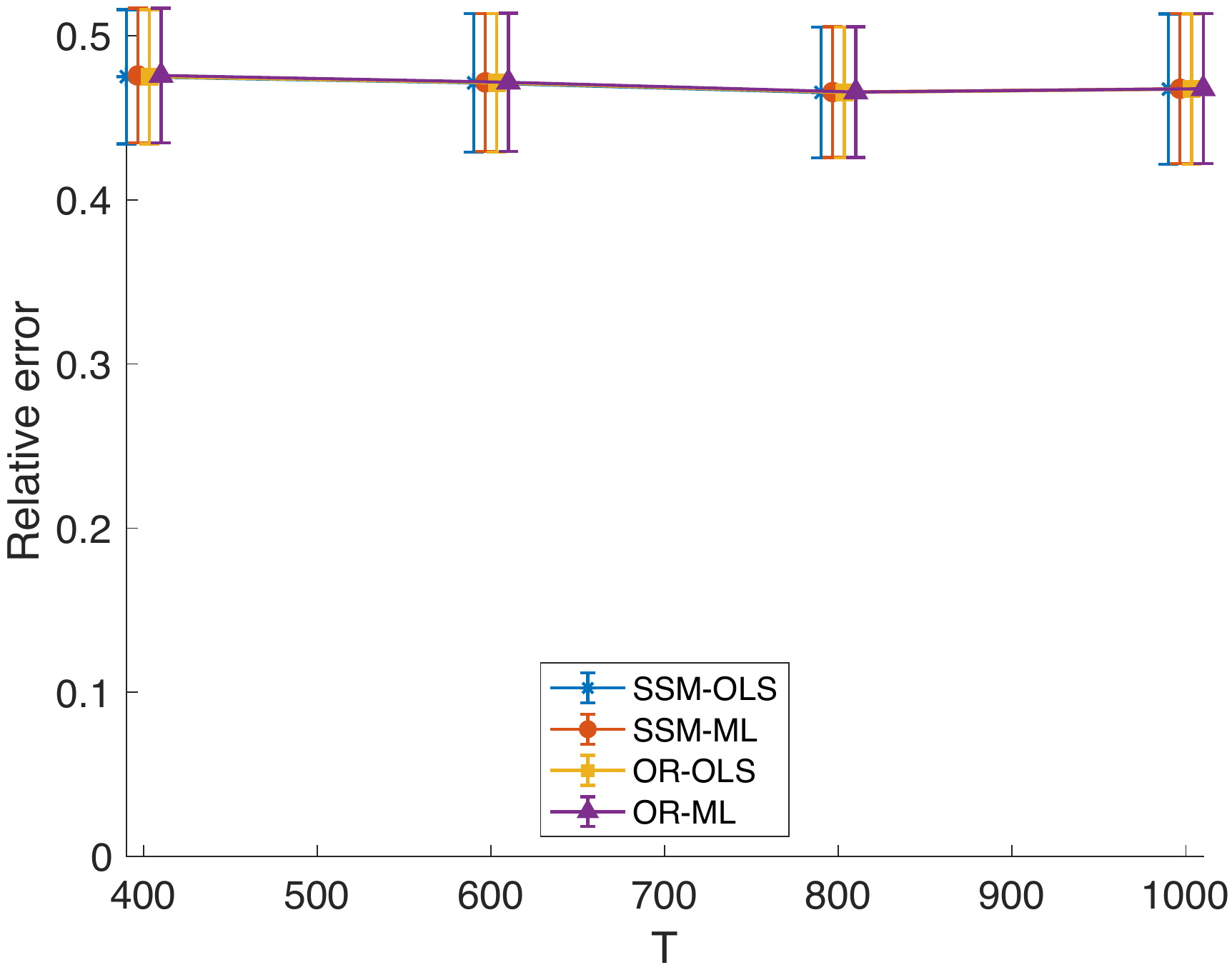}} &
 \hspace*{-3mm}
\parbox[c]{11em}{\includegraphics[width = .28 \textwidth, keepaspectratio]{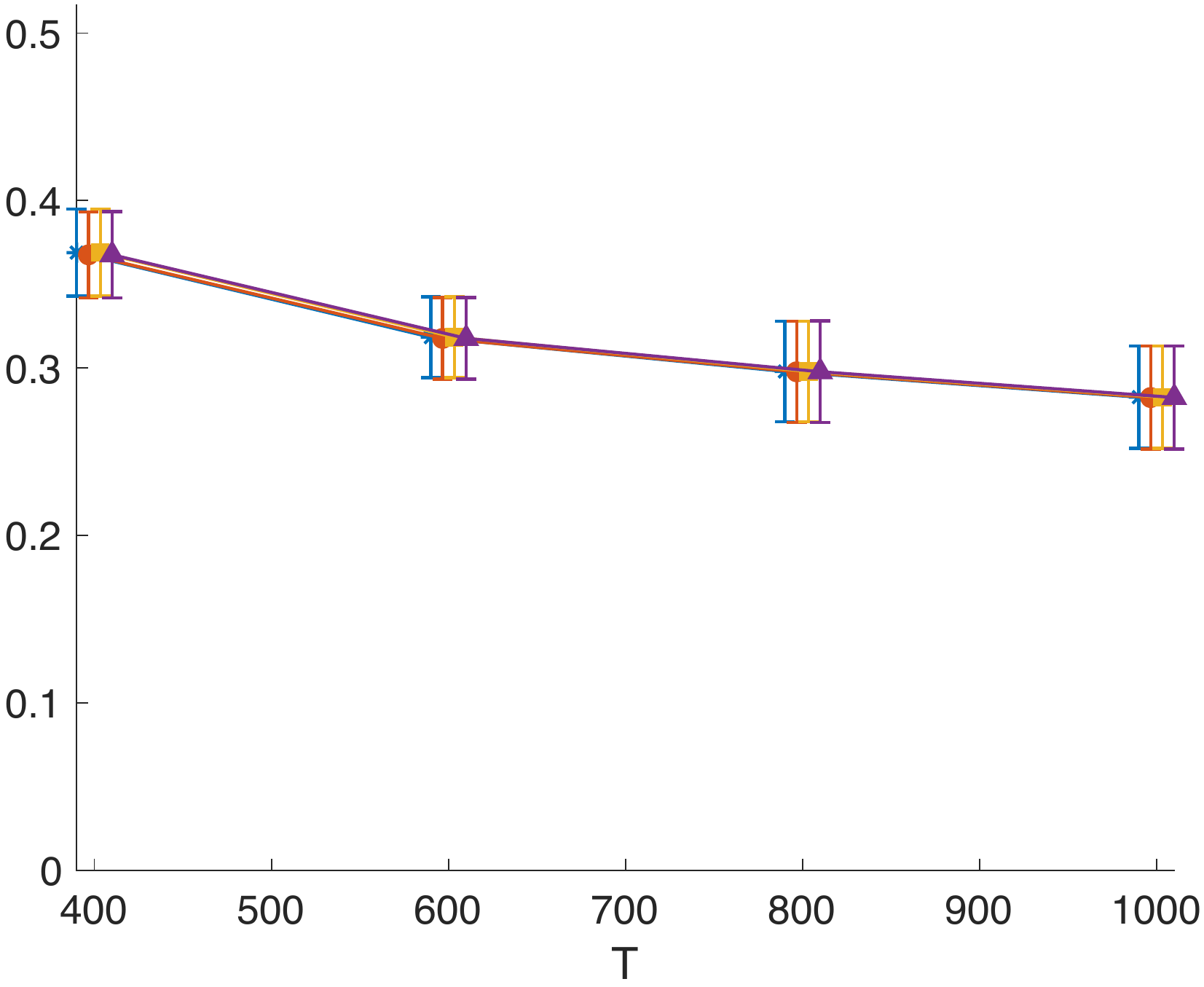}} &
\hspace*{-3mm}
\parbox[c]{11em}{\includegraphics[width = .28 \textwidth, keepaspectratio]{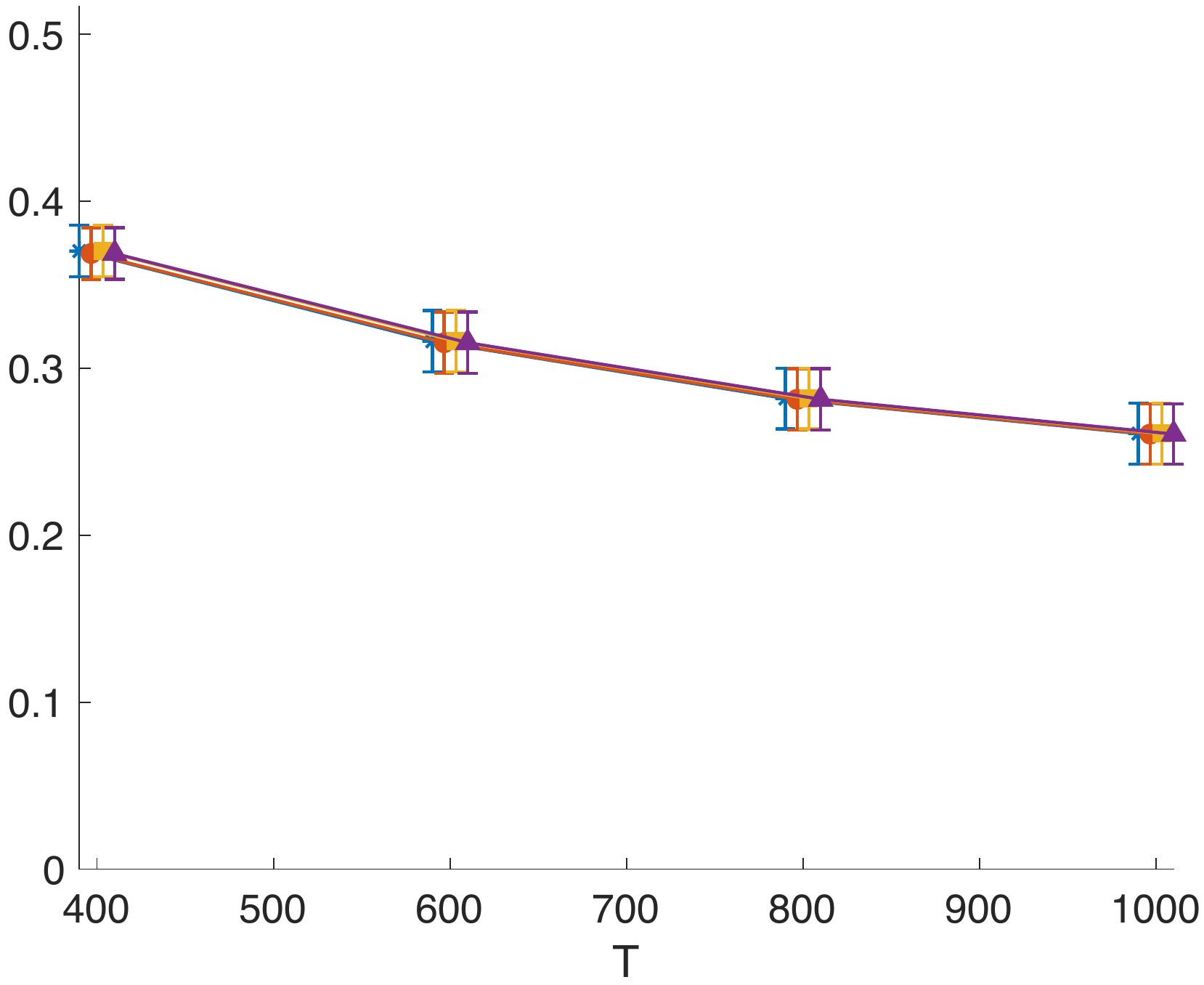}} \\
$\bZ$  \hspace*{3mm} & 
 \parbox[c]{11em}{\includegraphics[width = .28 \textwidth, keepaspectratio]{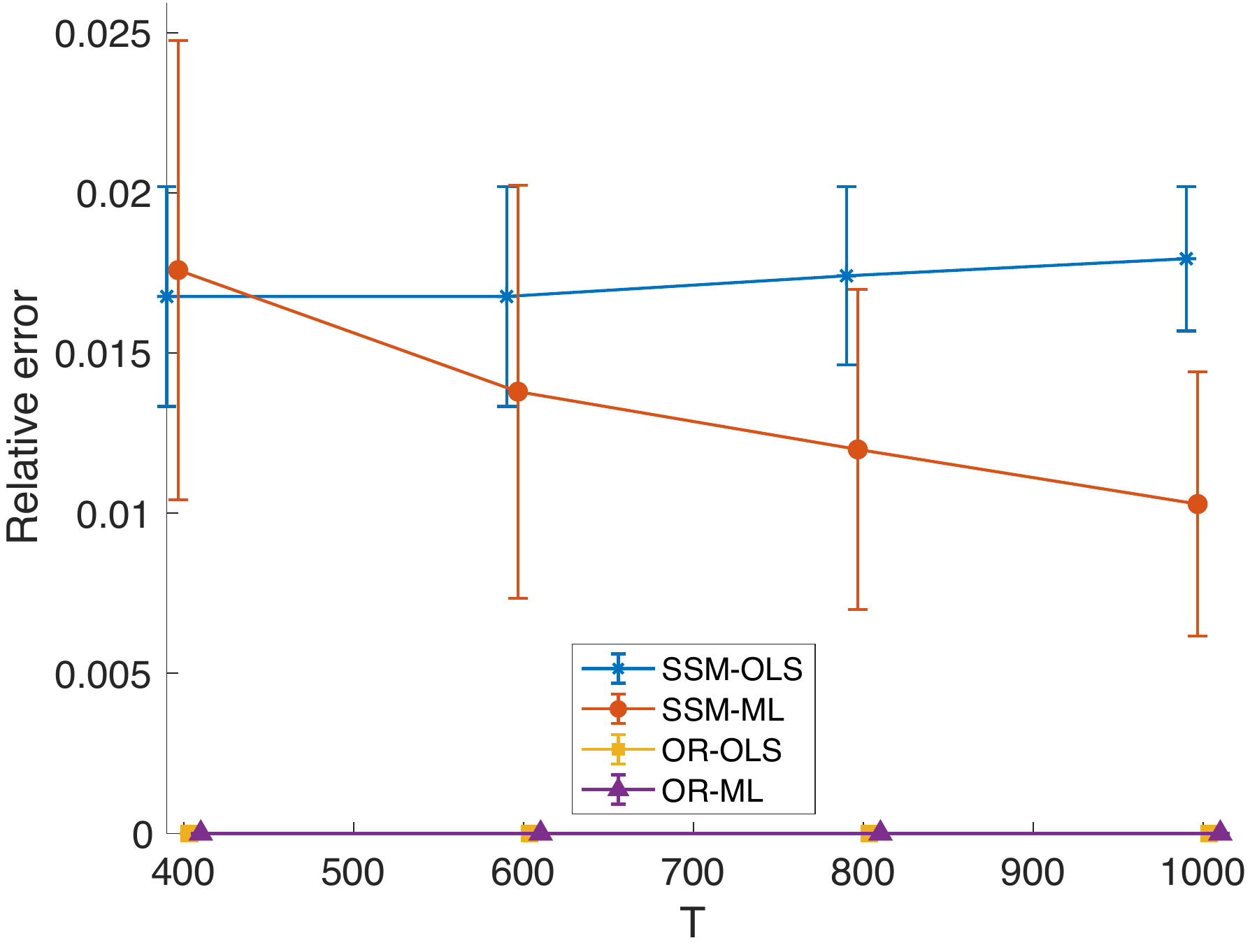}} &
 \hspace*{-3mm}
\parbox[c]{11em}{\includegraphics[width = .28 \textwidth, keepaspectratio]{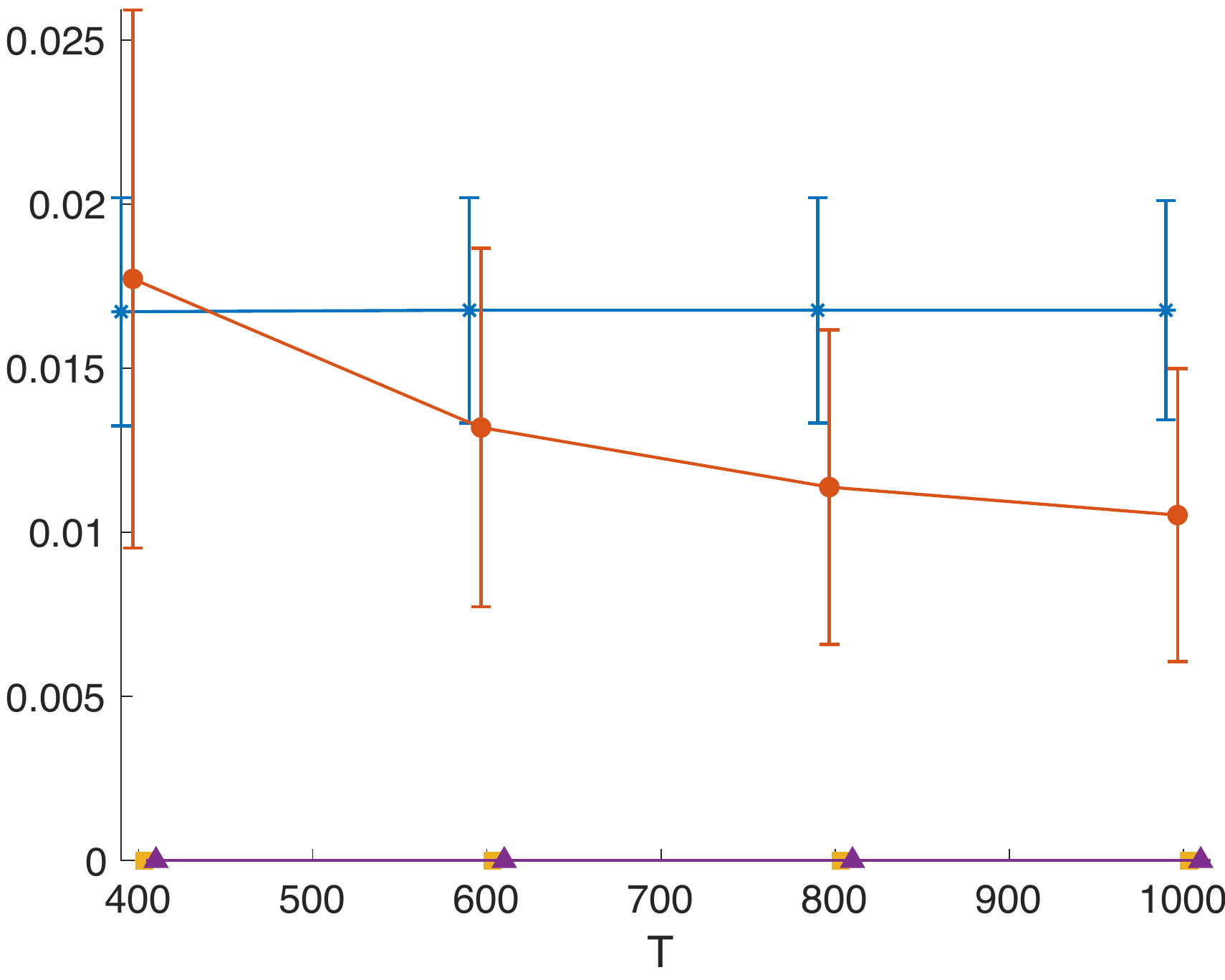}} &
\hspace*{-3mm}
\parbox[c]{11em}{\includegraphics[width = .28 \textwidth, keepaspectratio]{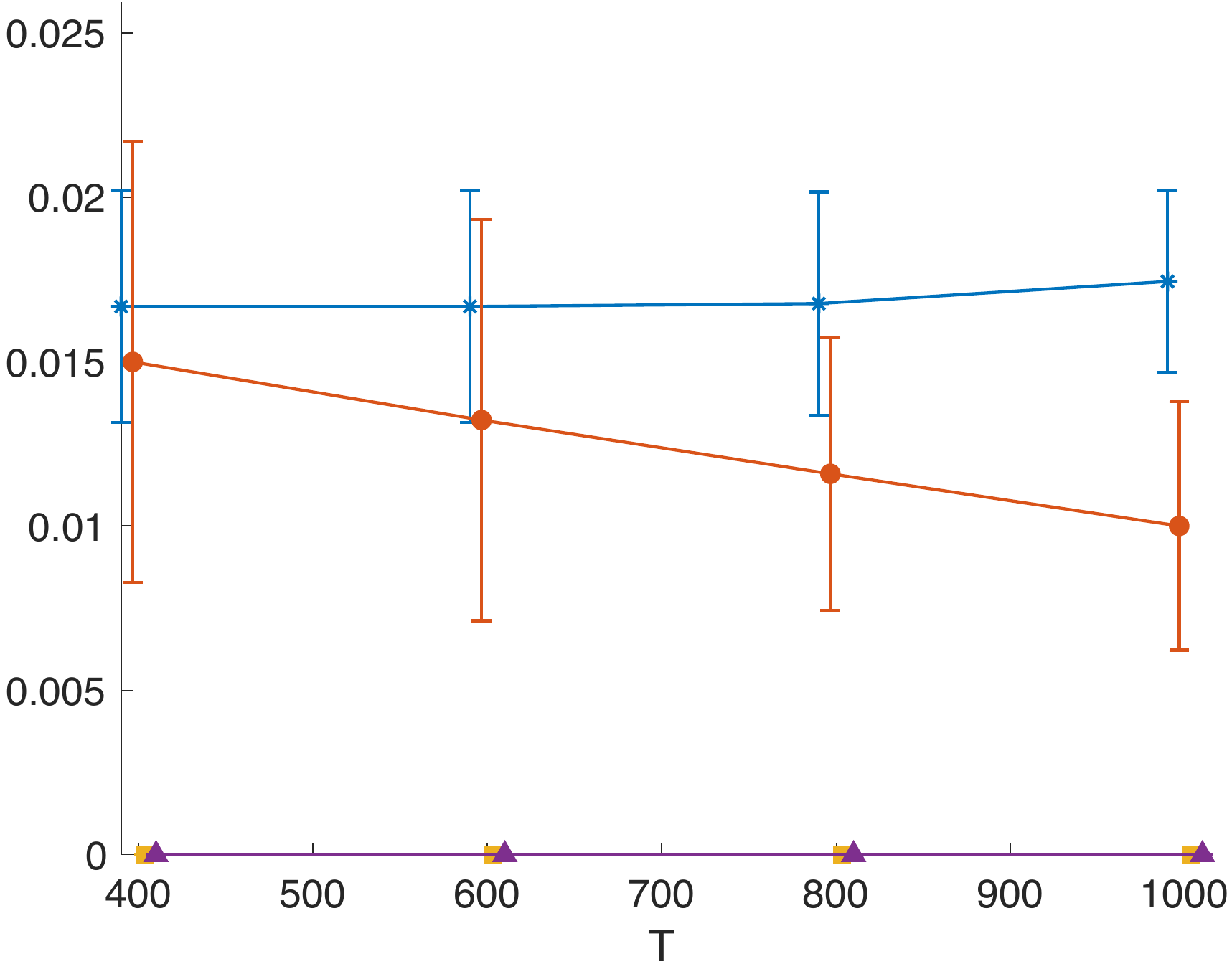}} \\
\end{tabular}
}
\caption{Accuracy of parameter estimation in the switching dynamics model \eqref{switch-dyn}.  
For each model parameter and dimension $(N,T)$, 500 simulations are conducted. 
For each estimation method, the median and median absolute deviation of the 
relative error are displayed as solid line and error bars.}
\label{fig: parameter estimation}
\end{center}
\end{figure*}

Figure \ref{fig: parameter estimation} shows the relative errors of  SSM-OLS, SSM-ML, OR-OLS, and OR-ML in model \eqref{switch-dyn}. 
Qualitatively similar results were found in models \eqref{switch-var}-\eqref{switch-obs} and are omitted here for the sake of space.   
SSM-ML has essentially the same accuracy as the oracles OR-OLS and OR-ML, 
which is not surprising given that this method estimates the regimes $(S_t)$ with high accuracy (top row of Figure \ref{fig: classification rate}). SSM-OLS estimates the observation parameters $\bC\bC'$ and $\bR$ 
as well as the other methods but it is less accurate in estimating the state parameters $\bA_{\ell j}$ and $\bQ_j$. 
The transition probability matrix $\bZ$ is very accurately estimated with 
a relative error between 0.01-0.02 for SSM-OLS and SSM-ML. 
(By construction, OR-OLS and OR-ML almost perfectly estimate this matrix.) 
Next, the estimation of $\bC\bC'$ is also very accurate (relative error between 0.03 and 0.06). 
The estimation of the $\bQ_j$ remains fairly accurate (relative error in the range $[0.05,0.1]$). 
Lastly, the estimation of the $\bA_{\ell j}$ and $\bR$ is overall mediocre (range $[0.3,0.48]$) 
although it  improves slowly as $T$ increases. For the $\bA_{\ell j}$, 
the relatively poor estimation may be partially explained by the error measure itself 
as outlined in the previous paragraph. In the case of $\bR$, 
the estimator $\widehat{\bR}$ shows non-negligible negative bias.

\setlength{\tabcolsep}{0pt}

\setlength{\tabcolsep}{6pt}


\paragraph{Estimation of stationary covariance and associated structures.}  

\begin{figure}[ht!]
\begin{center}
\hspace*{-7mm}
\begin{tabular}{ccc}
\includegraphics[width = .32 \textwidth]{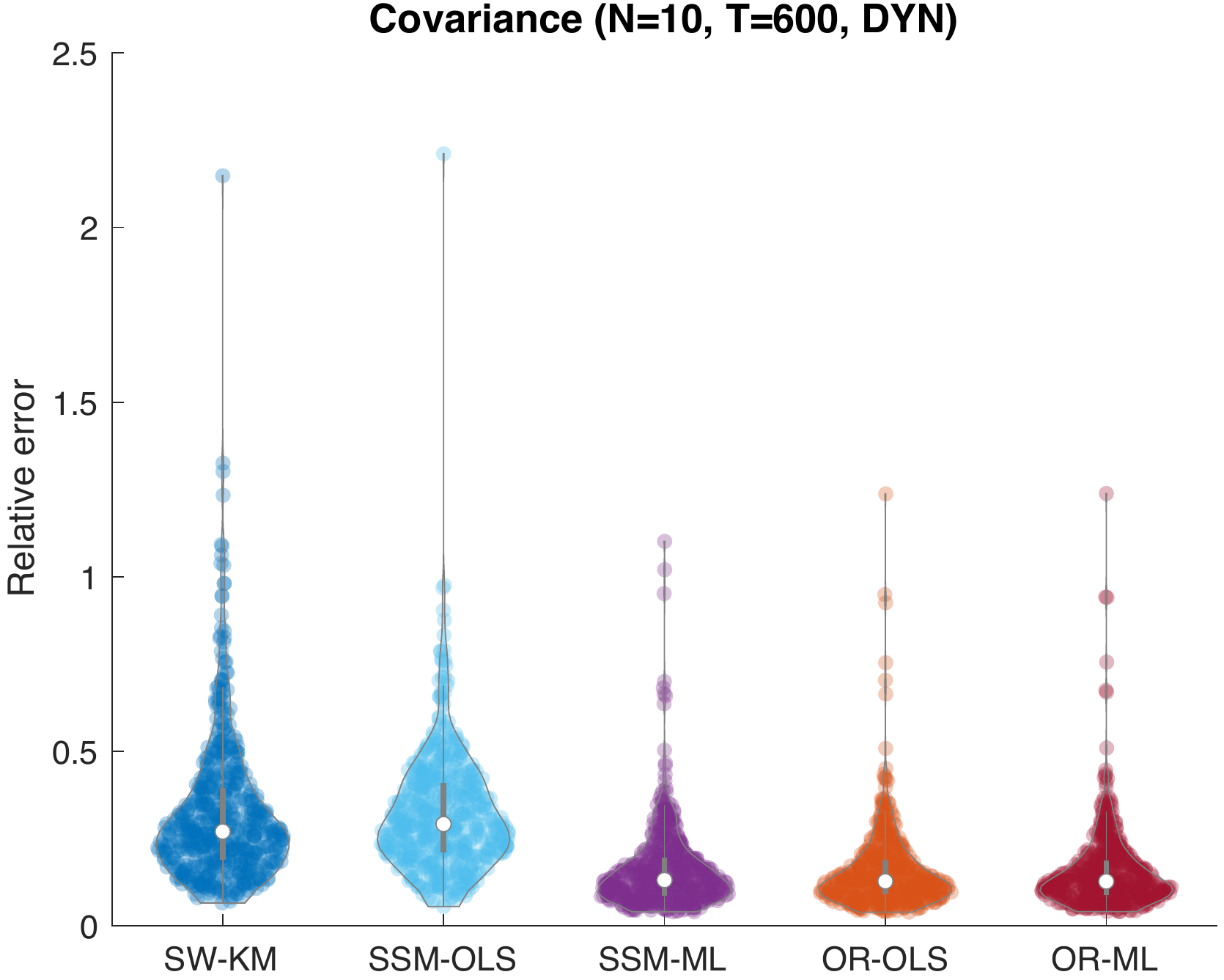} & 
\includegraphics[width = .32 \textwidth]{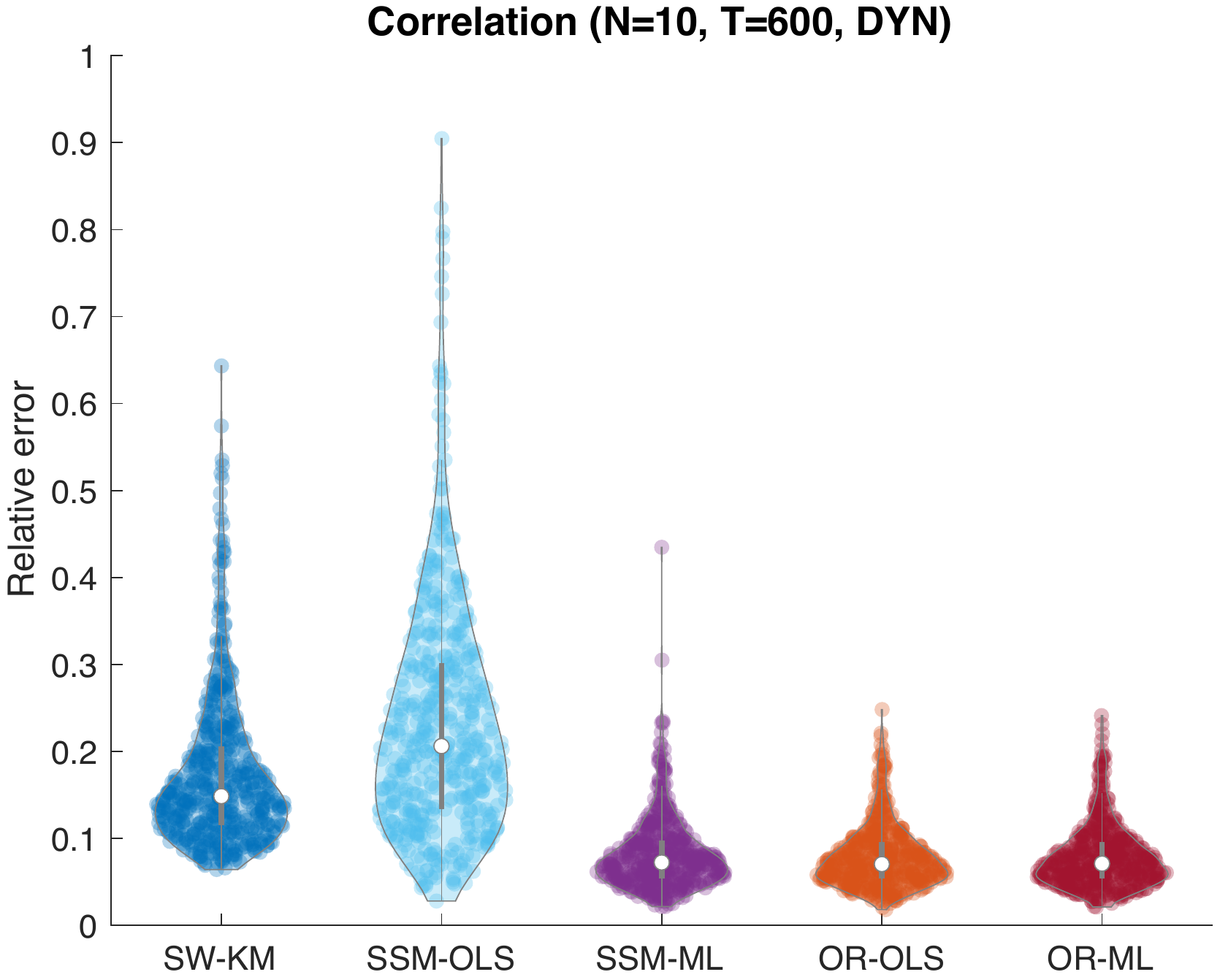} & 
\includegraphics[width = .32 \textwidth]{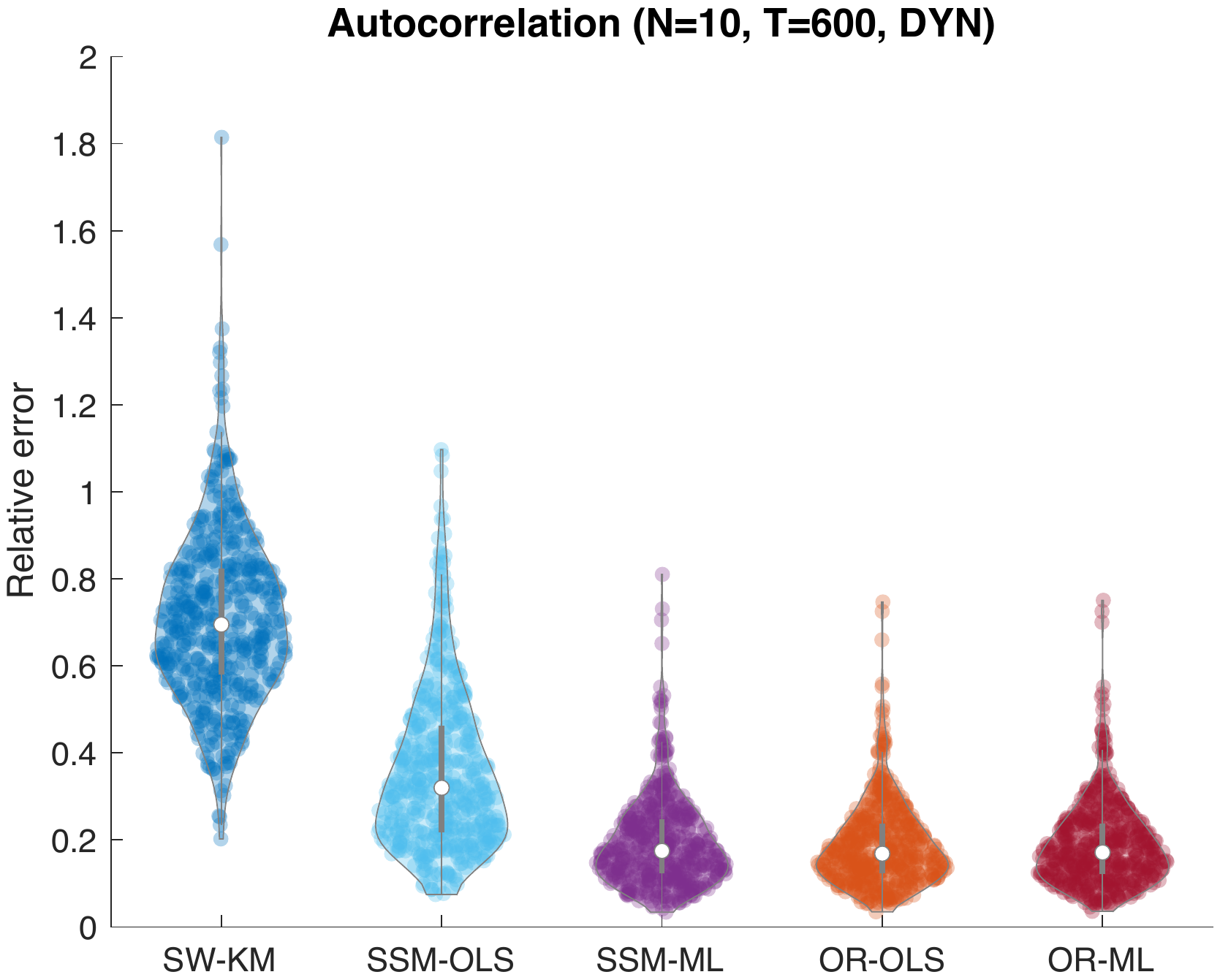} \\
\includegraphics[width = .32 \textwidth]{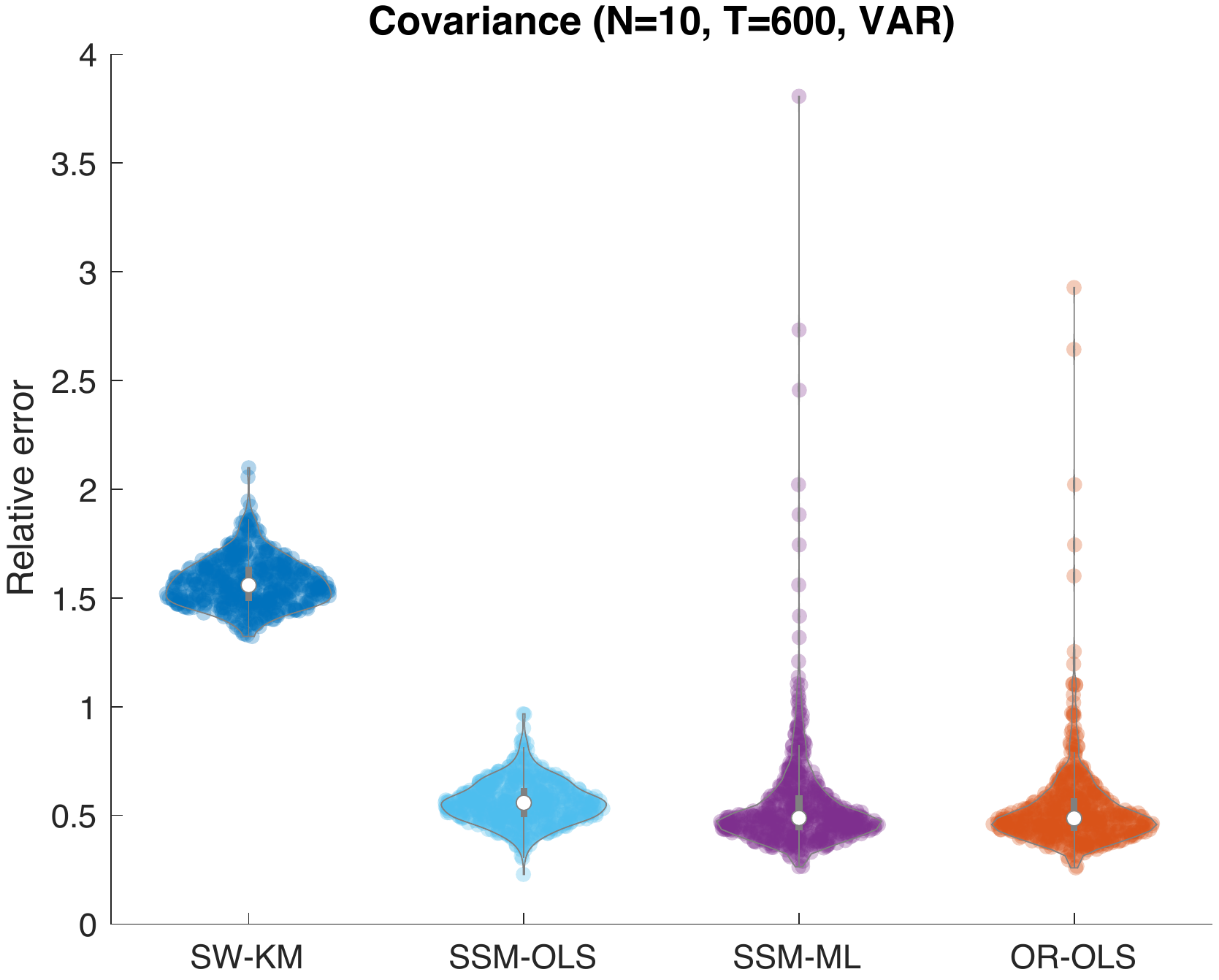} & 
\includegraphics[width = .32 \textwidth]{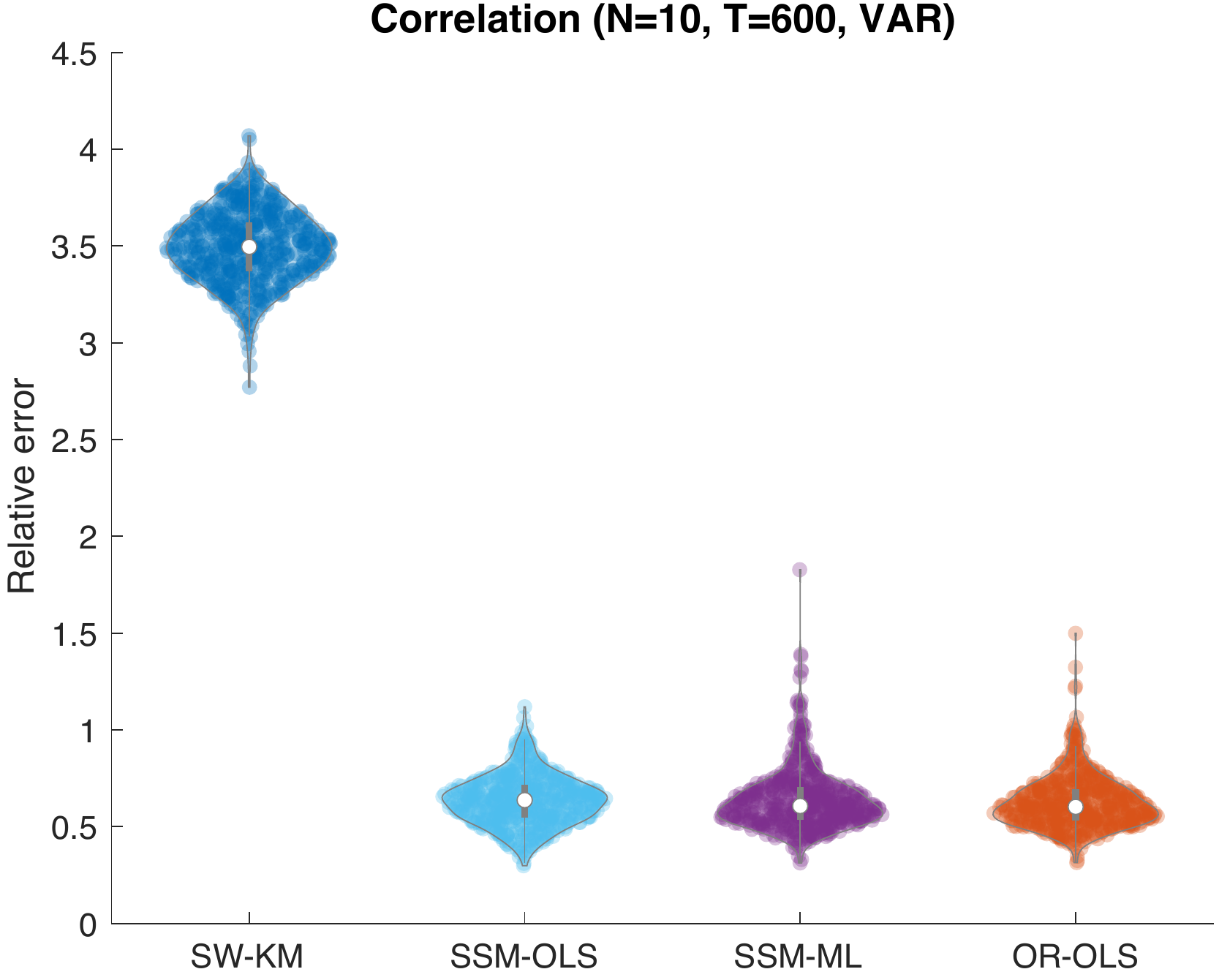} & 
\includegraphics[width = .32 \textwidth]{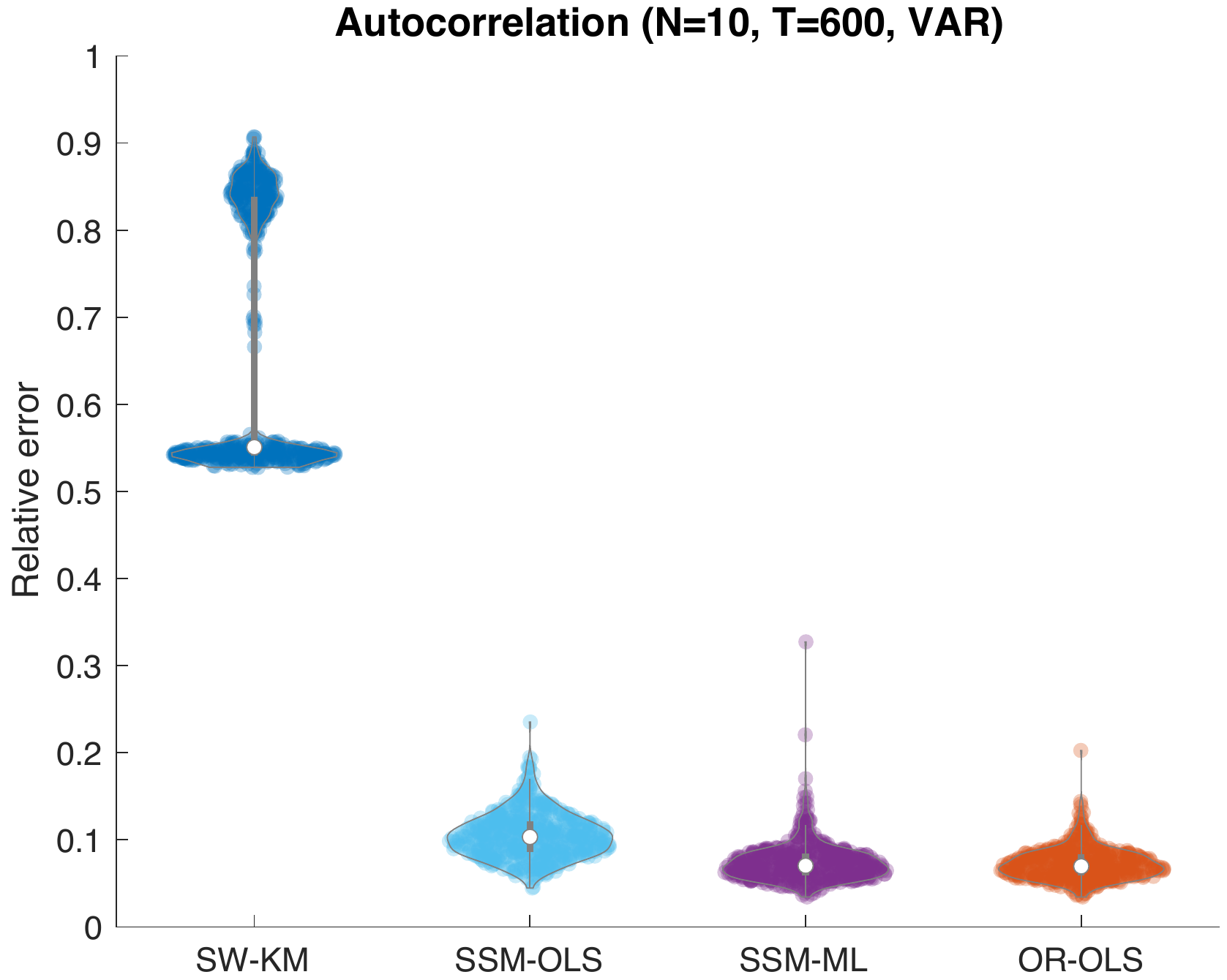} \\
\includegraphics[width = .32 \textwidth]{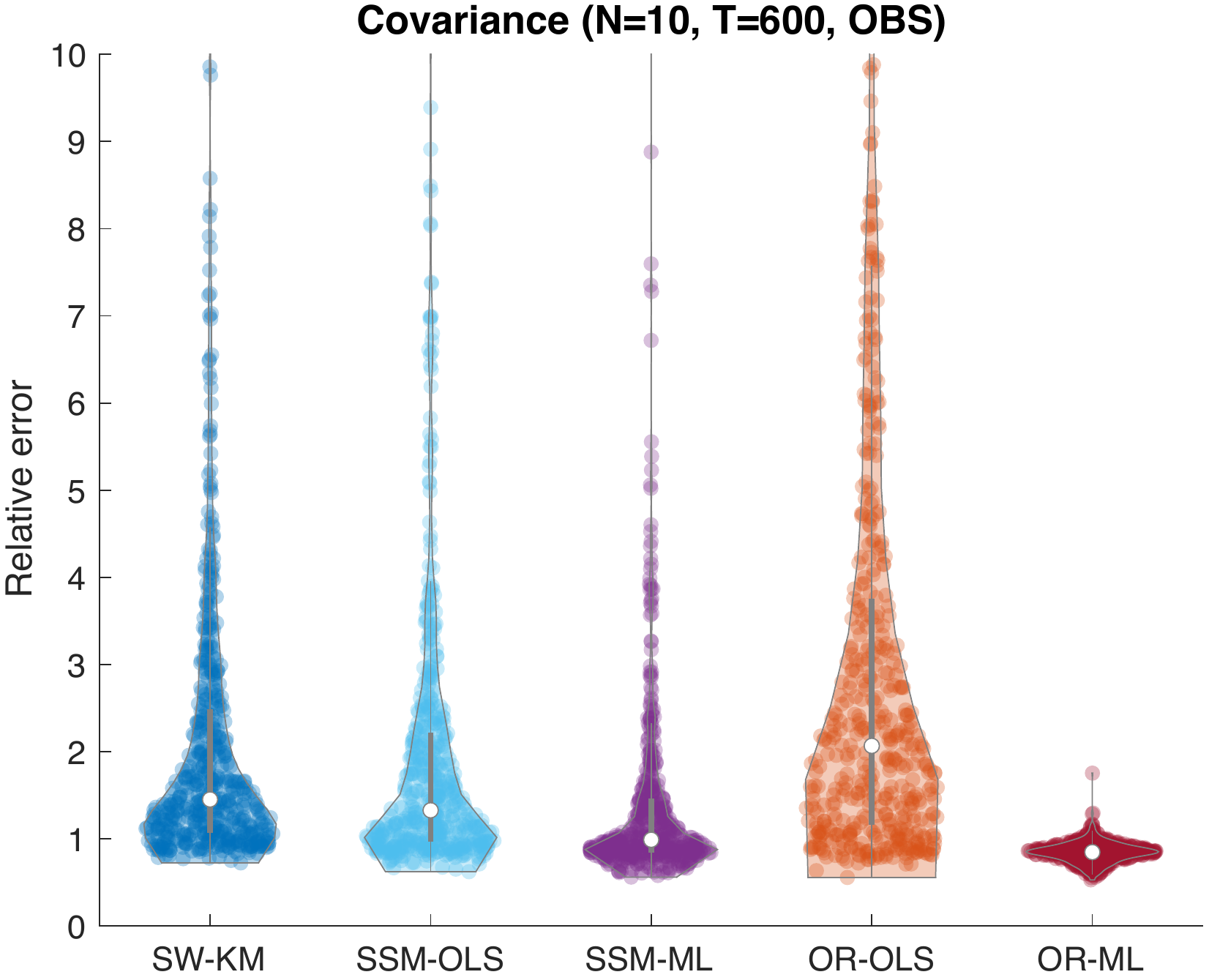} & 
\includegraphics[width = .32 \textwidth]{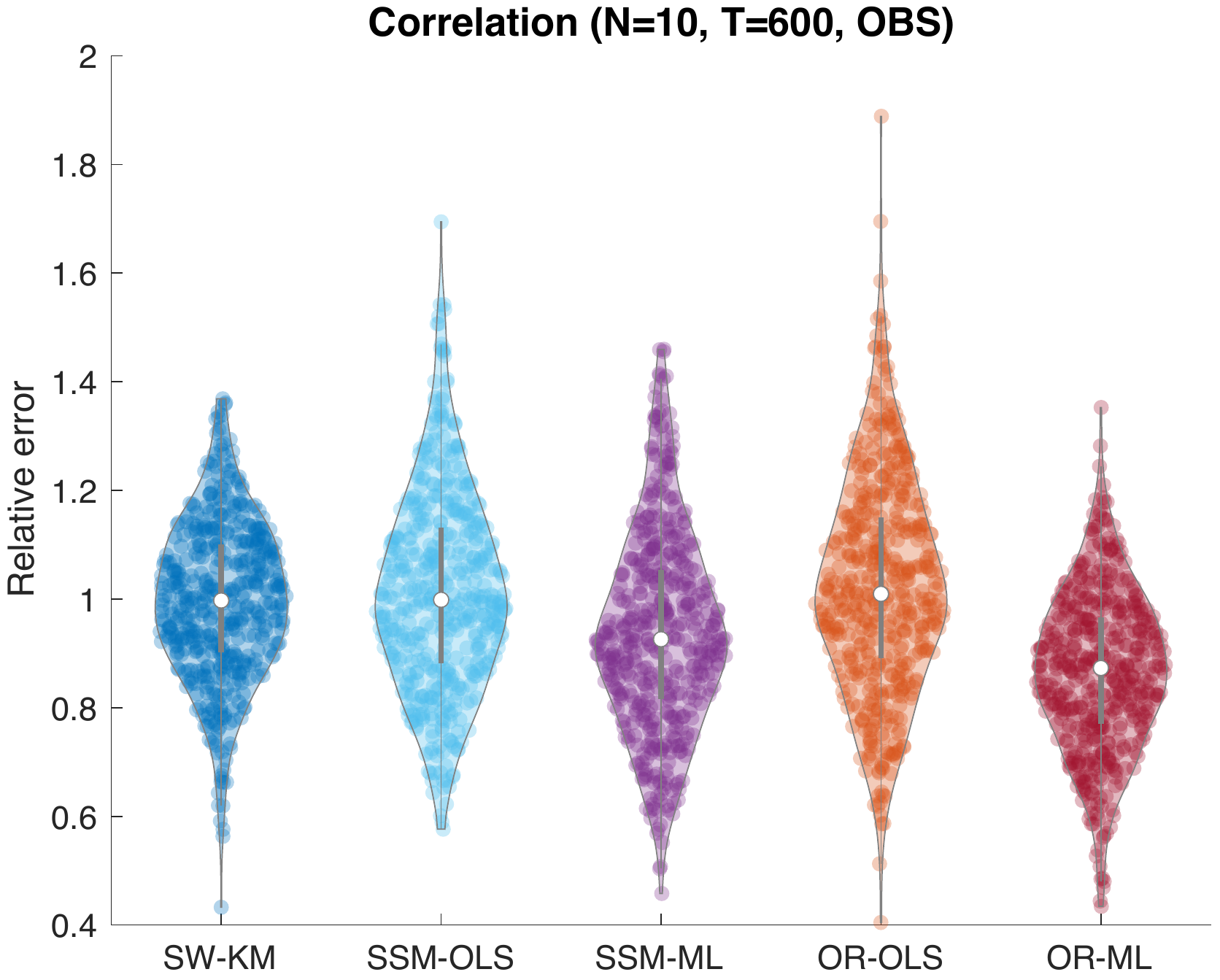} & 
\includegraphics[width = .32 \textwidth]{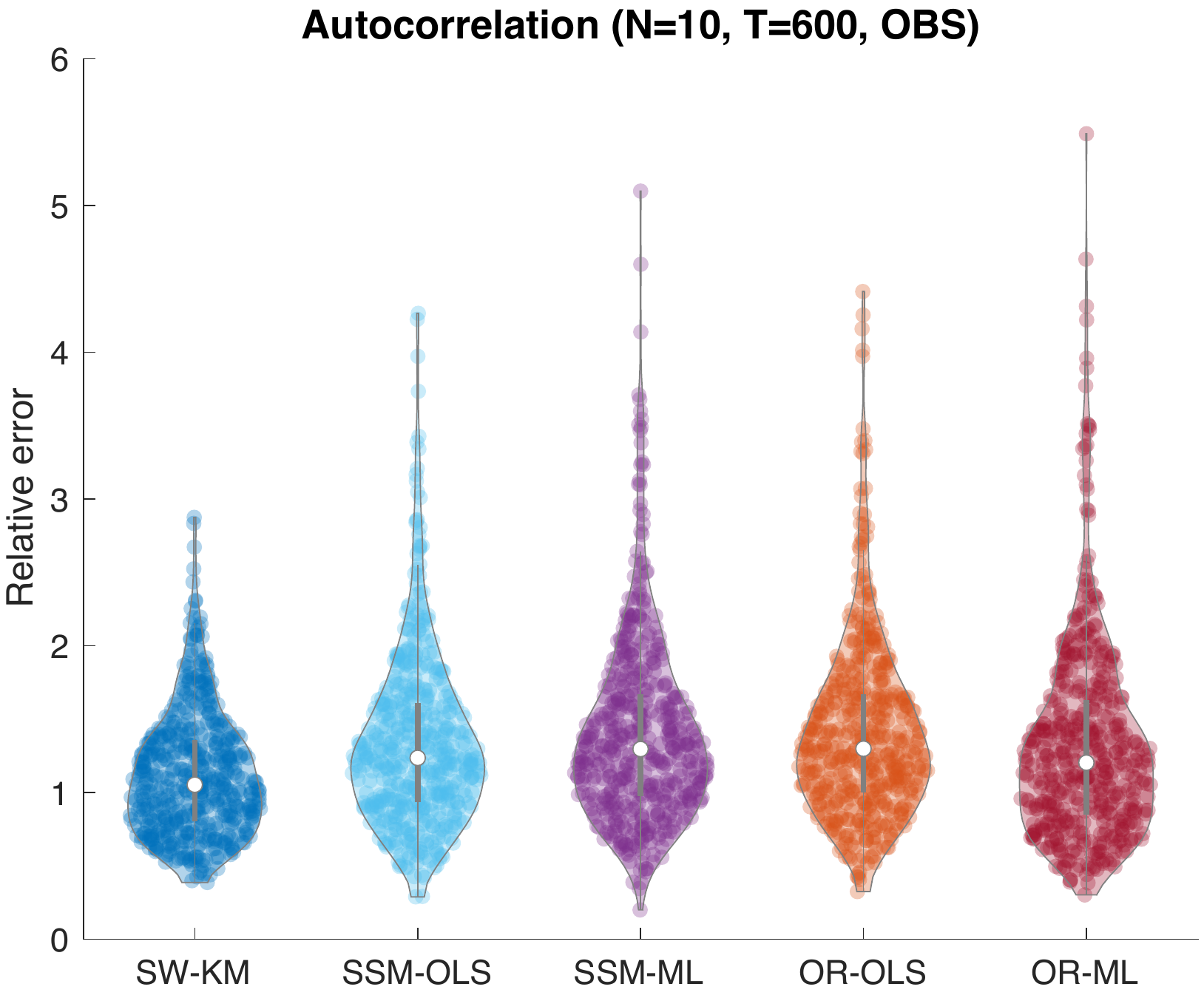} 
\end{tabular}
\caption{Accuracy in the estimation of stationary covariance, correlation, and autocorrelation structures
in models \eqref{switch-dyn}-\eqref{switch-var}-\eqref{switch-obs} (DYN, VAR, OBS). 
The distributions of relative errors over 200 simulations are displayed in the case $(N,T)=(10,600)$. Qualitatively similar
results hold  in most other simulation setups.}
\label{fig: covariance estimation}
\end{center}
\end{figure}

Figure \ref{fig: covariance estimation} displays results on the estimation of the stationary covariance matrices  $ \bSigma_{j}^{y} $, 
correlation matrices $\mathbf{R}^y_j$, and autocorrelation functions $\rho_{kj}^{y}$ for $ j=1,\ldots,  M$ and $k=1,\ldots, N$. 
As before, we measure estimation accuracy with the relative error across regimes, 
for example $  ( \sum_{j=1}^M \|  \widehat{\bSigma}_{j}^{y} -  \bSigma_{j}^{y} \|_{1,1}  ) 
 /  ( \sum_{j=1}^M \|    \bSigma_{j}^{y} \|_{1,1} )$, 
after matching estimates to parameters across regimes. 
 The autocorrelation functions $\rho_{kj}^{y}$ are estimated at lags $\ell=1,\ldots, 5$. 
 For reasons of space, we only present results in the case $(N,T)=(100,600)$.  
Results for other values of $(N,T)$ are qualitatively  similar. 
 
 As expected, the oracles OR-OLS and OR-ML achieve the best performance in virtually all setups. 
SSM-ML achieves nearly the same accuracy, 
which makes sense given its excellent performance in regime 
estimation for $N=10$ (left column in Figure \ref{fig: classification rate}). 
In comparison, the  relative error of SSM-OLS in models \eqref{switch-dyn}-\eqref{switch-var} 
 is higher on average and more variable.  
 In model \eqref{switch-dyn}, SW-KM performs slightly better than SSM-OLS and slightly worse than SSM-ML 
 with respect to  $ \bSigma_{j}^{y} $ and $\mathbf{R}^y_j$; 
 however it estimates $\rho_{kj}^{y}$ far worse than the other methods. 
 SW-KM also performs far worse than other methods and overall quite badly in model  \eqref{switch-var}. 
For all methods in model \eqref{switch-obs},  the estimation of model parameters $\btheta$ 
and $ \bSigma_{j}^{y} , \mathbf{R}^y_j , \rho_{kj}^{y} $ is highly unreliable. We ascribe this bad performance to the instability 
of the EM algorithm in this model (see discussion in Section \ref{sec: EM issues}).

 \paragraph{Bootstrap inference.} 
 Table \ref{table: bootstrap dyn} reports the pointwise coverage level 
 of bootstrap CIs  in  model  \eqref{switch-dyn}. 
To avoid visual clutter, standard errors are not displayed; 
they range from 1.5\% to 3.1\% 
for coverage levels between 75\% and  95\%. In other words, the observed coverage levels  
are not highly accurate but indicate if bootstrap CIs are in the general vicinity of the target level 90\%. 
Out of all combinations of inferential targets and model dimensions $(N,T)$ in the table, 
the observed coverage level falls within 5\% of the target level 90\% 
about 3 times out of 4  for the normal boostrap CIs 
and 2 times out of 3 for the percentile and basic bootstrap CIs. 
Other cases mostly result in undercoverage: 
overcoverage is rare and only occurs with the target $\bZ$.   
The noise covariance matrix $\bR$, 
whose point estimation is known to be inaccurate, 
is also poorly estimated with CIs.  
Regarding the projection matrix $\bC \bC'$, 
all three bootstrap CIs essentially attain nominal coverage 
 as soon as $T\ge 600$. 
For $T=1000$ and all $N$, the normal and percentile bootstrap CIs 
attain nominal coverage for all stationary covariance, correlation, 
and autocorrelation structures. 
For these two CI methods, 
given any fixed target and $N$,  
the observed coverage level 
appears to approach the nominal level 90\% as $T$ increases, although non-monotonically. 
Interestingly, for fixed $T$, the coverage level of these CIs  increases with $N$.

 \begin{table}[htp!]
\begin{center}
\scriptsize{
\begin{tabular}{ l  *{6}{c} | *{6}{c}}
\hline
Method &   COV & COR & ACF & $P_{\bC}$ & $\mathbf{Z}$ & $\bR$ 
 & COV & COR & ACF & $P_{\bC}$ & $\mathbf{Z}$ & $\bR$ \\
\hline
& \multicolumn{6}{c|}{$N= 10 ,\ T=400 $} &  \multicolumn{6}{c}{$N= 10 ,\ T=600 $} \\
\cline{2-13}
   percentile   &  87.5 &  87.8   & 86.3  & 88.3 & 92.8  & 57.5 &
         89.2  & 89.7 &  88.7  & 85.9  & 90.3 &  55.5 \\
    basic  &   87.1 &  87.2  & 83.1  & 88.7 & 84.3  &57.7 &
       88.8  & 89.1  & 86.6 &   86.1   & 83.0  & 55.6 \\
    normal  &    88.7 &   88.5  & 86.8  & 89.1 &  96.8 &  58.0 &
        89.8  &  89.8  & 89.1 &   86.3  & 92.3  &    55.6 \\
    \hline
& \multicolumn{6}{c|}{$N= 10 ,\ T=800 $} &  \multicolumn{6}{c}{$N= 10 ,\ T=1000 $} \\
\cline{2-13}
 percentile &    89.7 &  89.4  & 88.6 &  85.4  & 91.5   & 54.1 &
         87.0  & 86.9  &84.6 &  79.6 &  90.3  & 51.2 \\
   basic   &89.4   &88.9   &86.5  &85.8  & 85.1 &  54.5&
        86.9  & 87.1  & 83.2  & 79.9  & 88.0 &  51.6 \\
    normal   & 90.1  & 89.5  & 88.4  & 86.0  & 91.5   &54.7 &
       87.5  & 87.2  & 85.0  & 80.4 &  91.5  &51.6 \\
\hline
& \multicolumn{6}{c|}{$N= 50 ,\ T=400 $} &  \multicolumn{6}{c}{$N= 50 ,\ T=600 $} \\
\cline{2-13}
    percentile &    87.2 &  88.0   &86.4  & 90.1 &  94.3  & 79.7 &
    86.7  & 87.3 &  87.1  & 89.2  & 92.8  & 75.1 \\
    basic &  86.6   &88.3  & 85.1 &  90.0   &84.3   &79.6 &
     85.4  & 88.1  & 85.9   &89.1 & 87.8  & 75.0 \\
    normal    & 89.6  & 89.7 &   88.9   & 90.4 &   96.5  & 79.5 &
       88.0   & 89.3  & 88.5 &  89.5 &  95.5 &  75.0 \\
\hline
& \multicolumn{6}{c|}{$N= 50 ,\ T=800 $} &  \multicolumn{6}{c}{$N= 50 ,\ T=1000 $} \\
\cline{2-13}
      percentile&     89.9 &  89.7 &  89.8  & 88.4  & 92.3 &  72.5 &
           87.9  & 87.4  & 87.1 &  87.4    &89.5 &  68.8 \\
    basic  & 88.1 &  89.0  &  87.8   &88.3  & 88.3  & 72.4 &
       86.9  & 88.2   &86.7  & 87.2  & 85.3 &  68.8 \\
    normal &    89.7 &   90.1 &  89.5 &  88.5 &  93.8 &  72.3 &
       88.0  & 88.8 &  87.9  & 87.4 &  89.3   &68.7 \\
  \hline
& \multicolumn{6}{c|}{$N= 100 ,\ T=400 $} &  \multicolumn{6}{c}{$N= 100 ,\ T=600 $} \\
\cline{2-13}
  percentile &     86.5 &  87.3  & 86.4  & 89.6   &93.8  & 85.1 &
            87.5 &  87.7 &  87.0  &  88.8  &  88.5 &   82.7\\
    basic  &  85.3   &87.1 &  84.1   &89.5    & 81.0 &  84.9 &
       85.6 &   87.2 &  85.1 &  88.9  &   85  &  82.6\\
    normal   & 88.8 &  89.5  & 87.5  & 89.8  &  96.5 &  84.9 &
        87.8   &88.6   & 87.5  & 89.1   & 93.8   &82.6\\
\hline
& \multicolumn{6}{c|}{$N= 100 ,\ T=800 $} &  \multicolumn{6}{c}{$N= 100 ,\ T=1000 $} \\
\cline{2-13}
    percentile &      87.6   &88.1  & 87.0  & 88.8  &   86.5   & 81.6 &
   88.7  & 88.6  & 89.0  & 88.8  & 86.3   &80.4 \\
     basic  & 86.6   &88.2  & 86.2  & 88.8  & 83.3  & 81.6 &
    86.5    &87.8 &  87.2&    88.9  & 85.8 &   80.4 \\
    normal   & 88.3  &  89.2   &87.8 &  89.0     & 90.0  & 81.7   &    
88.1 &   88.5 &   88.8  & 89.1   &  89.0   &80.5 \\
\hline
\end{tabular}
}
\caption{Pointwise coverage level (\%) for parametric bootstrap confidence intervals in the switching dynamics model \eqref{switch-dyn}. Nominal level: 90\%. Number of simulations: 200 with $B=100$ bootstraps for each. 
Targets: stationary covariance matrices $\boldsymbol{\Sigma}^{y}_j$ (COV), correlation matrices $\bR_j^y$ (COR) and autocorrelation functions $\rho_{kj}^{y}$ (ACF), projection matrix $P_{\bC} = \bC\bC'$, and model parameters $\mathbf{Z}$ and 
$\mathbf{R}$. For each target, the average coverage over all target coefficients and simulations is reported.}
 \label{table: bootstrap dyn}
\end{center}
\end{table}%

Table \ref{table: bootstrap var} reports bootstrap results in the switching VAR model \eqref{switch-var}.  
 In this model, given the high dimension $O(MpN^2)$ of model parameters $\btheta$ relative to the number of available data $NT$,   
and given the suboptimal performance of SSM-ML in regime estimation (at least for $N=50,100$, see middle row of Figure \ref{fig: classification rate}), it is no surprise that the coverage level falls short of its target 90\% in many cases.     
In spite of this, the percentile boostrap CIs reliably cover the 
stationary covariance $\boldsymbol{\Omega}^y_j$  and correlation  $\mathbf{R}_j^y$
for most $(N,T)$ with coverage in $[.85,.95]$ in 20 out of 24 cases. 
Also, the normal bootstrap CIs show satisfactory
coverage  for all inferential targets except $\bZ$ when $N=10$ 
but become highly unreliable when $N=50,100$.

 \begin{table}[ht]
\begin{center}
\scriptsize{
\begin{tabular}{ l  *{6}{c} | *{6}{c}}
\hline
Method &   COV & COR & ACF & $\bA$ & $\bQ$ & $\bZ$ 
 & COV & COR & ACF & $\bA$ & $\bQ$ & $\bZ$ \\
\hline
& \multicolumn{6}{c|}{$N= 10 ,\ T=400 $} &  \multicolumn{6}{c}{$N= 10 ,\ T=600 $} \\
\cline{2-13}
 percentile  &90.1  & 95.5  & 91.0  & 99.1   &67.9  & 99.8 &
88.9   &94.1  & 90.5   &97.5 &  73.7  & 98.0  \\
basic &    81.3  & 73.7  & 80.9  & 85.2  & 82.3 &  80.3 &
88.1   &80.7  & 85.2 &  87.8  & 88.2   &83.3  \\
normal  & 89.7 &  81.3 &  88.6 &  88.6 &  89.2 &  93.8 & 
92.2 &  89.3  & 91.6 &  90.9  & 94.8  & 97.5 \\ 
    \hline
& \multicolumn{6}{c|}{$N= 10 ,\ T=800 $} &  \multicolumn{6}{c}{$N= 10 ,\ T=1000 $} \\
\cline{2-13}
 percentile &92.1  & 95.3   &91.9  & 96.8 &  82.3 &  97.0 &
90.1  & 94.2  & 88.8 &  94.7  & 83.4  & 94.8 \\
basic  &88.9    & 82.8 &  86.1  & 88.9  & 89.5  & 89.0 &
89.2   &83.9  & 87.5 &  88.7  & 90.4  & 86.0\\
normal &92.2   &89.2 &  92.4 &  91.8  & 95.2  & 95.5 &
91.0 &  89.6 &  92.0  & 91.1  & 95.9  & 94.5 \\
\hline
& \multicolumn{6}{c|}{$N= 50 ,\ T=400 $} &  \multicolumn{6}{c}{$N= 50 ,\ T=600 $} \\
\cline{2-13}
percentile  & 87.6  & 87.3  & 72.0  &  100.0  & 86.8 &  89.3 &
88.9 &  88.7 &   78.6 &  99.9 &  83.4 &  86.3  \\
    basic   & 66.0  & 43.6  &58.9  & 70.2  & 81.7   &39.3 &
      69.4 &  49.0  & 61.4  & 72.2  & 80.7 &   49.5\\
    normal &  84.4  & 44.7  & 69.4  & 81.5  & 94.9 &  52.3 &
     84.7   &50.0  &  72.6 &  80.4  &  94.5  &   64.0 \\
 \hline
& \multicolumn{6}{c|}{$N= 50 ,\ T=800 $} &  \multicolumn{6}{c}{$N= 50 ,\ T=1000 $} \\
\cline{2-13}
 percentile &     88.0 &  88.2 &   73.7  & 99.9  & 77.7  & 78.5 &
 88.7  & 89.1  & 78.3  & 99.9  & 73.0 &   77.3 \\ 
   basic &70.9  & 51.7  & 59.3  & 72.6 &  78.2  &  50.5 &
  73.4  & 57.7  & 64.0   & 73.6  & 74.5  &  52.5  \\ 
  normal  &85.4  & 52.4  & 71.5  &  76.5  &  92.9 &  59.5 & 
  86.8  & 58.7  & 75.6  & 76.1  & 89.3  & 63.3 \\ 
  \hline
& \multicolumn{6}{c|}{$N= 100 ,\ T=400 $} &  \multicolumn{6}{c}{$N= 100 ,\ T=600 $} \\
\cline{2-13}
 percentile& 84.6 &  84.2 &  58.5 &   99.8 &  85.2 &  81.5 &
       85.0  & 84.0 & 70.2   &99.9   & 85.5   &82.5 \\
basic  & 54.0 &  36.0 &  55.7 &  11.0 &   80.2  & 39.0 &
       51.5   &35.2  & 53.7   & 29.1  & 79.1 &  42.5 \\
  normal &  66.4  & 37.1   &59.4 &  14.1 &  89.0   &52.3 &
        69.6 &  37.4 &  63.0  & 43.3  & 91.0 &  59.5 \\
\hline
& \multicolumn{6}{c|}{$N= 100 ,\ T=800 $} &  \multicolumn{6}{c}{$N= 100 ,\ T=1000 $} \\
\cline{2-13}
percentile &  86.4  & 85.3  & 73.8  &   99.9 &  85.0 &   79.3 &
  87.1 &   86.3 &  73.2 &  99.9 &  84.1 &  81.0 \\
    basic    &  57.2 &   41.1  & 54.7  & 40.9  & 80.4     & 45.0 &
     63.3 &  46.6 &  56.9 &  48.9 &   80.5 &      46.3\\
    normal  &    74.6   &43.9   &65.4  & 53.5 &  92.8  & 59.4 &
    78.8   &49.3  & 69.7  & 62.2  & 94.0  & 63.0\\
\hline
\end{tabular}
}
\caption{Pointwise coverage level (\%) for parametric bootstrap confidence intervals in the switching VAR model \eqref{switch-var}. Nominal level: 90\%. Number of simulations: 200 with $B=100$ bootstraps for each. 
Targets: stationary covariance matrices $\boldsymbol{\Sigma}^{y}_j$ (COV), correlation matrices $\bR_j^y$
(COR), autocorrelation functions $\rho_{kj}^{y}$ (ACF), and model parameters $\bA_{\ell j}$, $\bQ_{j}$, and $\bZ$ ($\ell=1,\ldots,p,\, j=1,\ldots,M$).
For each target, the average coverage level over all coefficients and simulations is reported.}
\label{table: bootstrap var}
\end{center}
\end{table}%

We finally report attempts at nonparametric bootstrap in models \eqref{switch-dyn}-\eqref{switch-var}. 
We follow \cite{Airlane2013} who themselves adapt the bootstrap of \cite{Stoffer1991} to switching SSMs.
In short, the nonparametric bootstrap starts by simulating independently each regime $S_t^\ast$ 
according to the smoothed probabilities $P_{\widehat{\btheta}}(S_t = j| \by_{1:T}), \, j=1,\ldots, M$ for $t=1,\ldots, T$. 
After that, the residuals $(\hat{\mathbf{e}}_t)$  of the Kalman filtering step are standardized and resampled with replacement. The bootstrapped residuals are then used to generate bootstrap samples $(\by_t^\ast)$  via the innovation form of the SSM. (Please refer to the previous references for details.) 
In model \eqref{switch-dyn} we observe considerable undercoverage for all CI methods, inferential targets, and $(N,T)$. 
In model \eqref{switch-var} we observe strong overcoverage for the parameters $\bA_{\ell j}, \bQ_j, \bZ$ and covariances $\boldsymbol{\Omega}_j^y$ (99-100\% for all $N,T$) and undercoverage for the 
correlations $\bR_j^y$ (58-69\%) and autocorrelations  $\rho_{kj}^y$  (43-77\%). 
For the two latter quantities, the coverage appears to decrease as $T$ increases. 
We conjecture that one important cause for bad coverage in the nonparametric approach resides in the resampling method for the regimes $(S_t)$, which may fail to properly account for their variability. Indeed, the probabilities $P(S_t = j| \by_{1:T})$ are often either close to 0 or 1 so that the bootstrapped regimes $(S_t^{\ast})$ are almost always equal to the original estimates $(\widehat{S}_t)$ in \eqref{regime estimator}. In contrast the parametric bootstrap, by generating $(S_t^{\ast})$ as a proper Markov chain according to 
the estimated initial probabilities $\widehat{\boldsymbol{\pi}}$ and transition probabilities $\mathbf{\widehat{Z}}$, produces a good variety of regime sequences across  bootstrap replications.

 \paragraph{Summary of simulation results.}

\begin{itemize}

\item The maximum likelihood estimation method under study (SSM-ML) recovers the true regimes $(S_t)$ extremely well in model \eqref{switch-dyn}, not so well in model \eqref{switch-var} except when $N$ is small (this is due to the high dimension $O(MpN^2)$ of  $\btheta$), and fairly well in model \eqref{switch-obs} where the classification rate increases fairly rapidly with $T$. 
In most cases, the sliding window (SW-KM) and  EM initialization methods (SSM-OLS) estimate regimes with substantially worse accuracy.

\item  SSM-ML estimates the parameters $\bC $ and $ \bQ$ with high  accuracy in model \eqref{switch-dyn};  
its accuracy for $\bA $ and $\bR$ is fair to mediocre but improves as $T$ increases. In model \eqref{switch-var} this method estimates $\bA$ reasonably well for $N$ sufficiently large 
but its accuracy for $\bQ$ fluctuates widely with $(N,T)$. 
All methods perform very poorly in model \eqref{switch-obs}, 
including the oracles OR-OLS and OR-ML. This and the oscillatory behavior of the EM 
suggest that the filtering/smoothing approximations used in the EM 
are too crude for this model which contains many  more latent variables than model \eqref{switch-dyn} (about $M$ times as many). 
SSM-OLS estimates parameters with similar or worse accuracy than SSM-ML in almost all situations. 
The transition probability matrix $\bZ$ is estimated with very high precision in all cases, likely because its entries are either close to 0 or to 1.

\item  SSM-ML estimates well the stationary covariance, correlation, and autocorrelation structures 
$\bSigma_j^y, \bR_j^y, \rho_{kj}^y$  
 in model \eqref{switch-dyn} (average relative error in $[.05,.23]$ for all $(N,T)$).  
In model \eqref{switch-var} it estimates $\rho_{kj}^y$ accurately (average relative error in $[.09,.14]$ for all $(N,T)$) but not $\bSigma_j^y$ and $ \bR_j^y$. In the same model and for the same targets, the oracle estimator becomes much more accurate than SSM-ML as $N$ increases (by a factor at least 2 for $N=50,100$) whereas SSM-OLS has similar or worse performance than SSM-ML in most cases. In model \eqref{switch-obs} all methods estimate $\bSigma_j^y, \bR_j^y, \rho_{kj}^y$ poorly with average relative errors between 0.7 and 2.1.

\item The percentile, basic, and normal (parametric) bootstrap CIs have pointwise coverage levels reasonably close to nominal (within 5\%) for most targets and most $(N,T)$ in model \eqref{switch-dyn}. Their performance is good in model \eqref{switch-var} for $N=10$ (more so for the normal CIs) 
but not for $N=50,100$ where $T$ is comparatively too small to enable accurate estimation of regimes and 
model parameters. The coverage level is  poor in model \eqref{switch-obs}. The nonparametric bootstrap CIs suffer from considerable under- or overcoverage in almost all situations. 

\item The simulation results are optimistic in the sense that the estimation methods are based on the correct specification of the generative model. (Note that evaluating parameter estimation in a misspecified model would not make sense.) On the other hand, the signal-to-noise ratio used in the simulations are realistic for EEG applications. In this regard, the simulation results give a somewhat realistic view of the statistical accuracy that can be expected in practice.   
 
\end{itemize}


\section{Analysis of EEG data}
\label{sec: data analysis}

\subsection{Epileptic seizure data}

\begin{figure}[h]
\begin{center}
\includegraphics[width=.5\textwidth, keepaspectratio]{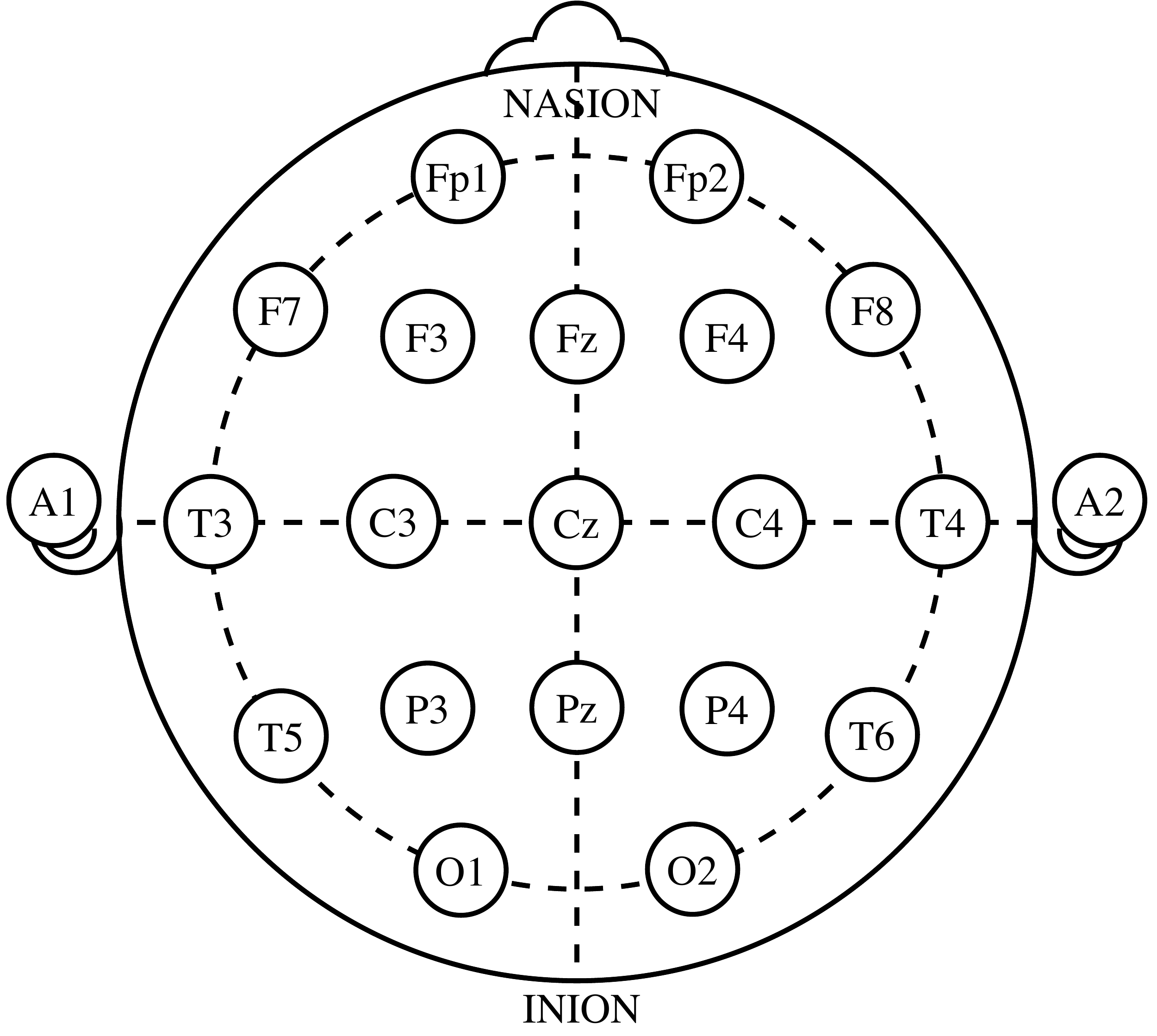}
\caption{International 10-20 system for standardized location of EEG electrodes. Recordings from 16 electrodes were used in the data analysis (all but A1, A2, Fz, Cz, and Pz).}
\label{fig: eeg 10-20}
\end{center}
\end{figure}

We first illustrate switching SSMs with EEG data recorded from a female patient  diagnosed with left temporal lobe epilepsy. The data were  collected at the Epilepsy Disorder Laboratory at the University of Michigan (primary investigator: Dr. Beth Malow, M.D.) and have been  analyzed by \cite{Ombao2001,Ombao2005}, \cite{Samdin2017}, \cite{Wang2018}, and \cite{Guerrero2021}. The data were recorded shortly before and during the occurence of an epileptic seizure (pre-ictal, ictal, and  late-ictal periods) with $N=16$ differential electrodes at the sampling rate 100\,Hz. The duration of the recording is 500\,s  which, after downsampling to 50\,Hz,  results in  $T=25000$ samples. Standard electrode locations are shown in Figure \ref{fig: eeg 10-20}.

\paragraph{Model selection.}

Models \eqref{switch-dyn}-\eqref{switch-var}-\eqref{switch-obs} were fitted to the data using a wide range of hyperparameters, namely  $1\le M \le 6$, $1\le p \le 8$, and $1\le r \le 16$. 
Not all models in the search domain were fitted as the space and time requirements of the EM algorithm 
preclude  taking large values of  $(M,p,r)$ simultaneously. 
 A notable exception is the switching VAR model \eqref{switch-var} which has no hidden state vector and thus requires far less storage and calculations.  Fitting this model to the epileptic seizure data only took a few minutes even for relatively large values of $M$ and $p$. Fortunately, model \eqref{switch-var} far surpasses models \eqref{switch-dyn}-\eqref{switch-obs} in  log-likelihood at the cost of a modest increase in model complexity. Accordingly, our search focused on this type of model and we are confident to have found the best model for both AIC and BIC over the entire initial range of models and hyperparameters.

\paragraph{Results.} 
 Table \ref{table: model selection} shows the two model fits with lowest  BIC
for each type of model \eqref{switch-dyn}-\eqref{switch-var}-\eqref{switch-obs} 
(respectively DYN, VAR, OBS in the table.)
The best model is a switching VAR model with $M=5$ regimes and autoregressive order $p=3$, which is a bit cumbersome to display. 
Instead, we present the fitted model \eqref{switch-var} with $M=3$ and $p=3$ 
($\log L = 3.02\cdot 10^4, \ \mathrm{AIC}=	-5.47\cdot 10^4,\ \mathrm{BIC} = -3.13 \cdot 10^4,\ \mathrm{MAPE} = 0.36$), 
which has easily interpretable regimes and whose time segmentation and regime-specific
stationary covariances, correlations, and autocorrelations closely approximate those of the best model. 
With respect to regime segmentation, the best model decomposes further the pre-ictal phase and late-ictal phase in 2 regimes each.

\begin{table}[h]
\begin{center}
\begin{tabular}{*{8}{c} } 
\hline
Model& $M$& $p$& $r$ & $\log L $ & AIC& BIC & MAPE \\ \hline
VAR&5&3&16&4.45 & -7.96 &-4.16 & 0.35 \\
VAR&4&3&16&3.85 & -6.95&-3.88 &0.36 \\
DYN & 5& 3 & 15 & 3.49 & -6.09 & -2.43 & 0.36  \\
DYN & 4& 3 & 15 & 2.73 & -4.71 & -1.71 &0.36 \\
OBS & 3 & 5 & 8 & -9.29 &18.9 &20.3 & 0.53 \\
OBS & 3 & 6 & 8 &  -9.26 &18.9&20.4 & 0.56 \\
\hline
\end{tabular}

\end{center}
\caption{Model selection for the epileptic seizure data. 
The log-likelihood, AIC and BIC scores are scaled by $10^4$.
MAPE stands for the mean absolute prediction error \eqref{MAPE}. 
Lower AIC, BIC, and MAPE values indicate better fit to the data.}
\label{table: model selection}
\end{table}

\begin{figure}[!ht]
\begin{center}
\includegraphics[width=1 \textwidth, keepaspectratio]{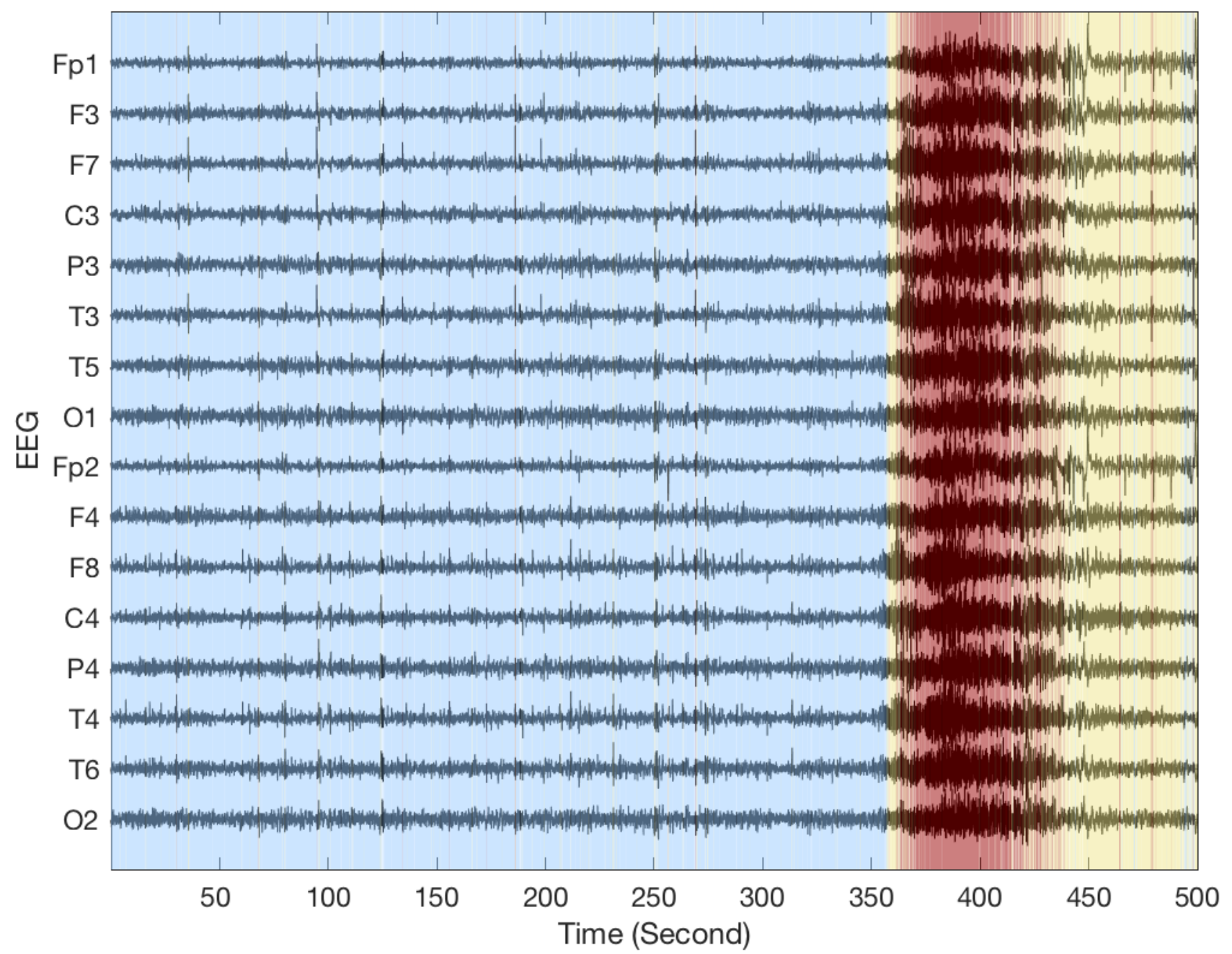}
\caption{Estimated functional connectivity regimes for the epileptic EEG data. The first regime (blue) mostly corresponds to the pre-ictal phase, the second (red) to the ictal phase, and the third (yellow) to the late-ictal phase.}
\label{fig: FC regimes}
\end{center}
\end{figure}

Figure \ref{fig: FC regimes} shows that this model segments the EEG time series in intervals that closely match the pre-ictal, ictal, and late-ictal phases. However, a closer look reveals recurrence of regimes: 
for example, regime 1, which is completely dominates the pre-ictal phase, returns at the very end of the time series (late-ictal  phase). Also, regime 3, which occupies the majority of the late-ictal phase, 
occurs regularly during the pre-ictal phase and even more  strongly just  before the ictal phase and at its beginning.

In the rest of the subsection we study dynamic functional connectivity in terms of 
the regime-specific covariance matrices $\widehat{\bSigma}_j^y$, 
correlations $\widehat{\bR}_j^y$, and autocorrelations $\widehat{\rho}_{kj}^y$ 
 defined in \eqref{eq: stationary covariance} and thereafter.

\begin{figure}[!ht]
\begin{center}
\includegraphics[width = 0.32 \textwidth]{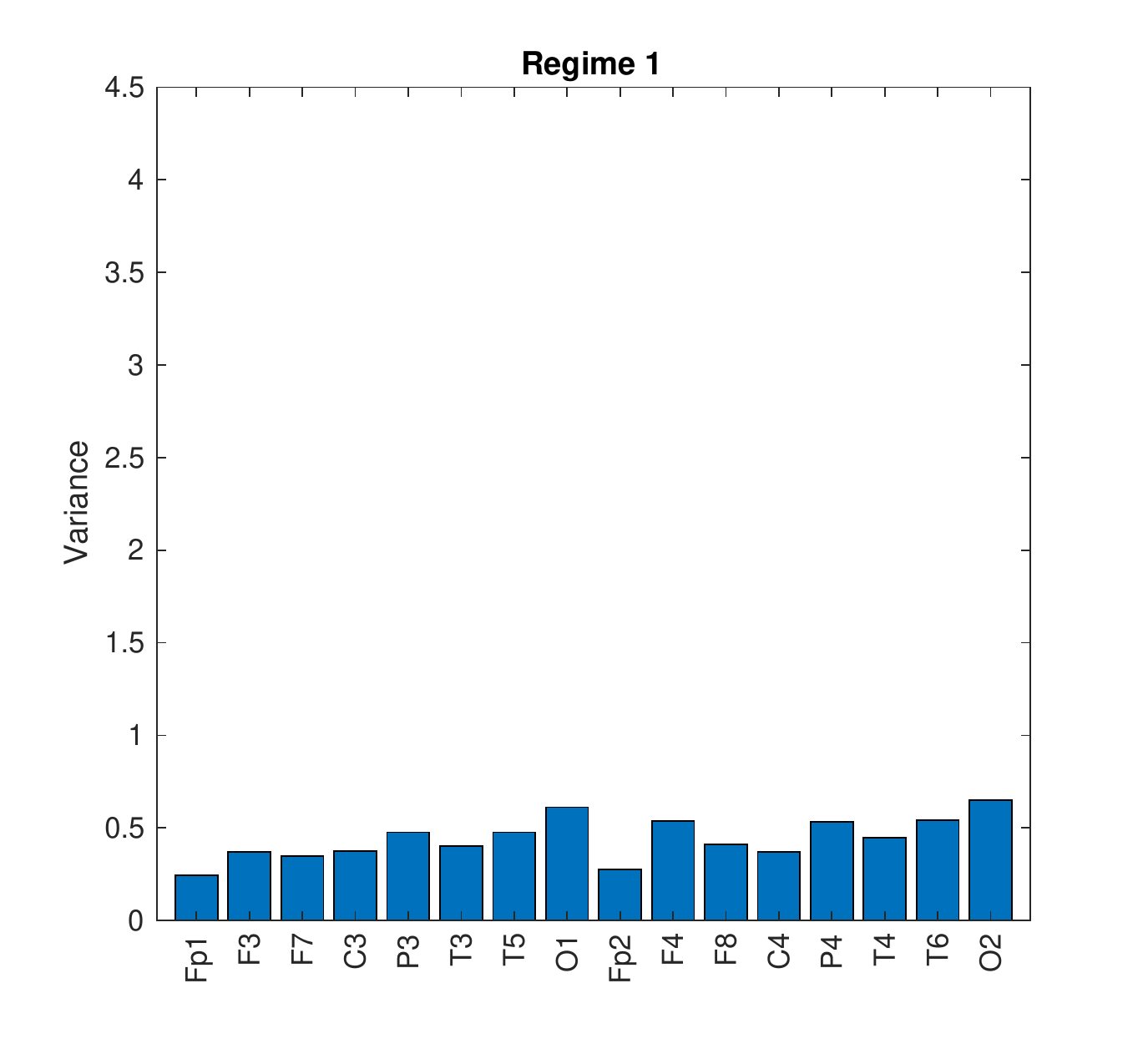}
\includegraphics[width = 0.32 \textwidth]{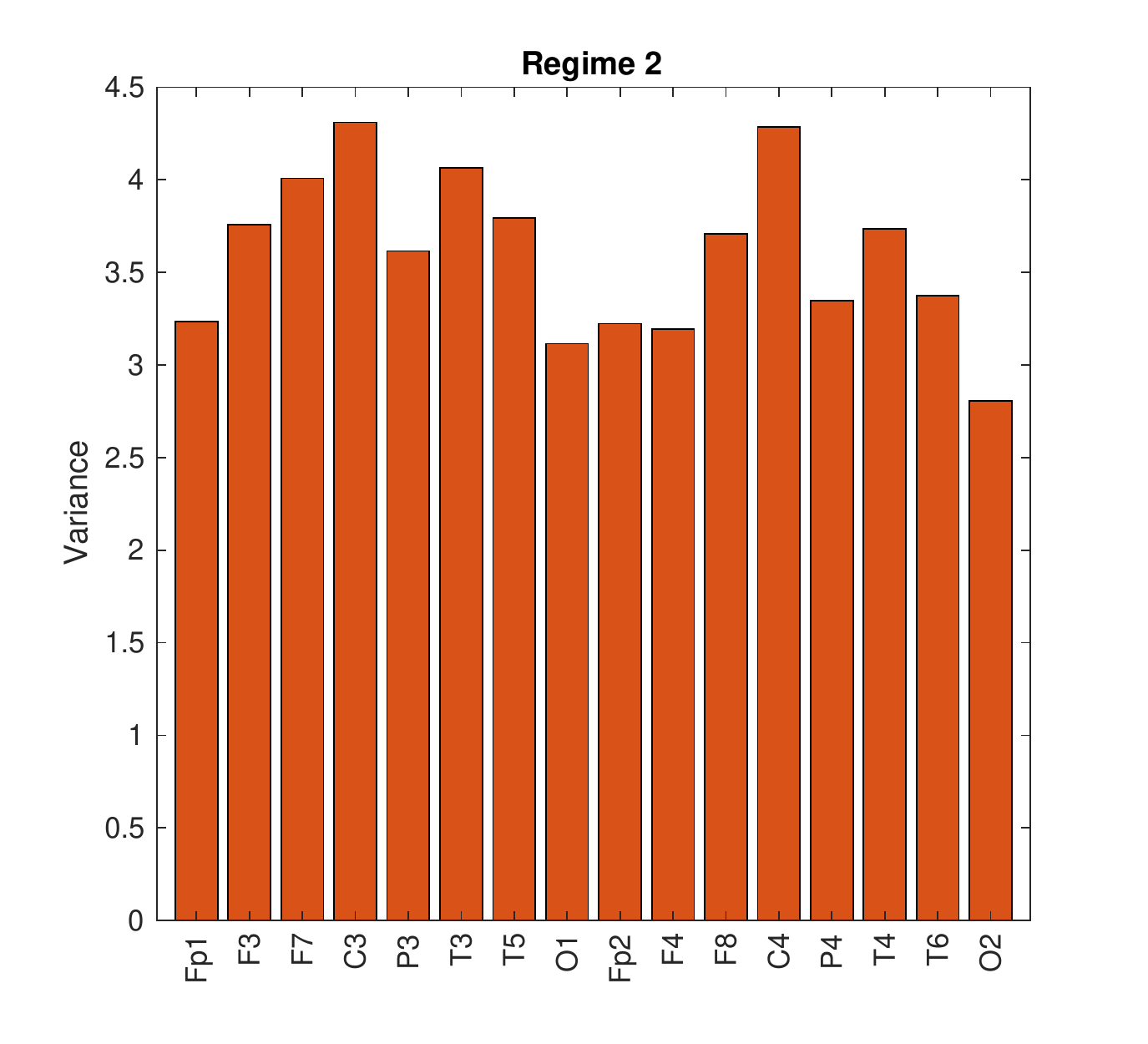}
\includegraphics[width = 0.32 \textwidth]{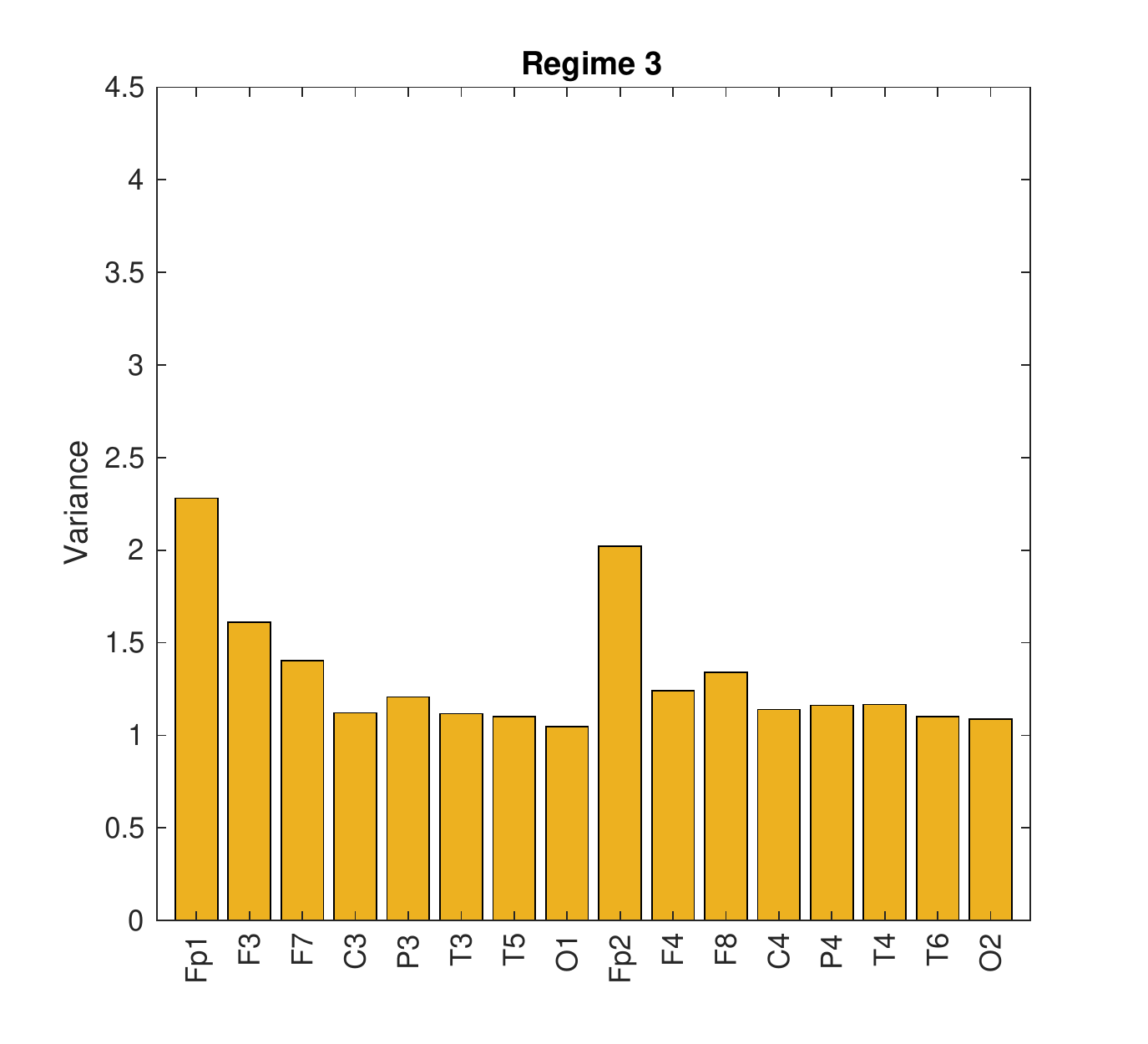} \\
\includegraphics[width = 0.32 \textwidth]{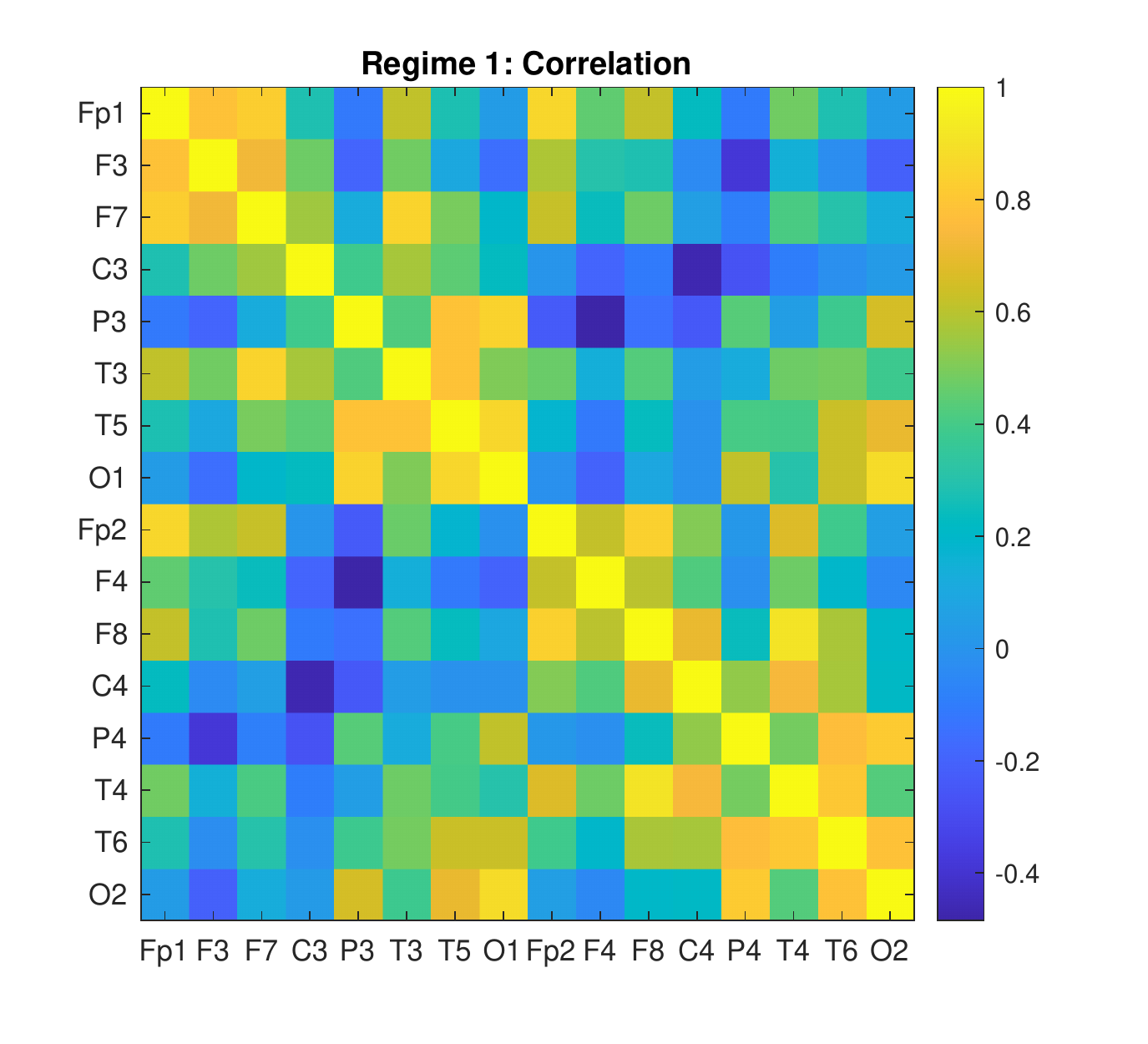}
\includegraphics[width = 0.32 \textwidth]{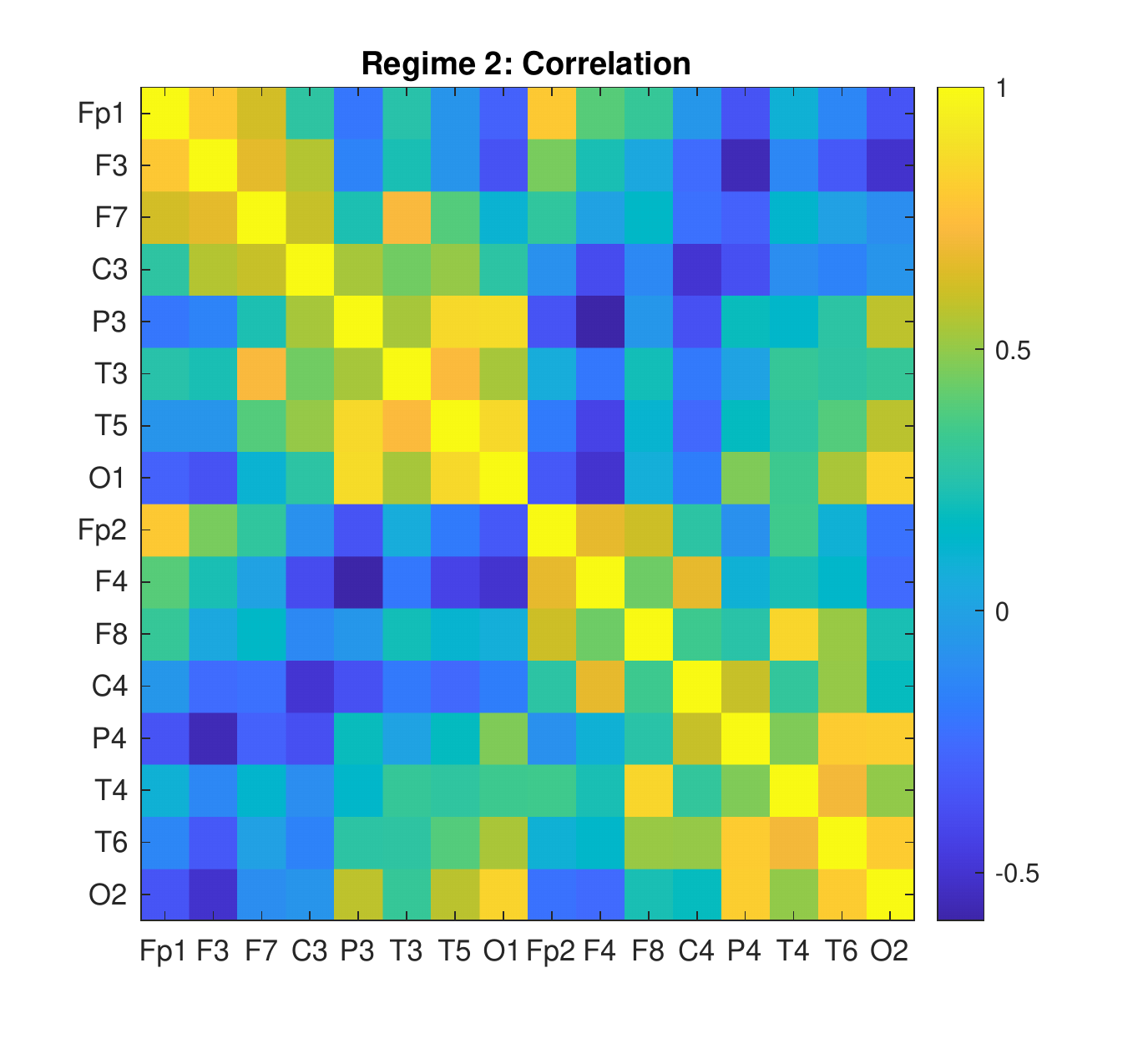}
\includegraphics[width = 0.32 \textwidth]{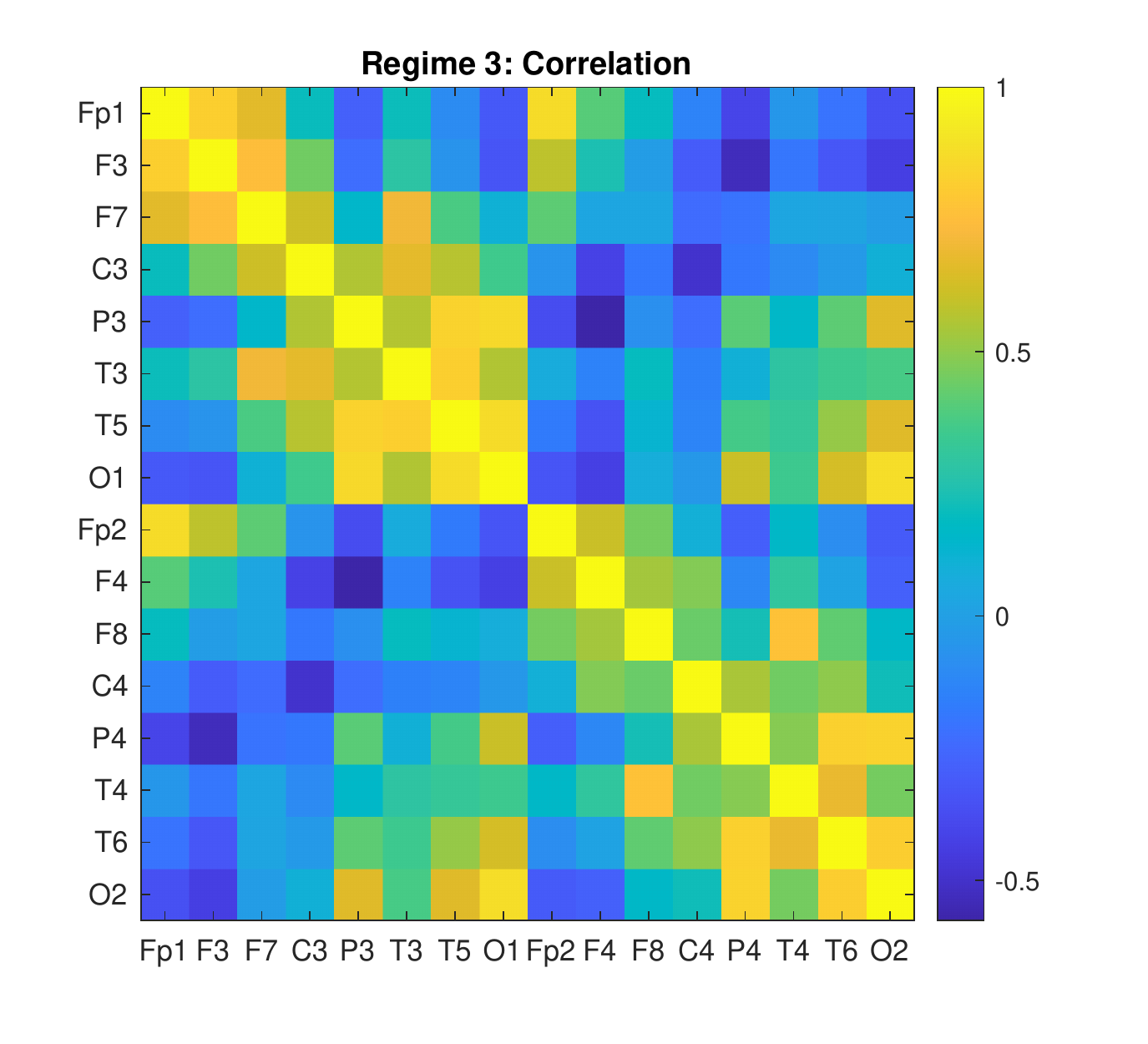}

\caption{Estimated variance (top row) and correlation (bottom row)  by regime (1: pre-ictal, 2: ictal, regime 3: late-ictal).} 
\label{fig: variance correlation epi}
\end{center}
\end{figure}

Figure \ref{fig: variance correlation epi} presents the variance and correlation estimates for each regime. Expectedly, the  variance in EEG activity is much higher during the ictal period (regime 2) than either before ( regime 1) or in the late-ictal period (regime 3). 
The overall variance level goes back down in the late-ictal phase but remains higher than its pre-ictal level. 
The estimated correlation matrices are fairly similar across the three regimes, suggesting that the overall correlation structure is not excessively affected by the seizure. Clustering is apparent among neighboring electrodes.

\begin{figure}[!ht]
\begin{center}
\includegraphics[width = 0.32 \textwidth]{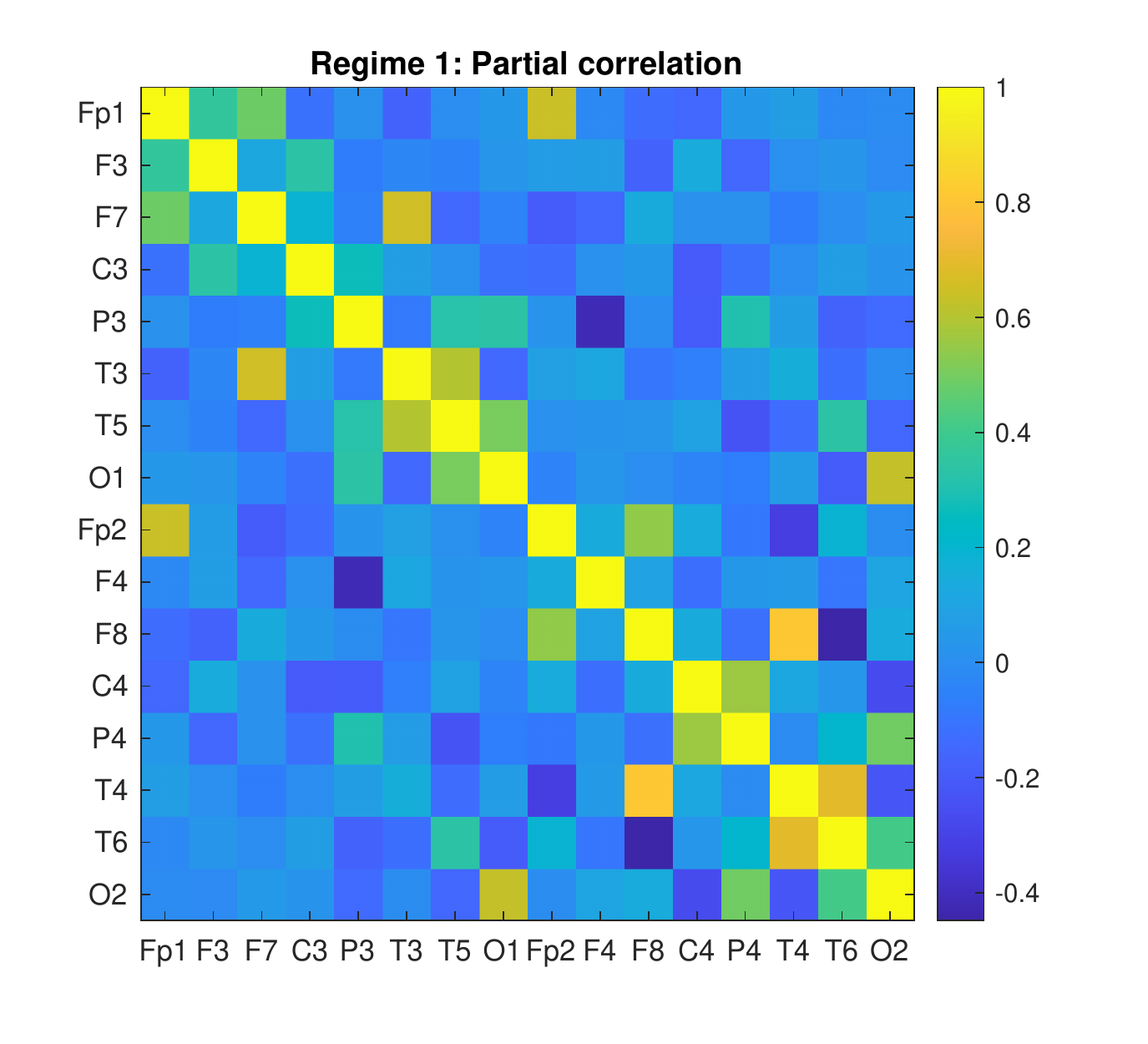}
\includegraphics[width = 0.32 \textwidth]{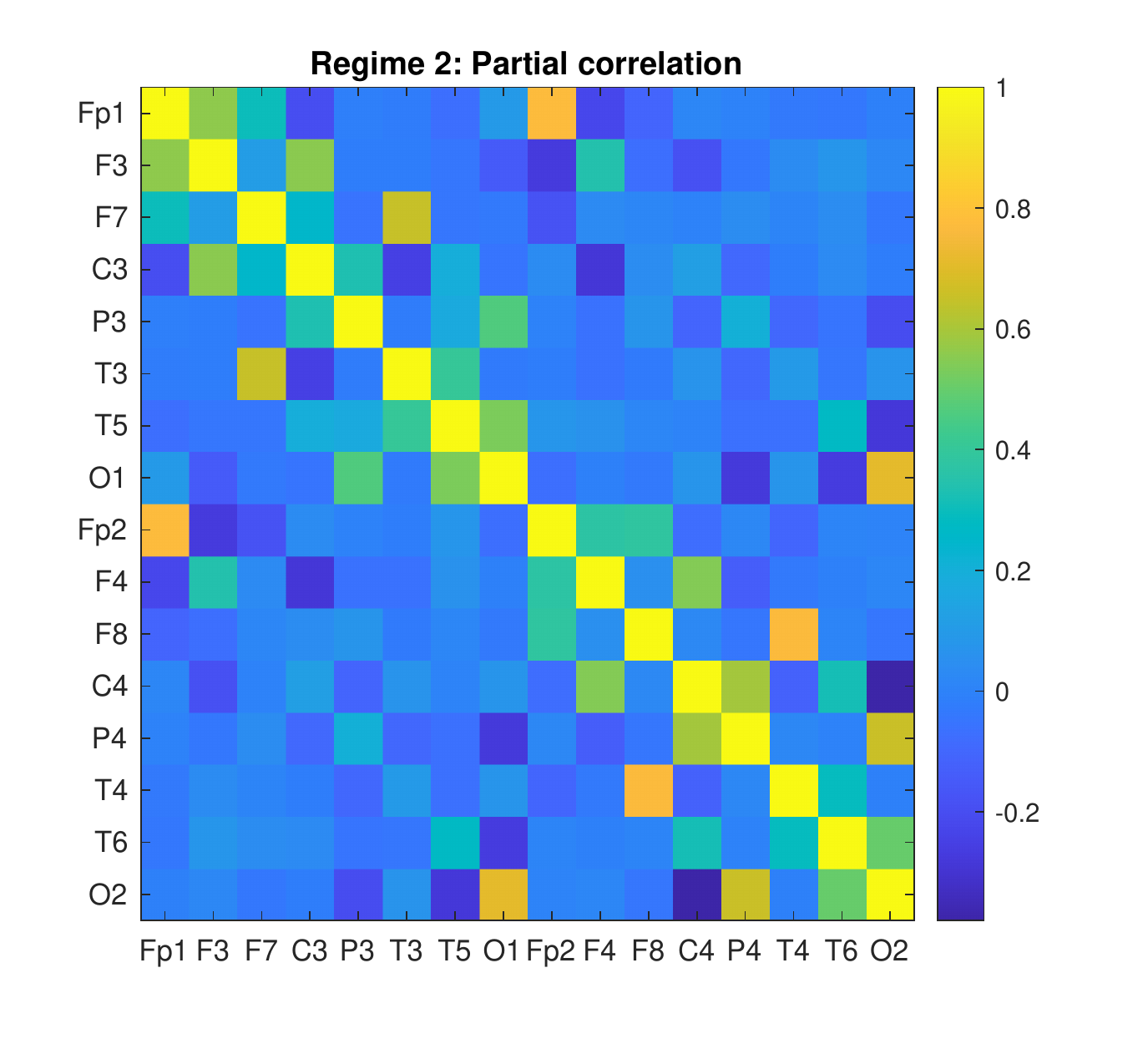}
\includegraphics[width = 0.32 \textwidth]{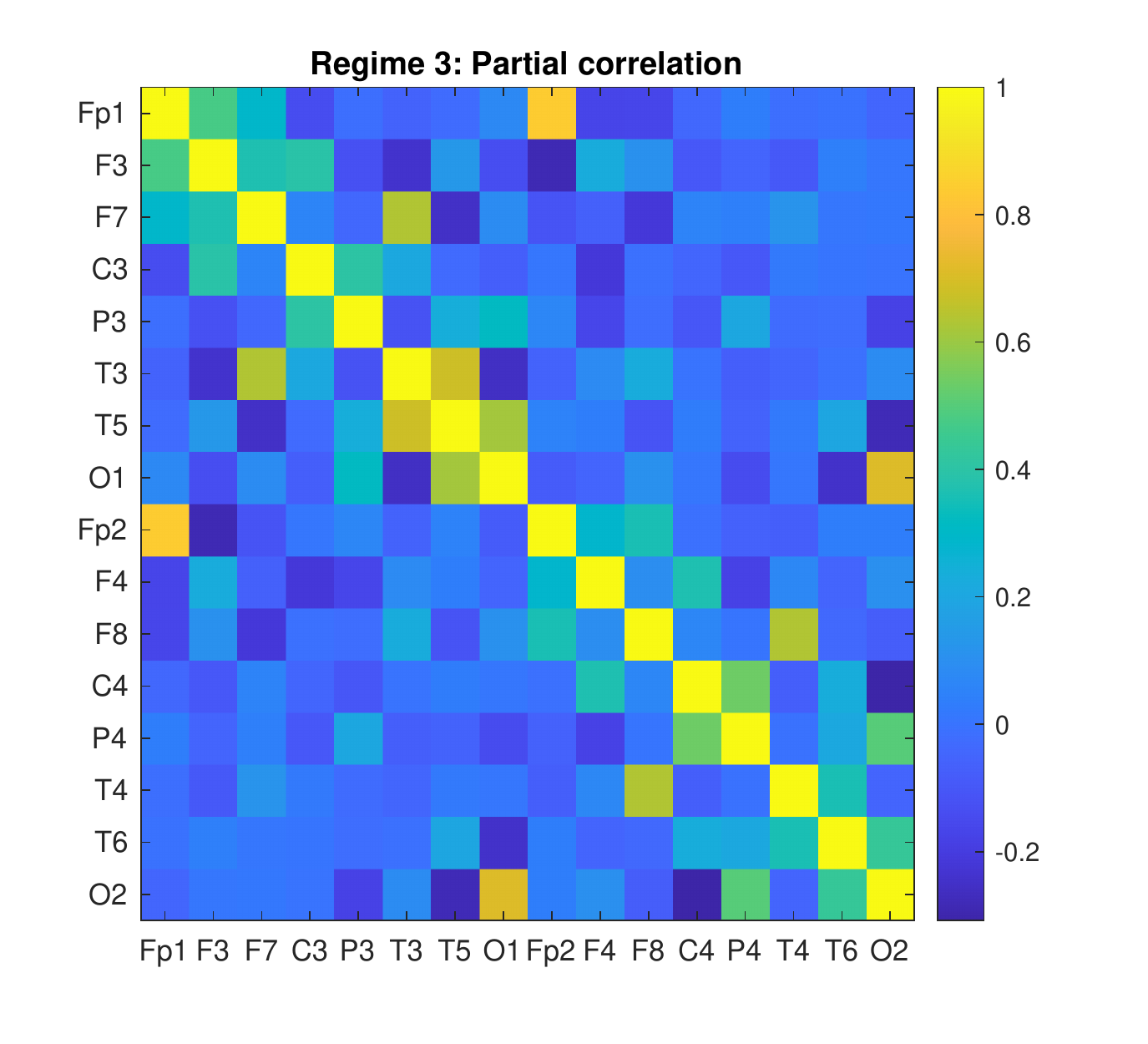} \\
\hspace*{-4mm}
\includegraphics[width = 0.32 \textwidth]{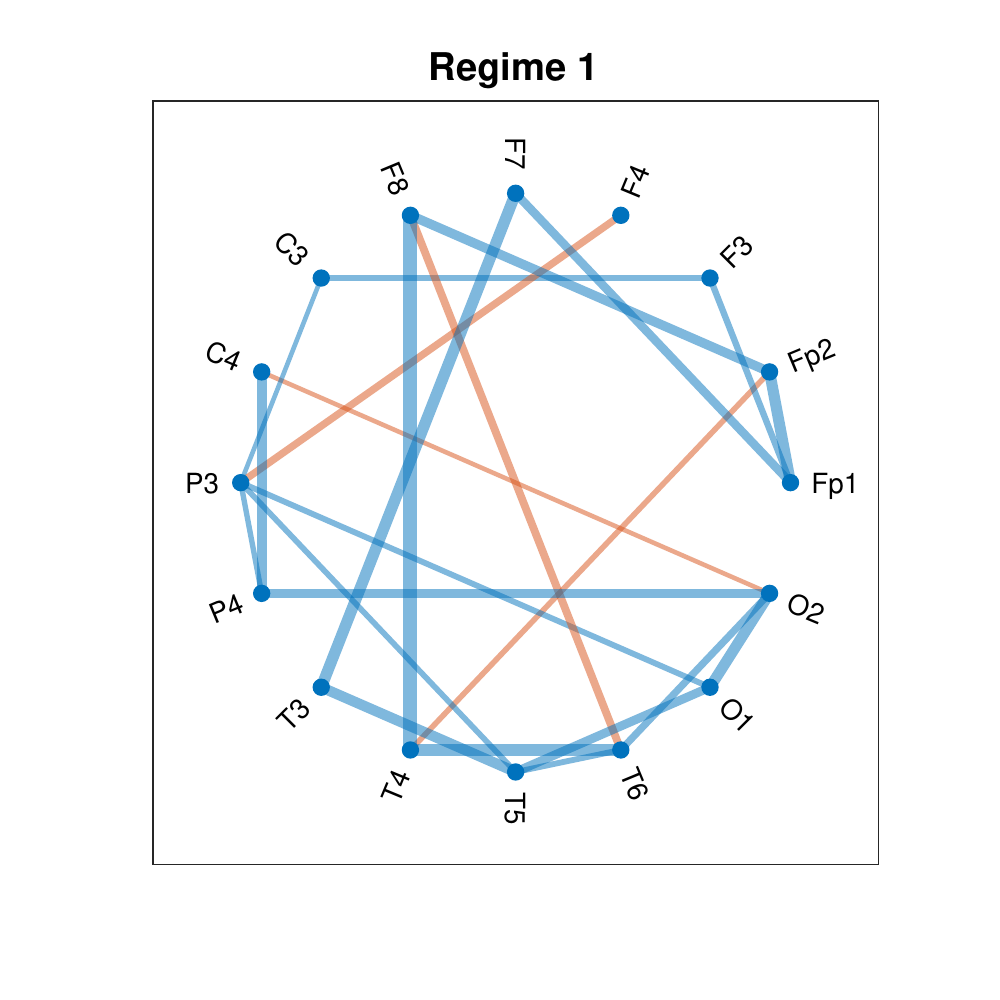}
\includegraphics[width = 0.32 \textwidth]{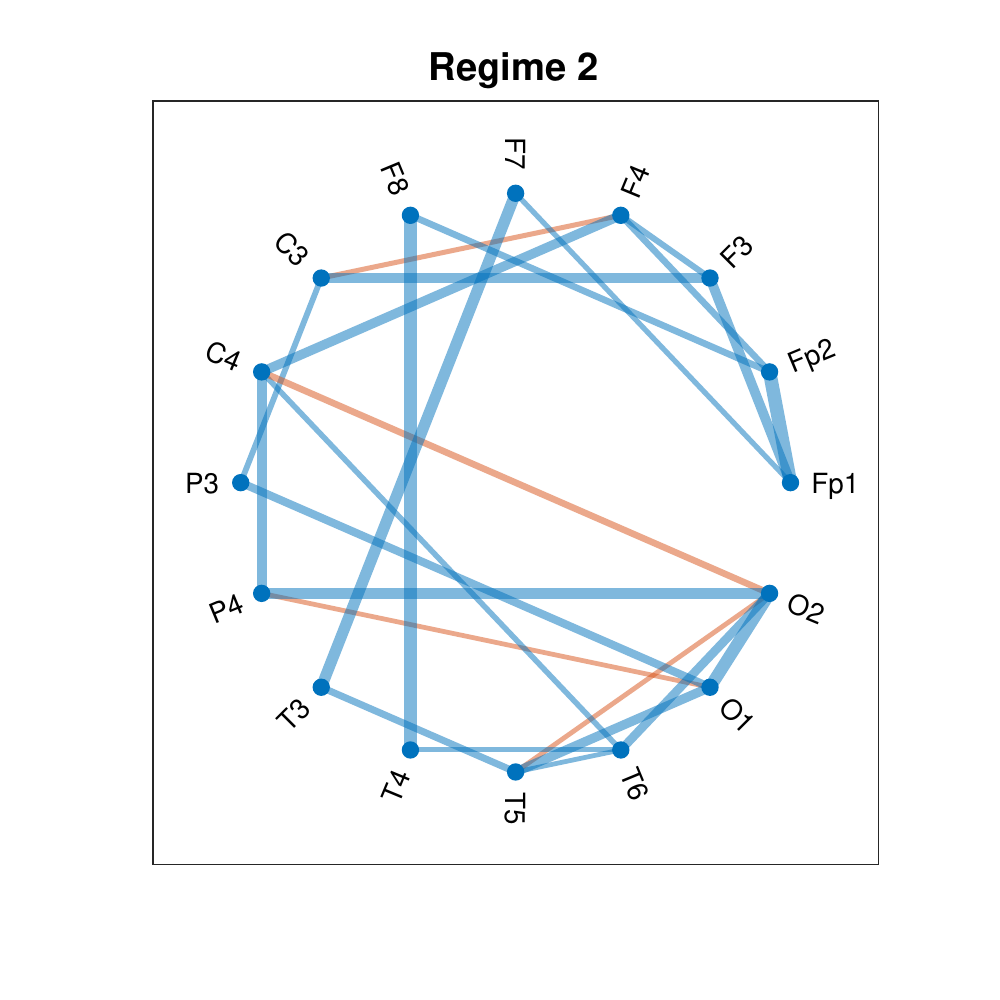}
\includegraphics[width = 0.32 \textwidth]{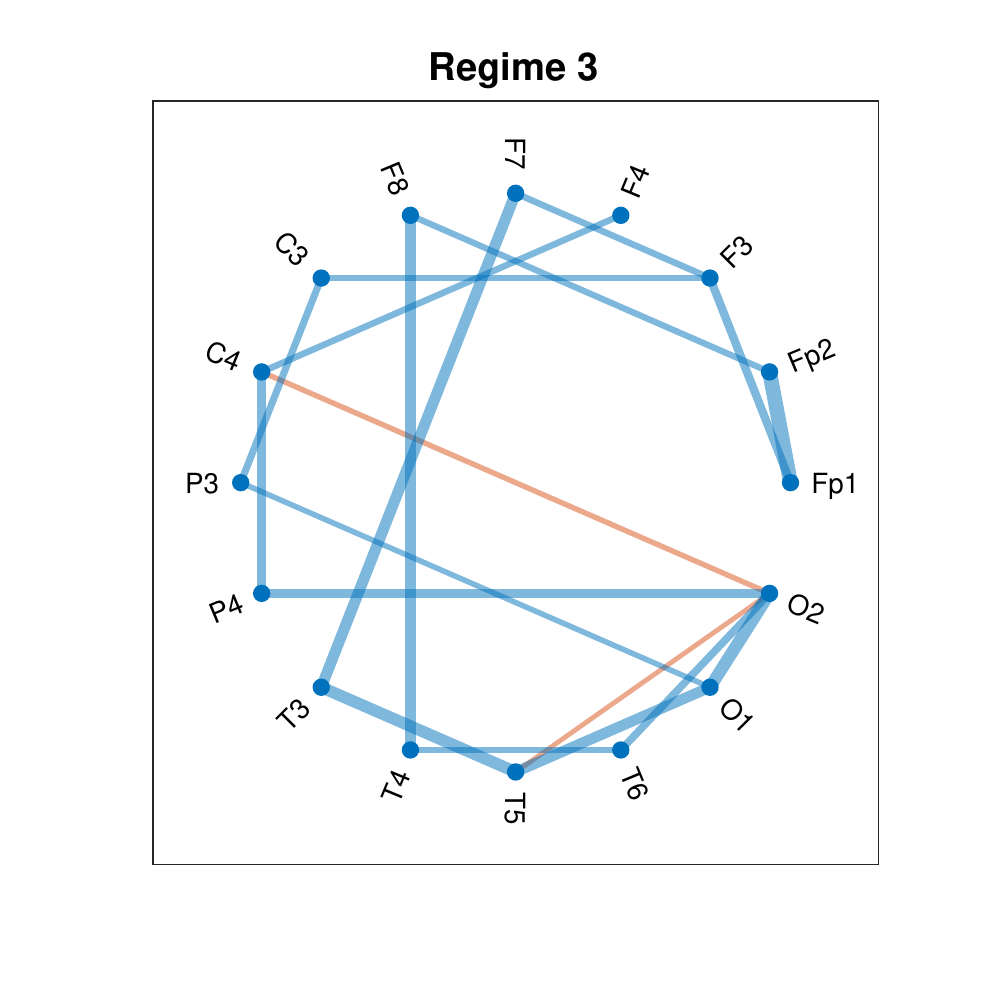}
\caption{Estimated partial correlation by FC regime (1: pre-ictal, 2: ictal, 3:  late-ictal). 
Top row: partial correlation matrices. Bottom row: associated FC graphs with  edge significance 
inferred by bootstrap. 
The edge width indicates the magnitude of the partial correlation and the color indicates it sign 
 (blue: positive, orange: negative).} 
\label{fig: partial correlation epi}
\end{center}
\end{figure}

Figure \ref{fig: partial correlation epi} shows the partial correlation between electrodes, i.e., the correlation that is specific to each pair of electrodes and not shared with other electrodes. In the top row, the partial correlation coefficients seem to increase in magnitude from regime 1 (pre-ictal) to regime 3 (late-ictal). In particular, patterns of anticorrelation progressively emerge in the left hemisphere during the ictal and  late-ictal phase, e.g.,   between T3 and F3, F7 and T5, and T3 and O1. 
The bottom row displays FC graphs for each regime based on significant partial correlations.
The graphs were obtained as follows. 
First, the MLEs $(\widehat{\bA}_j , \widehat{\bQ}_j) $ ($1\le j \le M$) were bootstrapped  $B=2000$ times as in Section \ref{sec: bootstrap}, from which bootstrap replicates of the covariance matrices $\widehat{\bSigma}_j^y$ and associated partial correlation matrices were deduced as in Section \ref{sec: simulations}. Then bootstrap percentile confidence intervals (CI) were constructed for each partial correlation coefficient at the pointwise level 99.5\%. Due to the large number of CIs (360 per regime), 
this produced a simultaneous confidence level between 62\% and 65\% for each regime. Partial correlations whose CIs did not contain values less than 0.2  (in absolute value) were deemed significant (this corresponds to testing for a magnitude greater than 0.2). 
As can be seen, the FC graphs are mostly stable across regimes. A comparison with Figure \ref{fig: eeg 10-20} indicates that, unsurprisingly,  most of the significant partial correlations take place between nearby or adjacent electrodes.  This is in contrast with full correlation patterns which can extend between distant electrodes (Figure \ref{fig: variance correlation epi}). The number of edges in the FC graph  increases very slightly during the ictal phase and decreases more substantially in the late-ictal phase.

\begin{figure}[!ht]
\begin{center}
\includegraphics[width = 0.32 \textwidth]{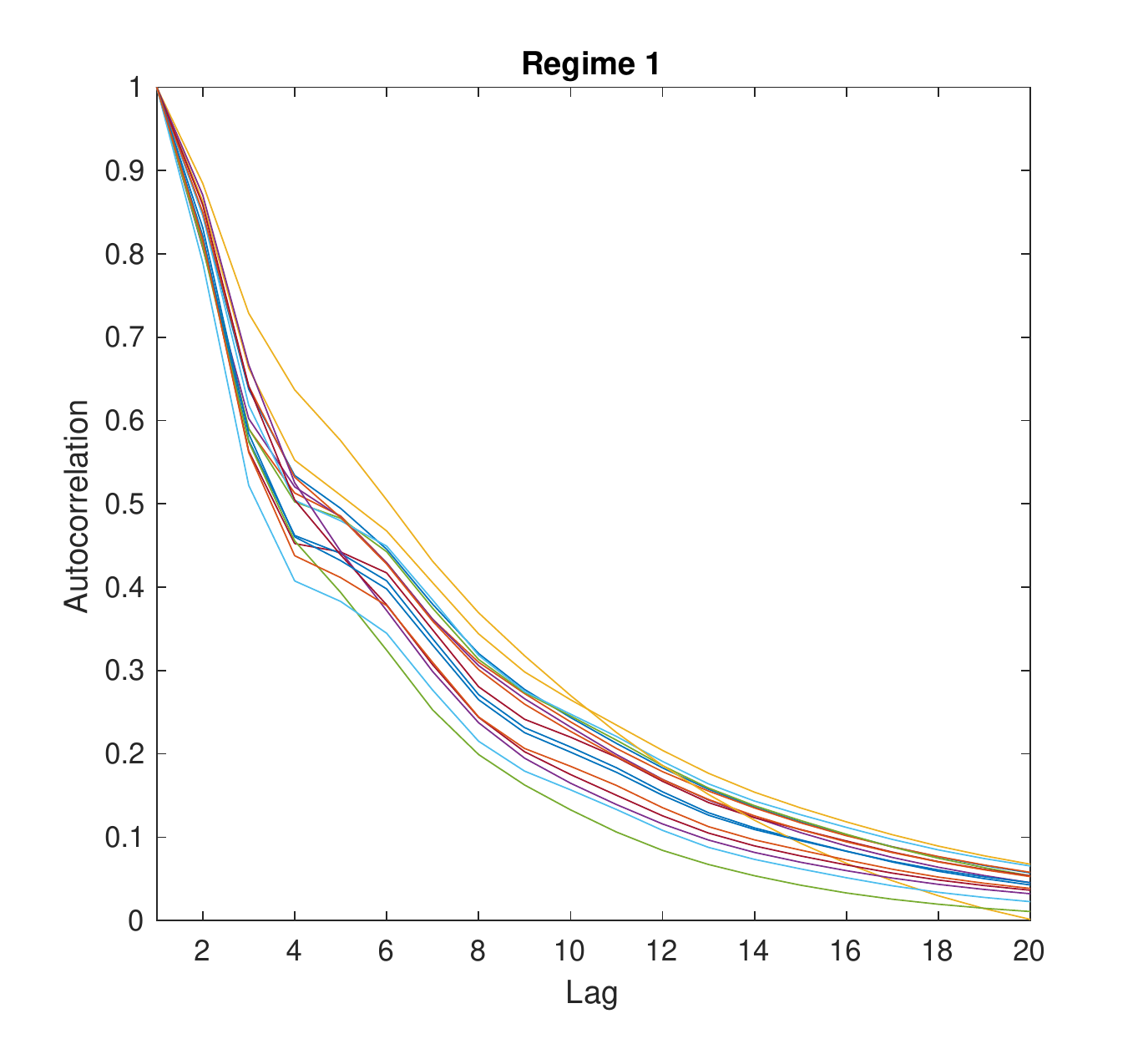}
\includegraphics[width = 0.32 \textwidth]{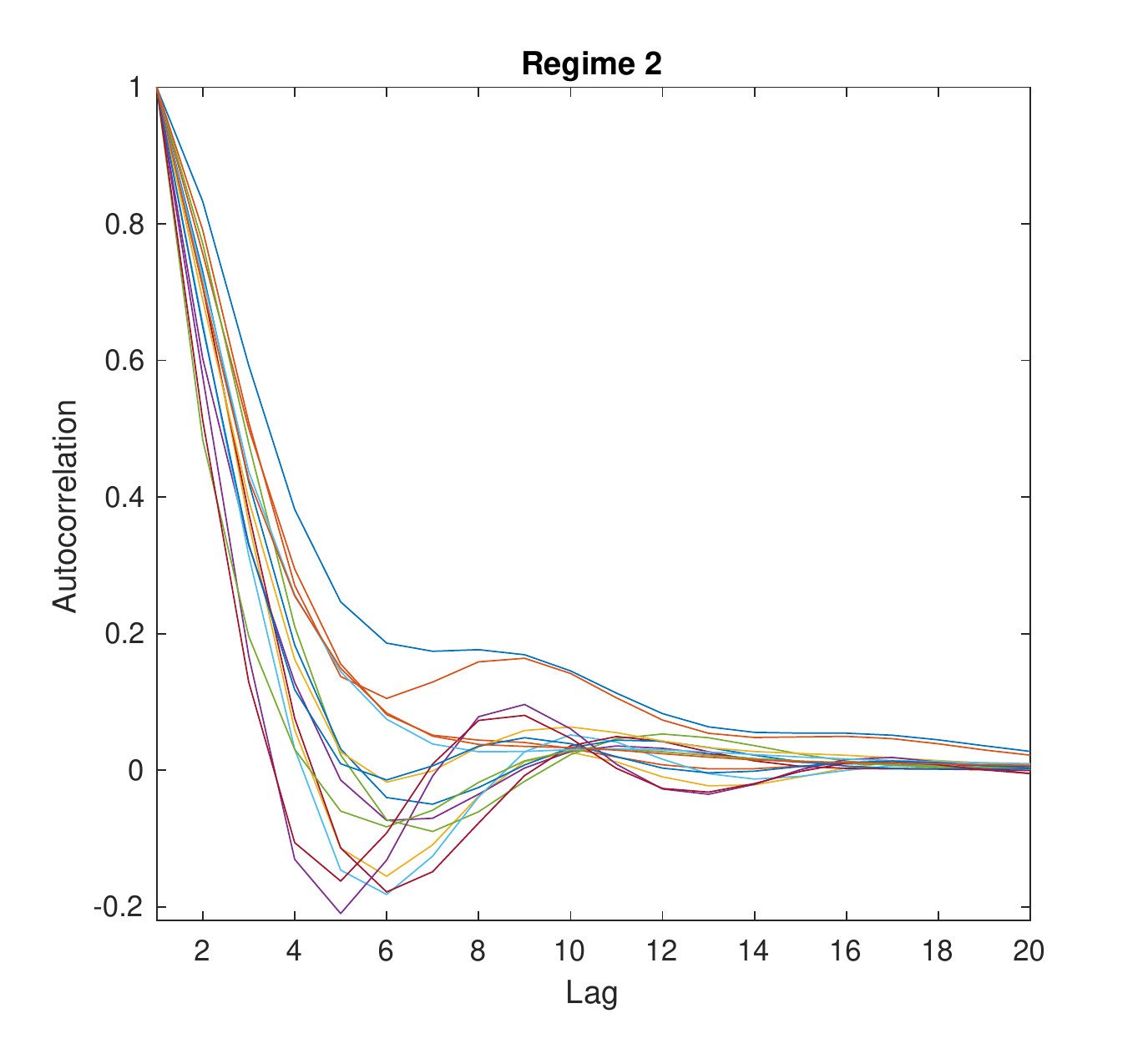}
\includegraphics[width = 0.32 \textwidth]{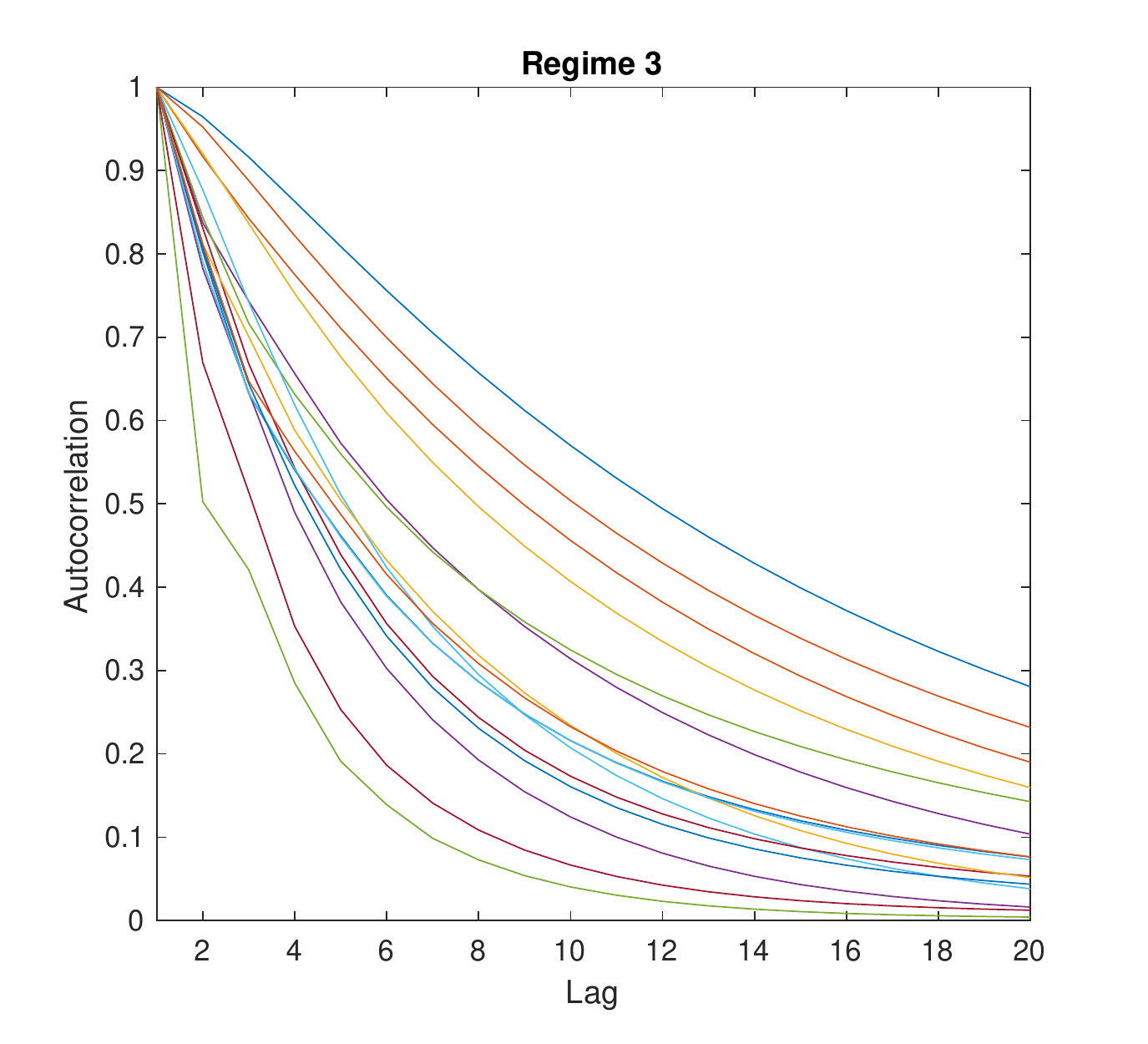} \\
\caption{Estimated autocorrelation functions for all EEG channels  
(regime 1: pre-ictal, regime 2: ictal, regime 3:  late-ictal).}
\label{fig: lagged correlation}
\end{center}
\end{figure}

Figure \ref{fig: lagged correlation} displays the autocorrelation functions (ACF) of each electrode in each regime. 
The fact that the ACFs decrease much faster  in regime 2 (ictal period) than in regimes 1 and 3 suggests that during the epileptic seizure, the ability of the brain to keep track of its past activity and to maintain temporal integration is greatly reduced. %
Another remarkable contrast is that most ACFs exhibit damped oscillations in regime 2, with negative peaks between lags 5 and 7 (100-140\,ms) whereas the ACFs show a monotonic decrease under regimes 1 and 3. This points to more complex brain dynamics and more heterogeneity in space. Specifically,  channels  Fp1, Fp2, F3, P4, and O2 show a strictly or nearly monotonic decrease; 
T4, F8, T5, F7, T3 show strong oscillations and one or more negative peaks. 
Finally,  the ACFs are much more scattered in the  late-ictal period (regime 3) than in the pre-ictal period (regime 1). 
This suggests that following the seizure, different brain regions return to their baseline condition at different rates. 
Interestingly, channels at the front and left side of the scalp (Fp1, Fp2, F3, F7, P3, and C3) 
exhibit slowly decreasing ACFs whereas T4 and C4 on the opposite (right) side show very rapidly decreasing ACFs. This lateralization may be correlated with the fact that the seizure  originated in the left temporal area.

In comparison to previous studies that divided the same data in two regimes 
(pre-ictal/ictal or ictal/nonictal), this analysis distinguishes at least three regimes (in fact, five) 
mostly (but not fully) aligned with the pre-ictal, ictal, and  late-ictal phases. 
As a result, it uncovers stark contrasts between the pre- and  late-ictal phases. 
For example, it shows that the EEG signal in the  late-ictal phase has higher amplitude 
and spatially more heterogeneous dynamics than during the pre-ictal phase. 
It also provides new evidence that statistically defined FC regimes 
recur before, during, and after the epileptic seizure, 
in other words they extend over several clinical phases.  
Further, the present analysis complements previous work in the spectral domain by revealing essential temporal characteristics such as the damped oscillations in the ACF during the ictal phase. 
Finally, because it considers all available EEG channels rather than a subset of channels specifically associated with the seizure episode as in previous work, this study not only highlights what changes in  (partial correlation) FC graphs across regimes but also what is preserved.

\subsection{Brain-computer interface data}

In this section we apply switching SSMs to continuous EEG data from a motor imagery experiment \citep{Blankertz2007}.
These data appeared in the Brain-Computer Interface (BCI) competition IV as Dataset 1 \citep{Tangermann2012} and 
 are publicly available at  \url{http://www.bbci.de/competition/iv}. 
 While the data were used for a classification task in the competition, our aim here is to quantify variations in dFC between subjects and sessions. For space reasons, we limit our exposition to the covariance matrices $\bSigma_{j}^y$ associated with each FC regime. These matrices display richer variations than the associated correlation matrices $\bR^y_j$ and  autocorrelation functions 
$\rho_{kj}^{y}$.

\paragraph{Data description.} EEG data were continuously recorded from four healthy subjects using $N=59$ electrodes most densely distributed over sensorimotor areas. (The BCI competition also included artificial EEG data which  we do not analyze here.) 
The original signals were sampled at 1000\,Hz and bandpass-filtered between 0.05 and 200\,Hz. 
For the BCI competition, they were lowpass-filtered with a Chebyshev filter of order 10 (type II, stopband ripple 50\,dB down, stopband edge frequency 49\,Hz) and downsampled to 100\,Hz by averaging consecutive  blocks of 10 samples. 
To remove non-brain signals and speed up computations, we further bandpass-filtered the data to the frequency range $[8,25]$\,Hz (finite impulse response filter of minimum order with linear phase) and downsampled them to 50\,Hz by averaging consecutive blocks of 2 samples. The time series of all EEG channels were centered, scaled to unit variance, and extreme observations outside $[-10,10]$ were clipped to this interval.  

Throughout the session motor imagery was performed without feedback. For each subject two classes of motor imagery were selected from the three classes: left hand, right hand, and foot. In the first two runs (\emph{calibration data}), arrows pointing left, right, or down were presented as visual cues on a computer screen. Cues were displayed for a period of 4\,s during which the subject was instructed to perform the cued motor imagery task. These periods were interleaved with 2\,s of blank screen and 2\,s with a fixation cross shown in the center of the screen. 
Then 4 runs followed (\emph{evaluation data}) in which the motor imagery tasks were cued by soft acoustic stimuli (words left, right, and foot) for periods of varying length between 1.5 and 8\,s. The end of the motor imagery period was indicated by the word stop. Intermitting periods also had a varying duration of 1.5 to 8\,s. 
The calibration and the evaluation parts of a session lasted about 32 and 41 minutes, respectively.  
For convenience, we refer hereafter to calibration and evaluation as two separate sessions.

\vspace*{-3mm}

\paragraph{Model selection.} After fitting a number of switching SSMs to the calibration and evaluation data of the 4 subjects (labeled  A, B, F,  G in the BCI competition), we retained the switching dynamics model \eqref{switch-dyn} with hyperparameters $M=5$, $p=3$, and $r=5$. We chose model \eqref{switch-dyn} over model \eqref{switch-obs} because it requires less computation time while delivering similar estimates. The choice $r=5$ is based on the fact that for each of the 8 datasets (4 subjects times 2 sessions), between 93.5\% and 96.6\% of the data variations take place in 5 dimensions as shown by singular value decomposition. Also, in all 8 cases, the choice $M=5$ adequately captures the most frequent FC regimes as well as a few rarer ones. Larger values of $M$ only add regimes with short dwell time (5\% of the  session duration or less) and pathological characteristics, e.g.,    extreme variance or exceedingly long autocorrelation. Finally, the choice $p=3$ was sufficient to capture the main aspects of temporal dependence (typically, damped oscillations in the ACF) but small enough to avoid overfitting.

\begin{figure}[htp]
\begin{center}
\hspace*{-8mm}
\setlength{\tabcolsep}{2pt}
\begin{tabular}{|c | c | c | c|}
\hline
\multicolumn{1}{|c}{\includegraphics[width = .27 \textwidth, keepaspectratio]{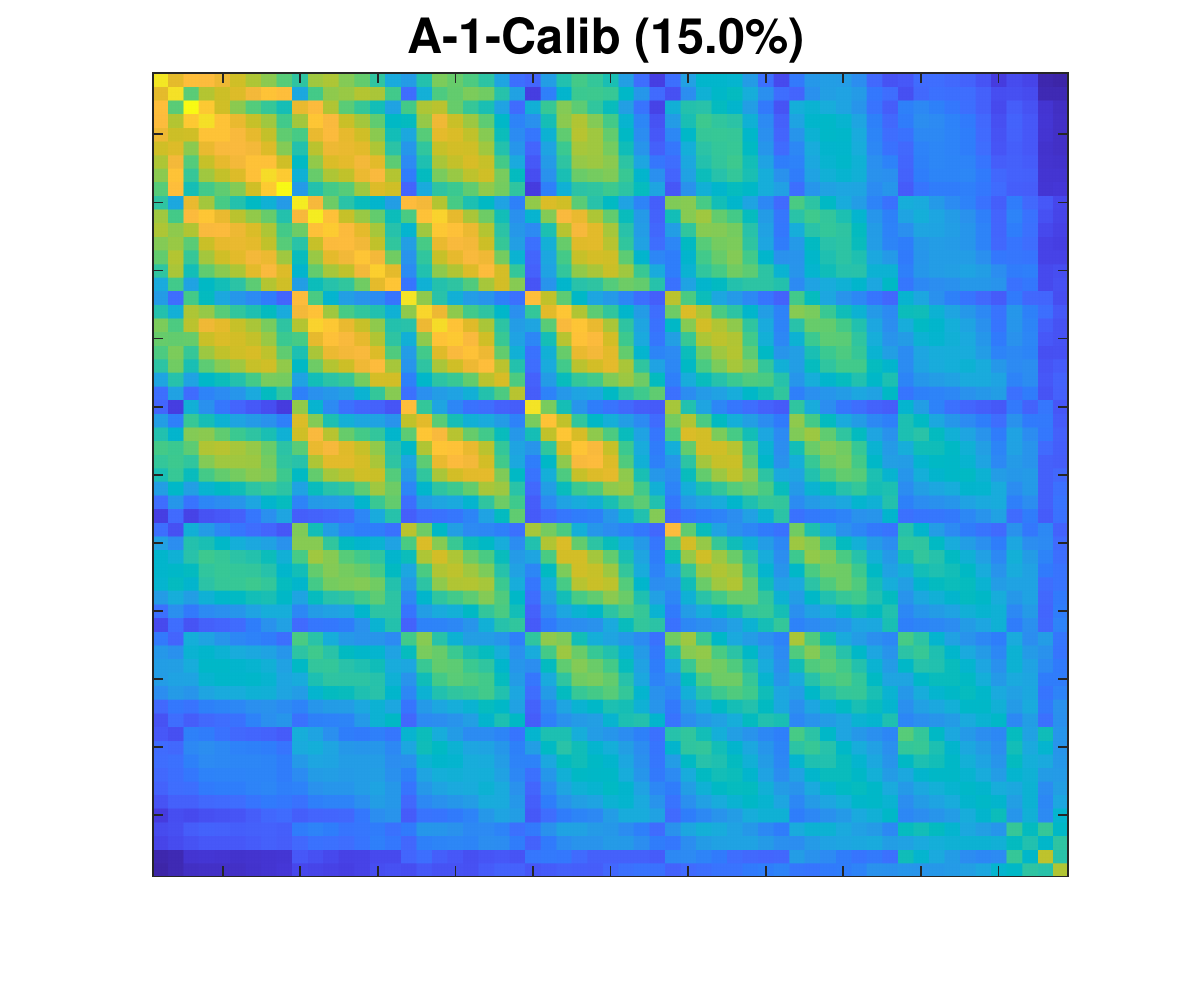}} & 
\multicolumn{1}{c}{\includegraphics[width = .27 \textwidth, keepaspectratio]{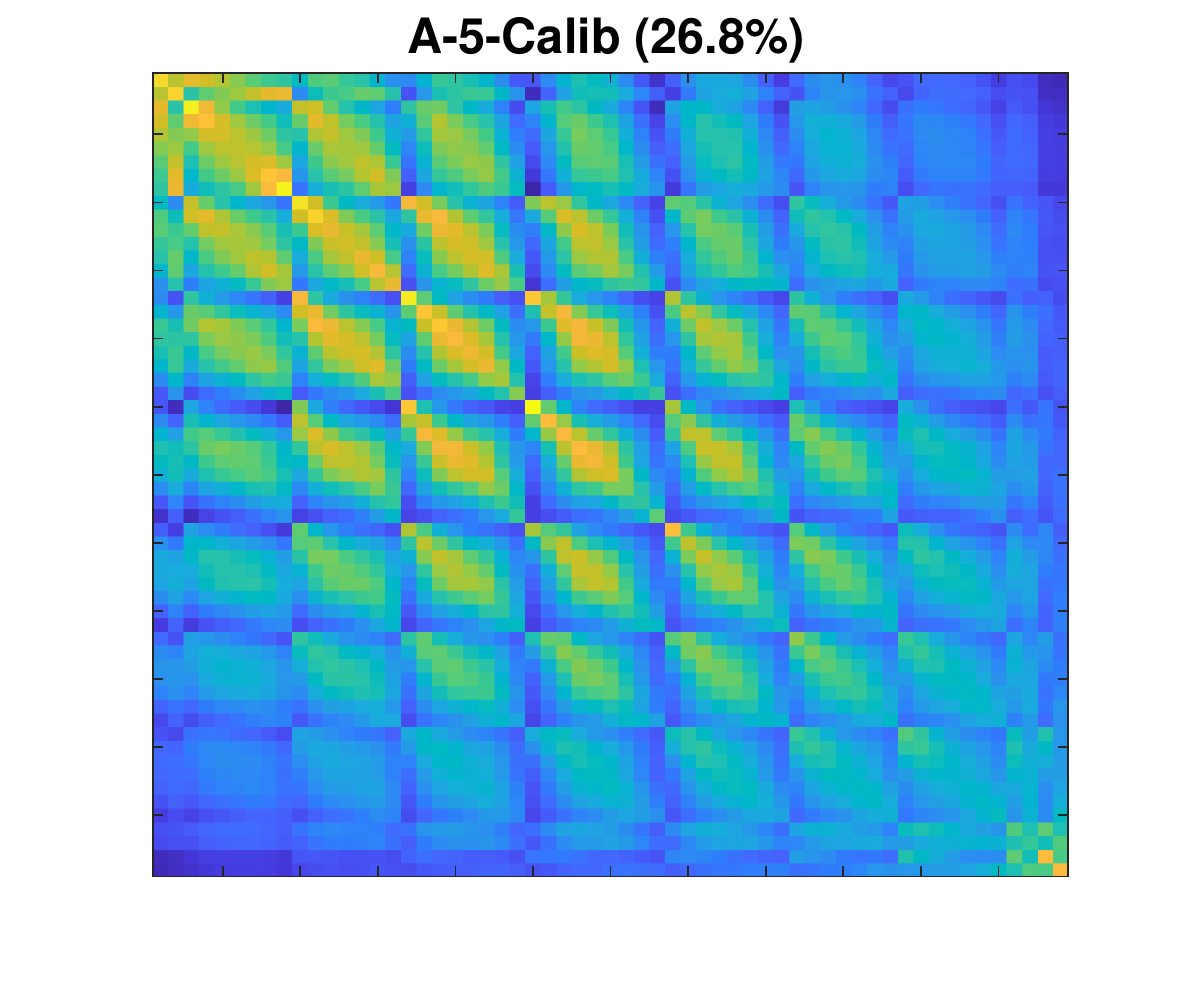}}  & 
\multicolumn{1}{c}{\includegraphics[width = .27 \textwidth, keepaspectratio]{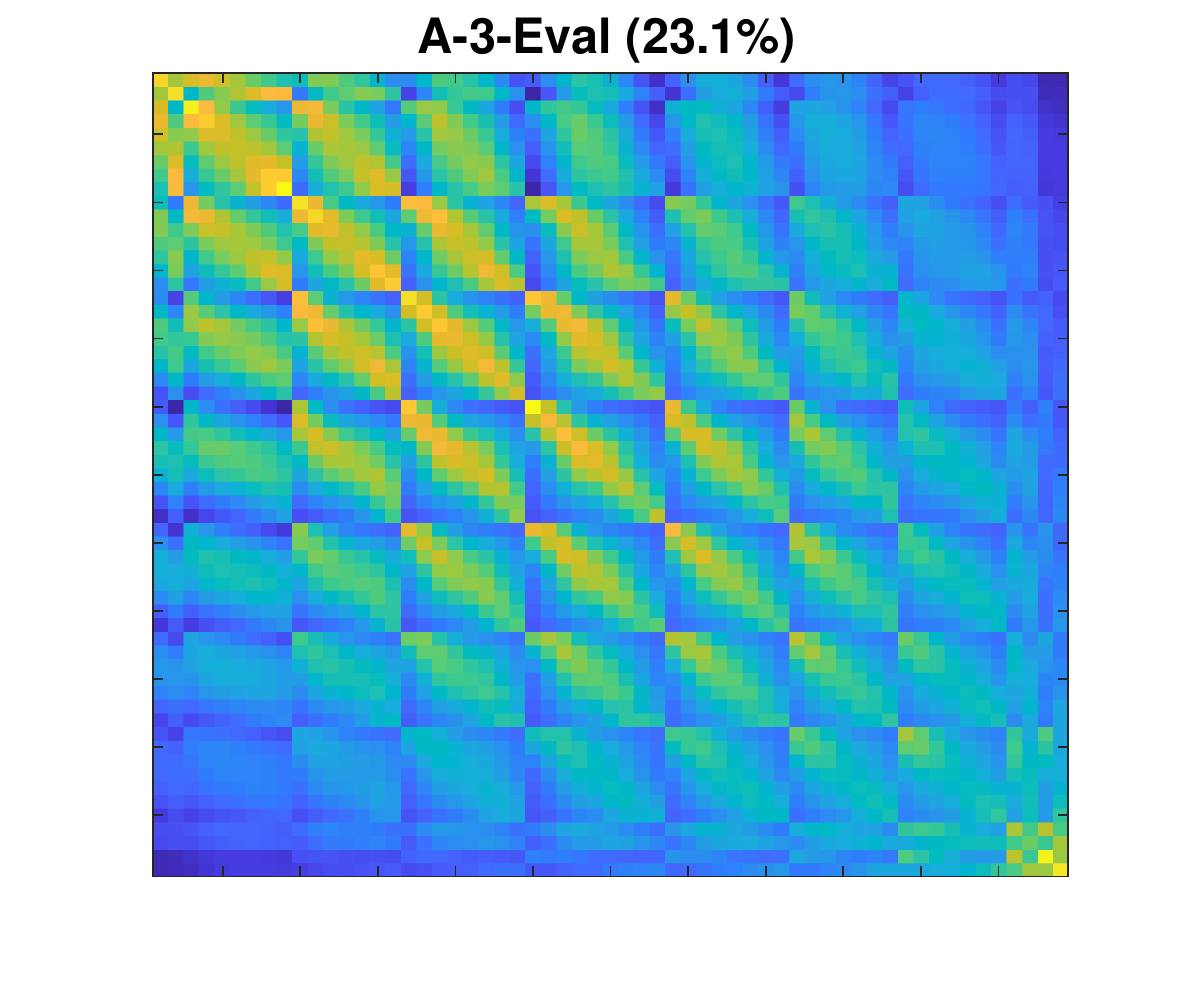}}  & 
\multicolumn{1}{c|}{\includegraphics[width = .27 \textwidth, keepaspectratio]{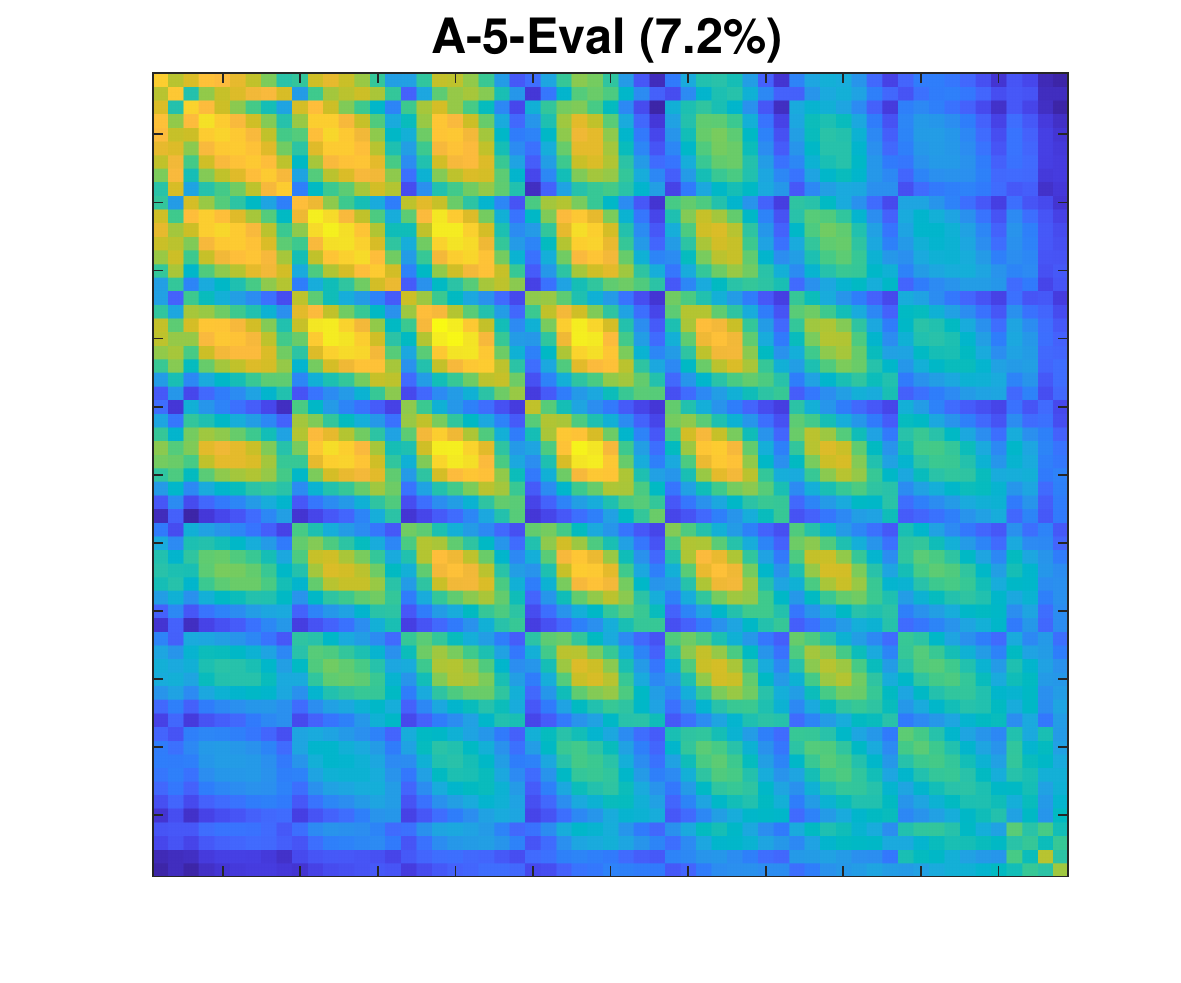}}  \\ 
\hline
\multicolumn{1}{|c}{\includegraphics[width = .27 \textwidth, keepaspectratio]{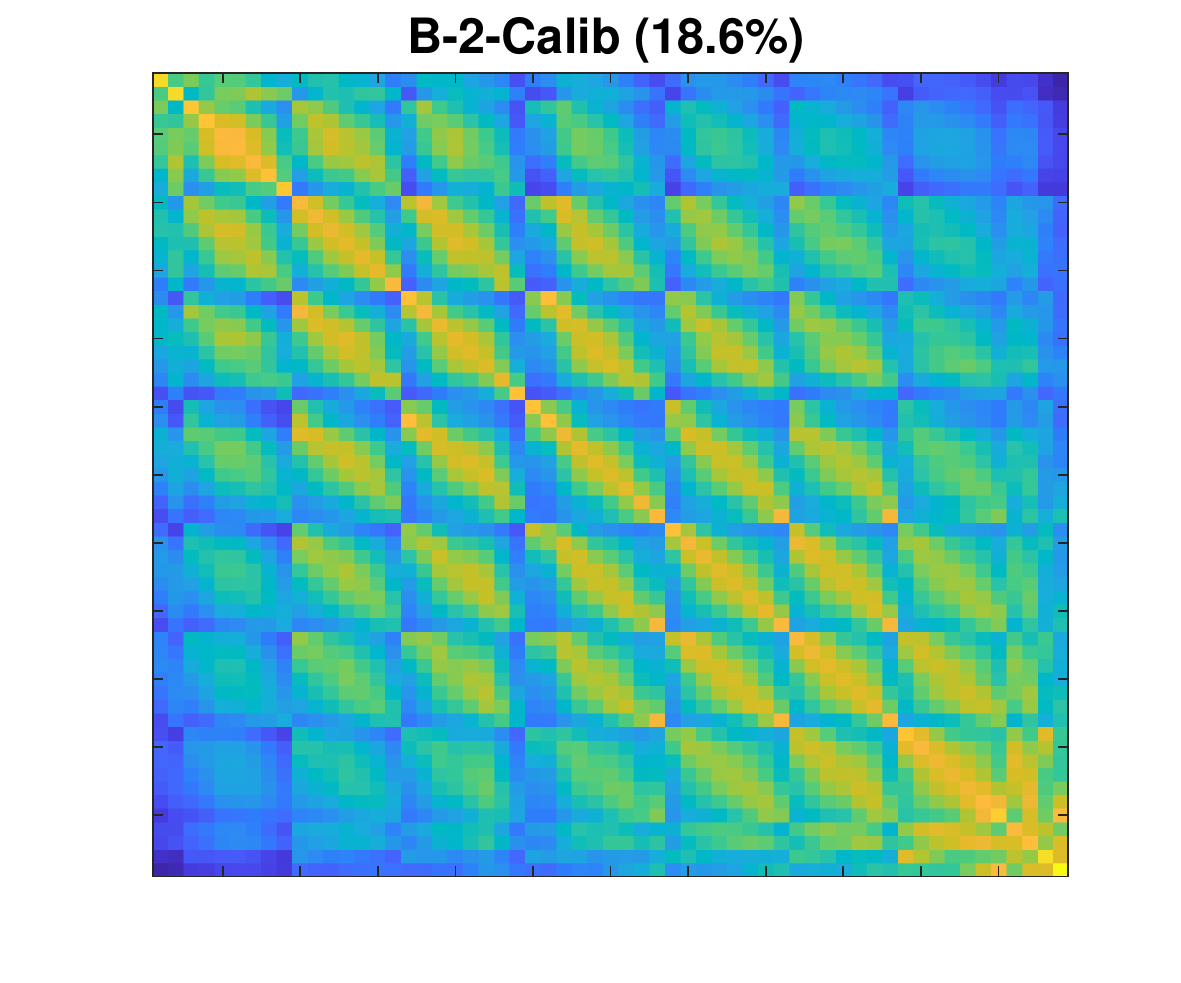}}  & 
\multicolumn{1}{c}{\includegraphics[width = .27 \textwidth, keepaspectratio]{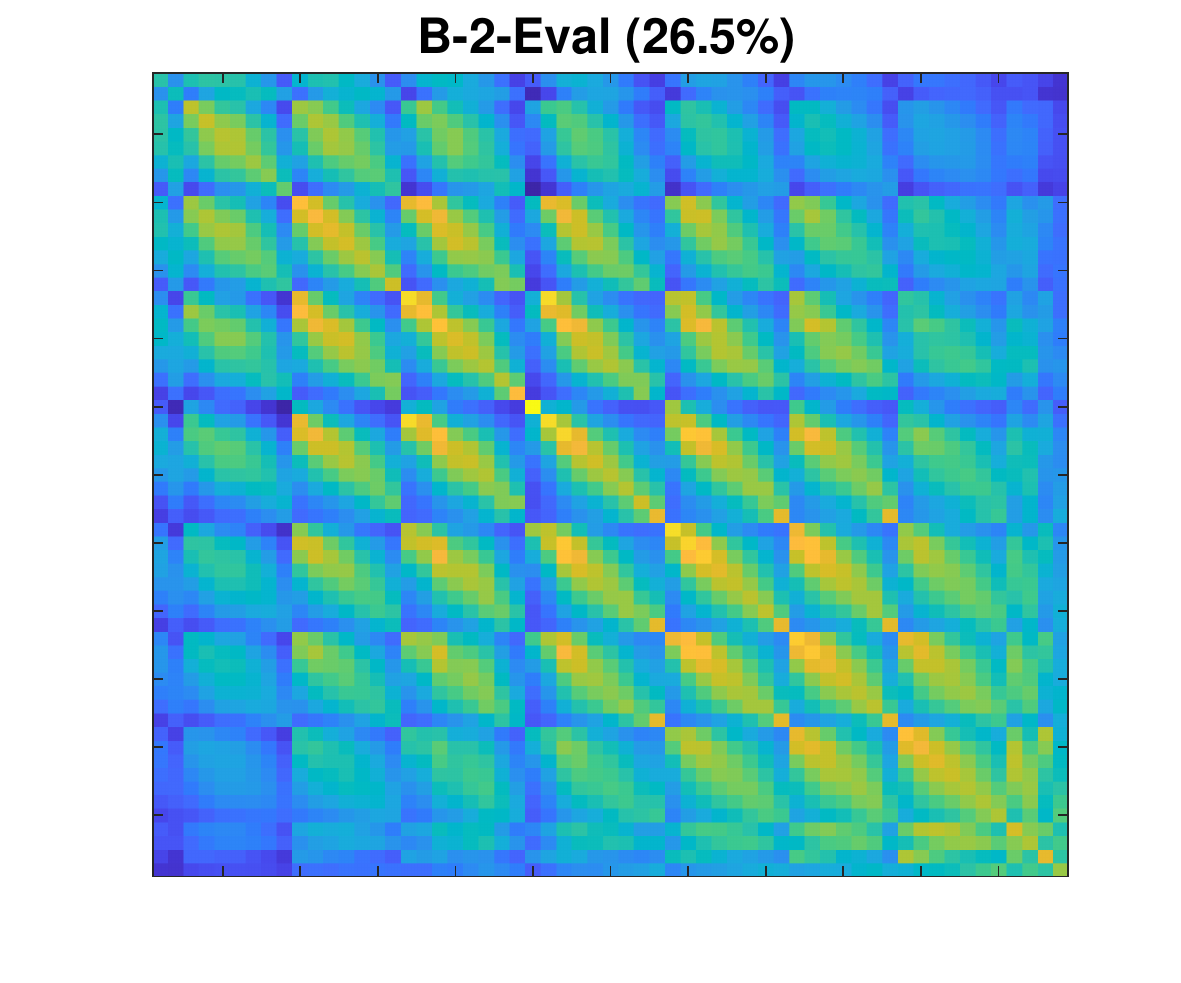}} & 
\multicolumn{1}{c}{\includegraphics[width = .27 \textwidth, keepaspectratio]{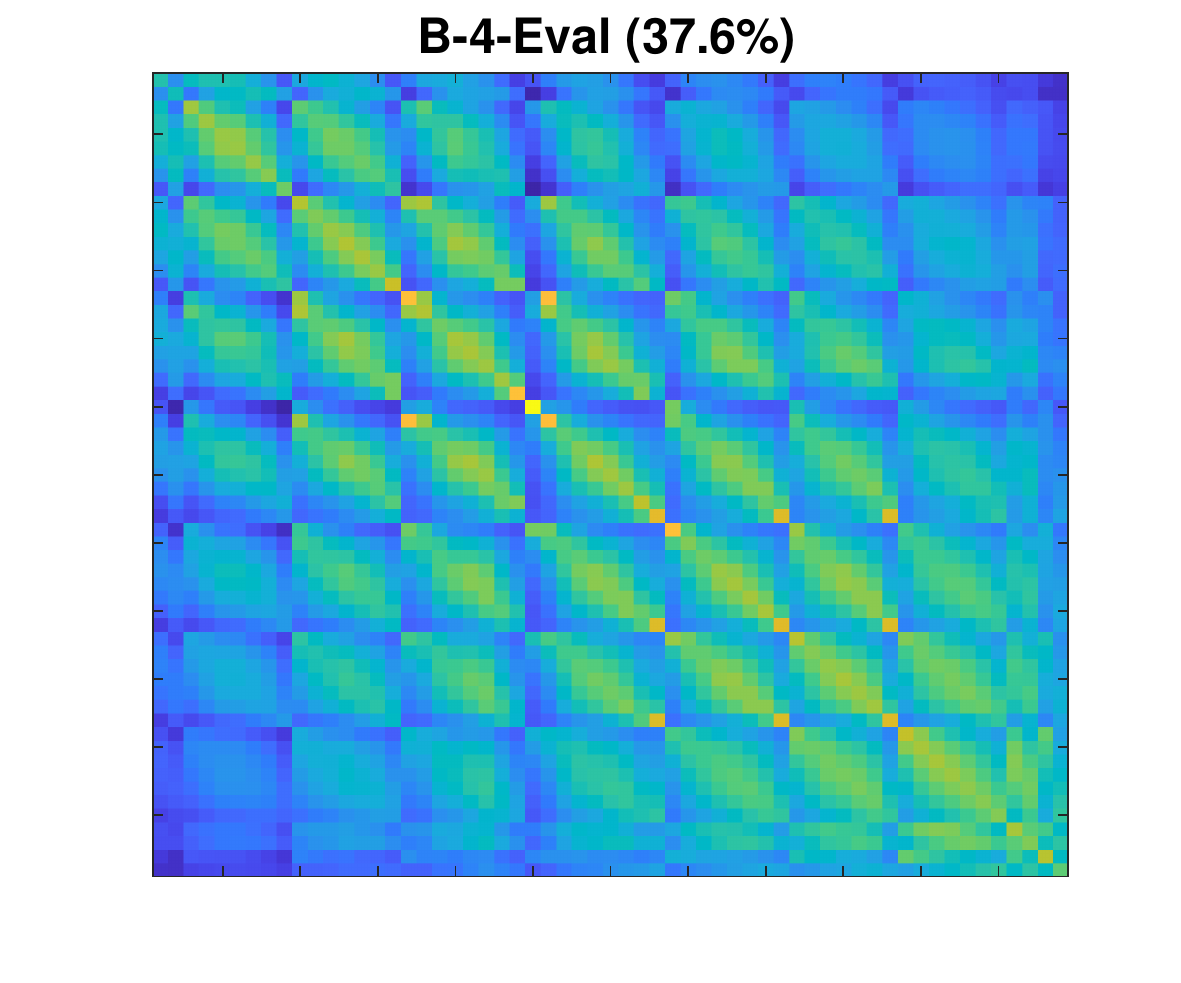}} & 
\multicolumn{1}{c|}{\includegraphics[width = .27 \textwidth, keepaspectratio]{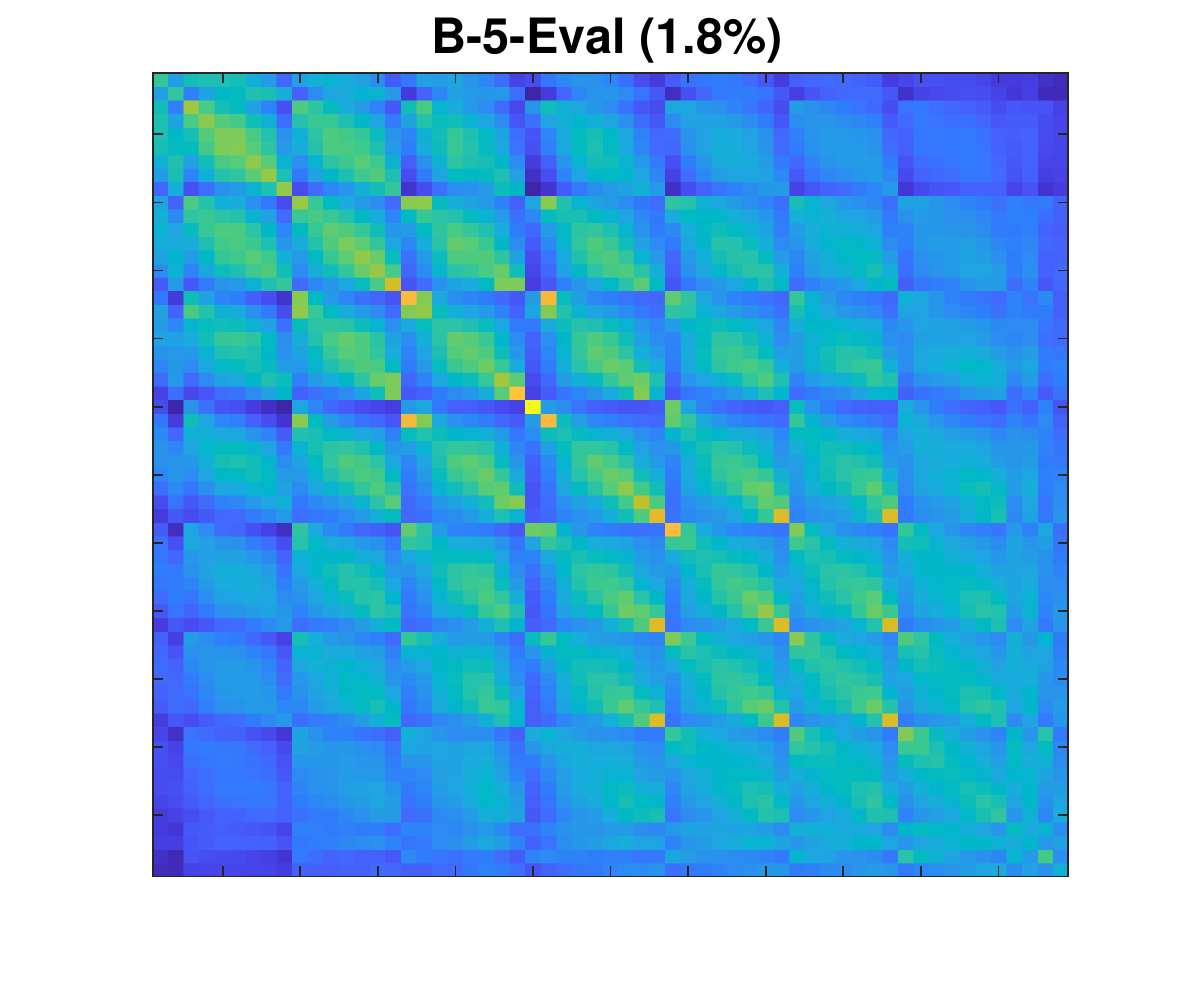}} \\ 
\hline
\includegraphics[width = .27 \textwidth, keepaspectratio]{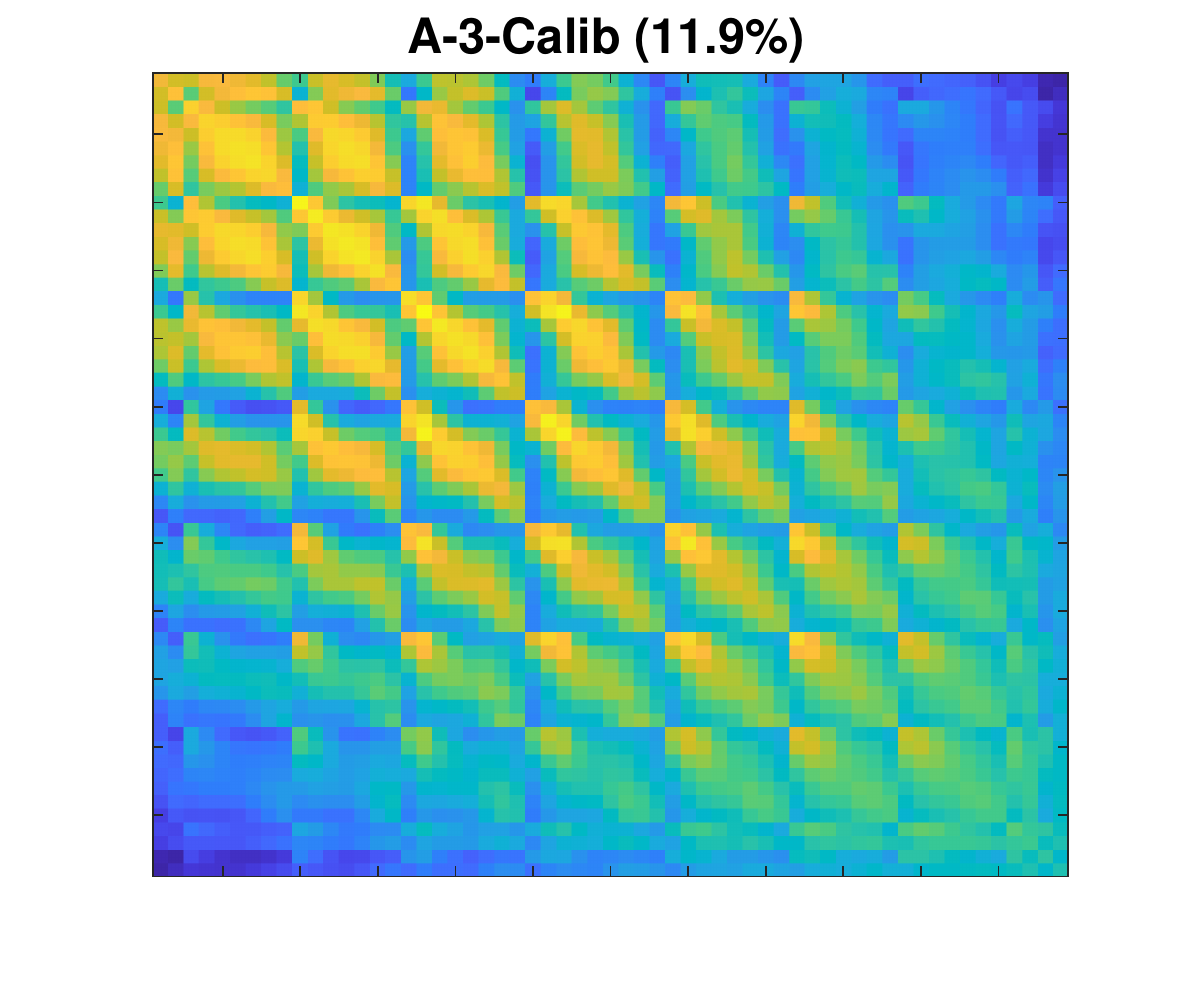} &
\includegraphics[width = .27 \textwidth, keepaspectratio]{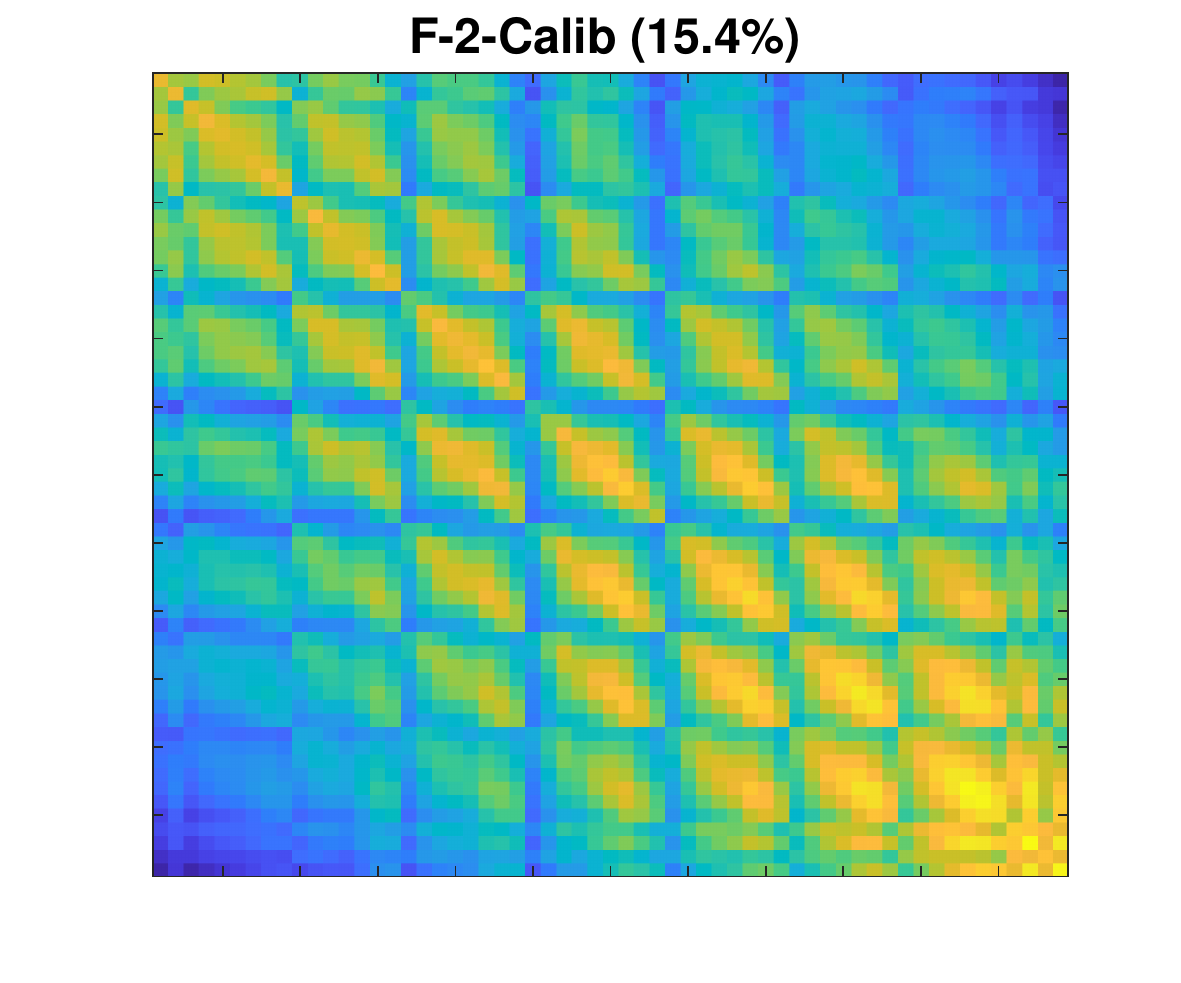}  &
\includegraphics[width = .27 \textwidth, keepaspectratio]{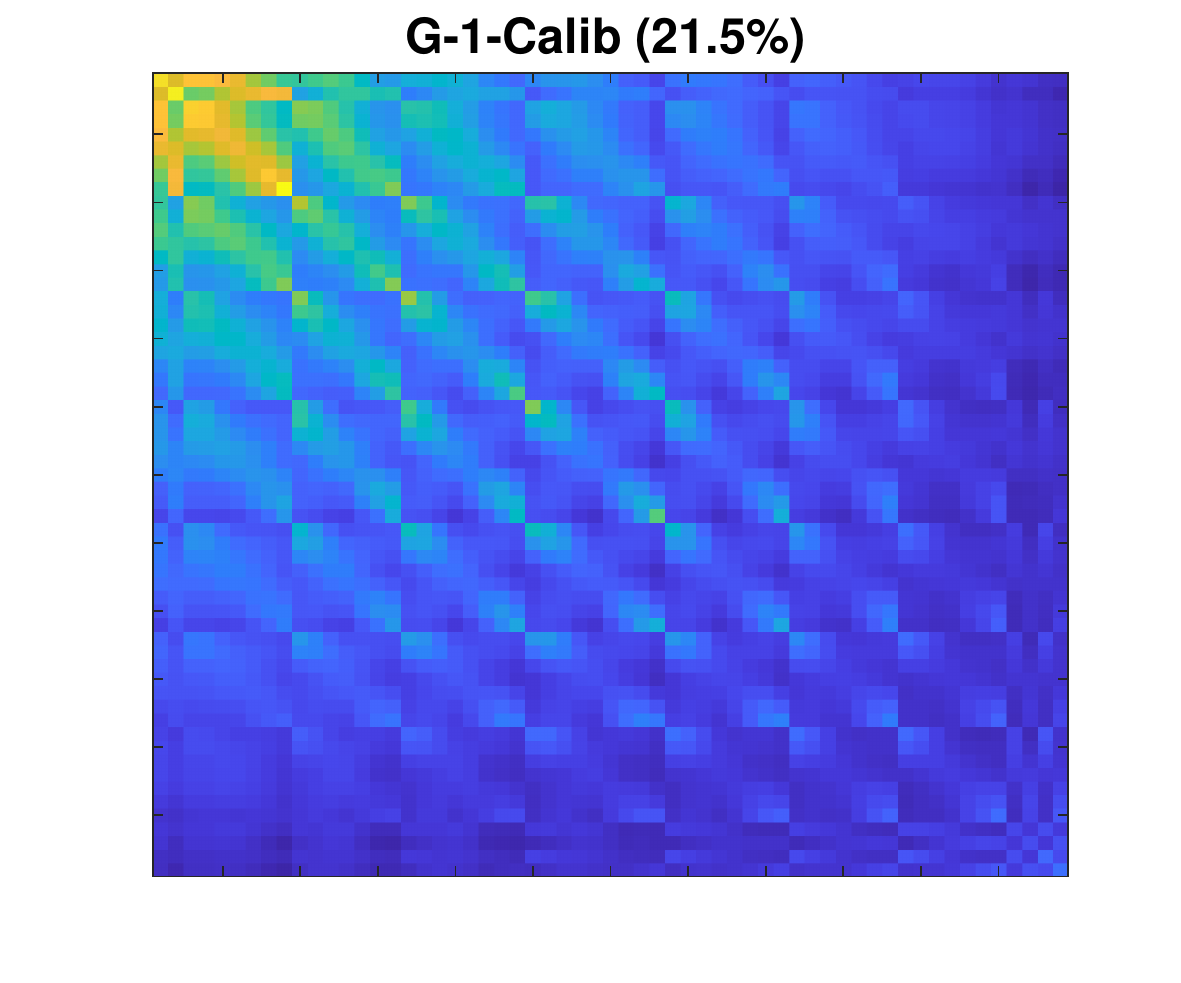}  &
\includegraphics[width = .27 \textwidth, keepaspectratio]{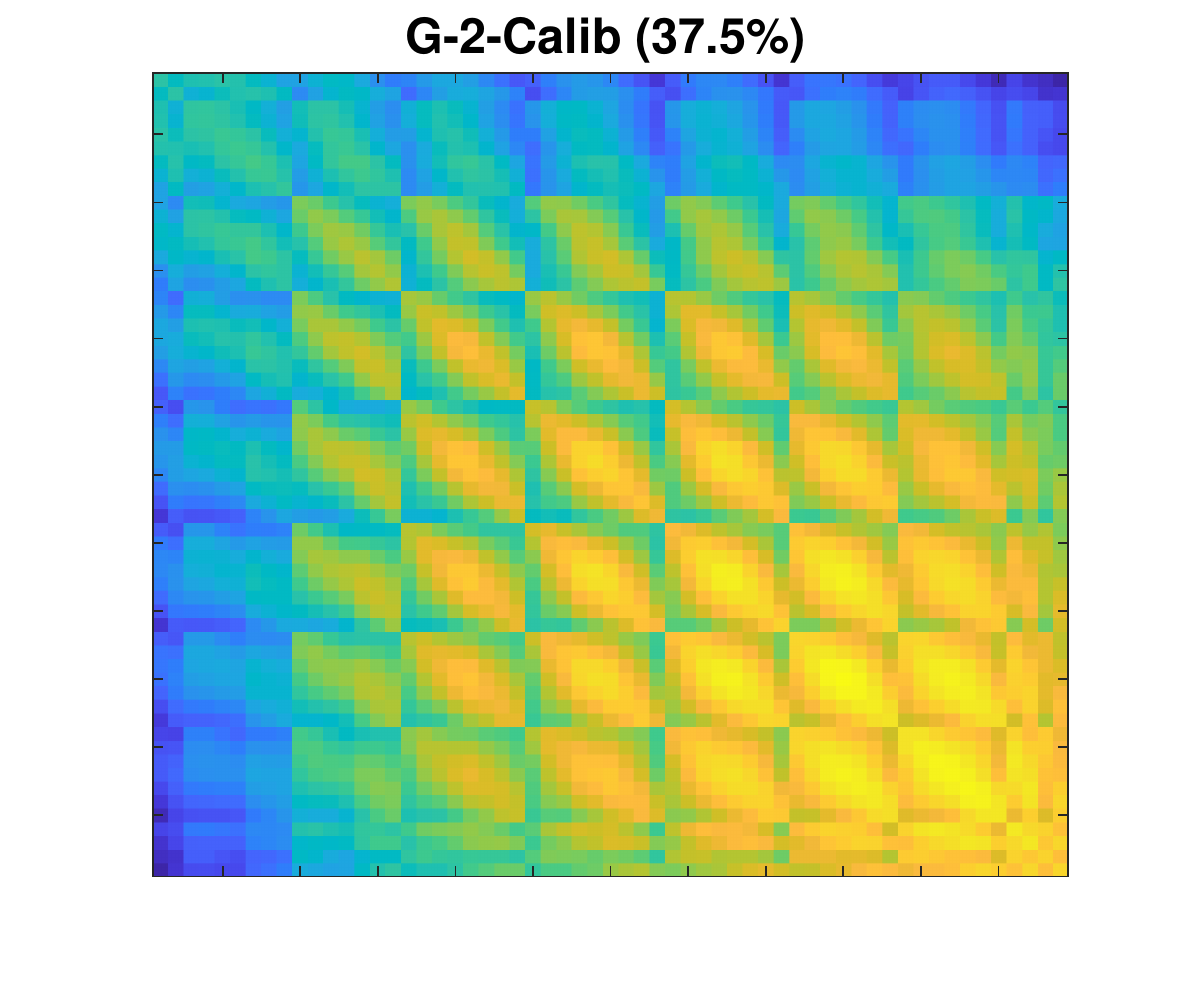} \\
\includegraphics[width = .27 \textwidth, keepaspectratio]{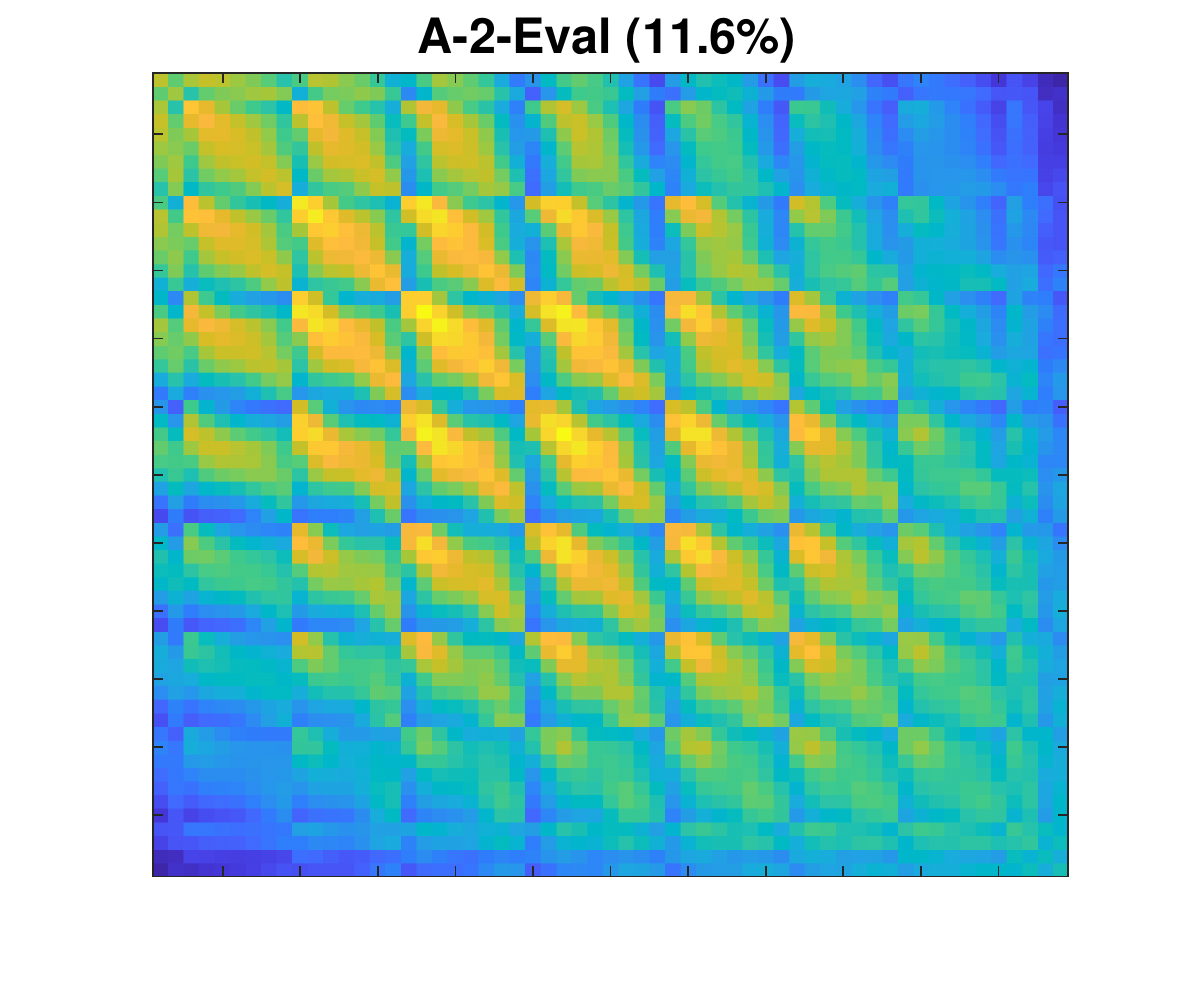} &
\includegraphics[width = .27 \textwidth, keepaspectratio]{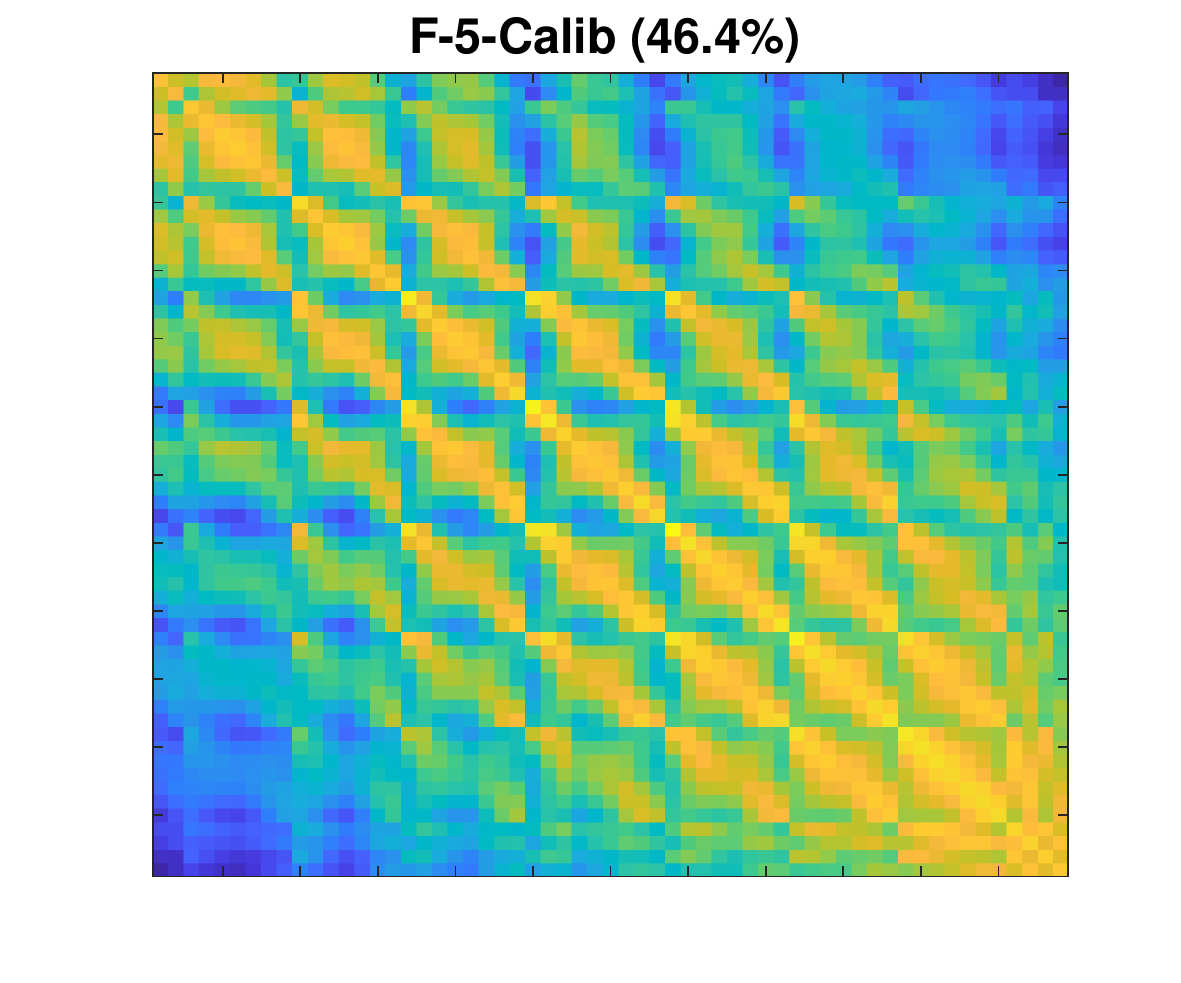} &
\includegraphics[width = .27 \textwidth, keepaspectratio]{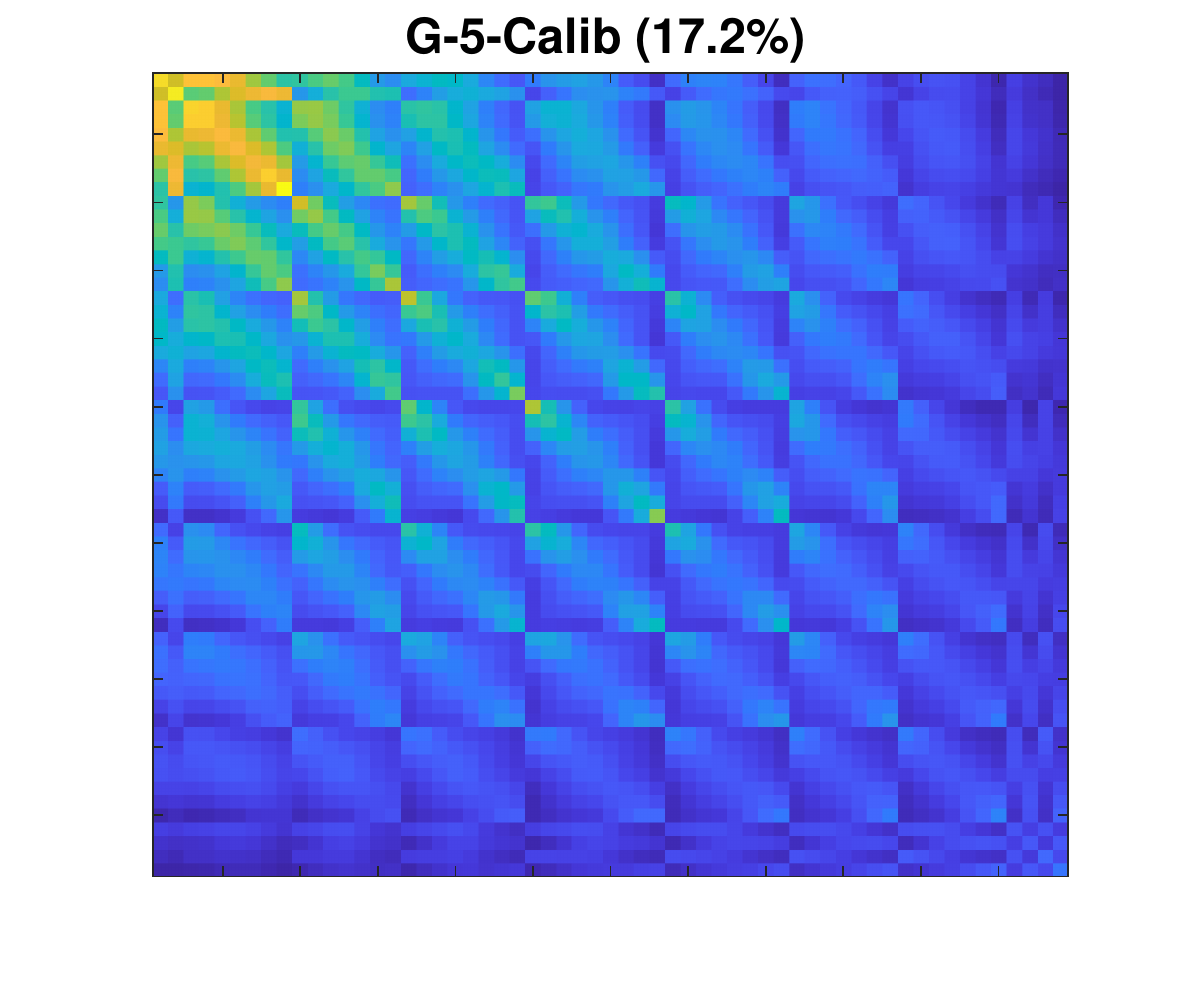}  & 
\includegraphics[width = .27 \textwidth, keepaspectratio]{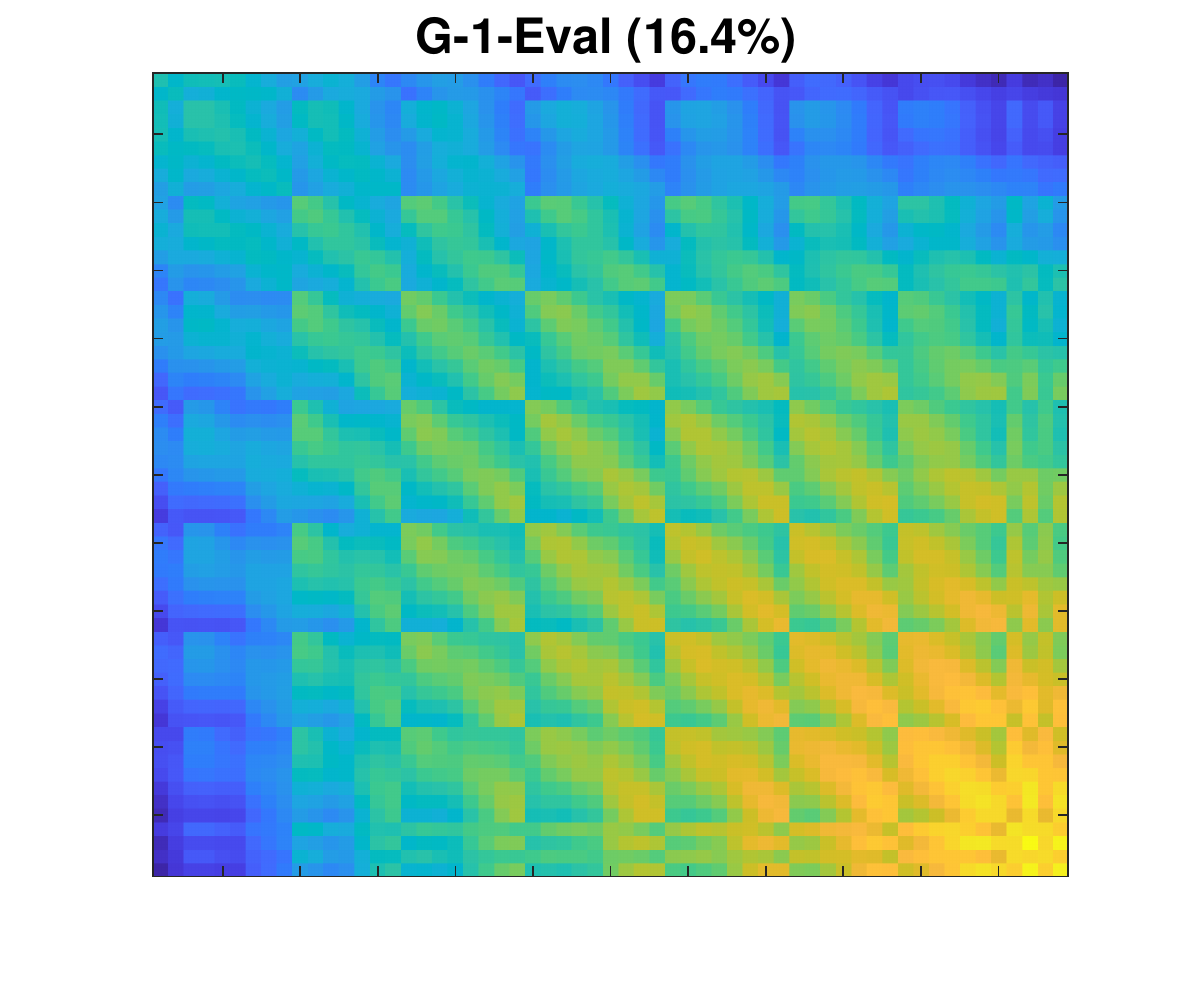} \\
\includegraphics[width = .27 \textwidth, keepaspectratio]{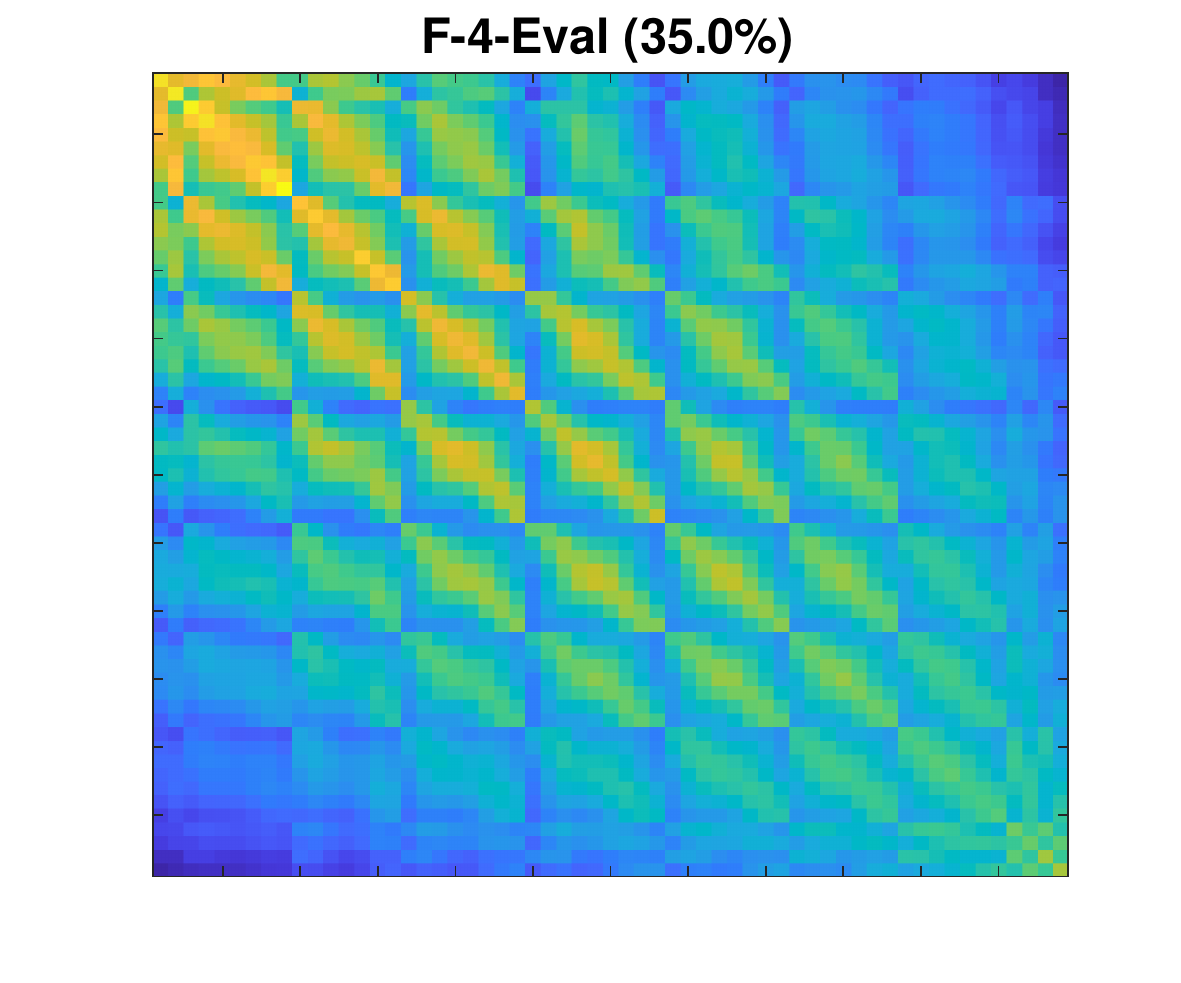} &
\includegraphics[width = .27 \textwidth, keepaspectratio]{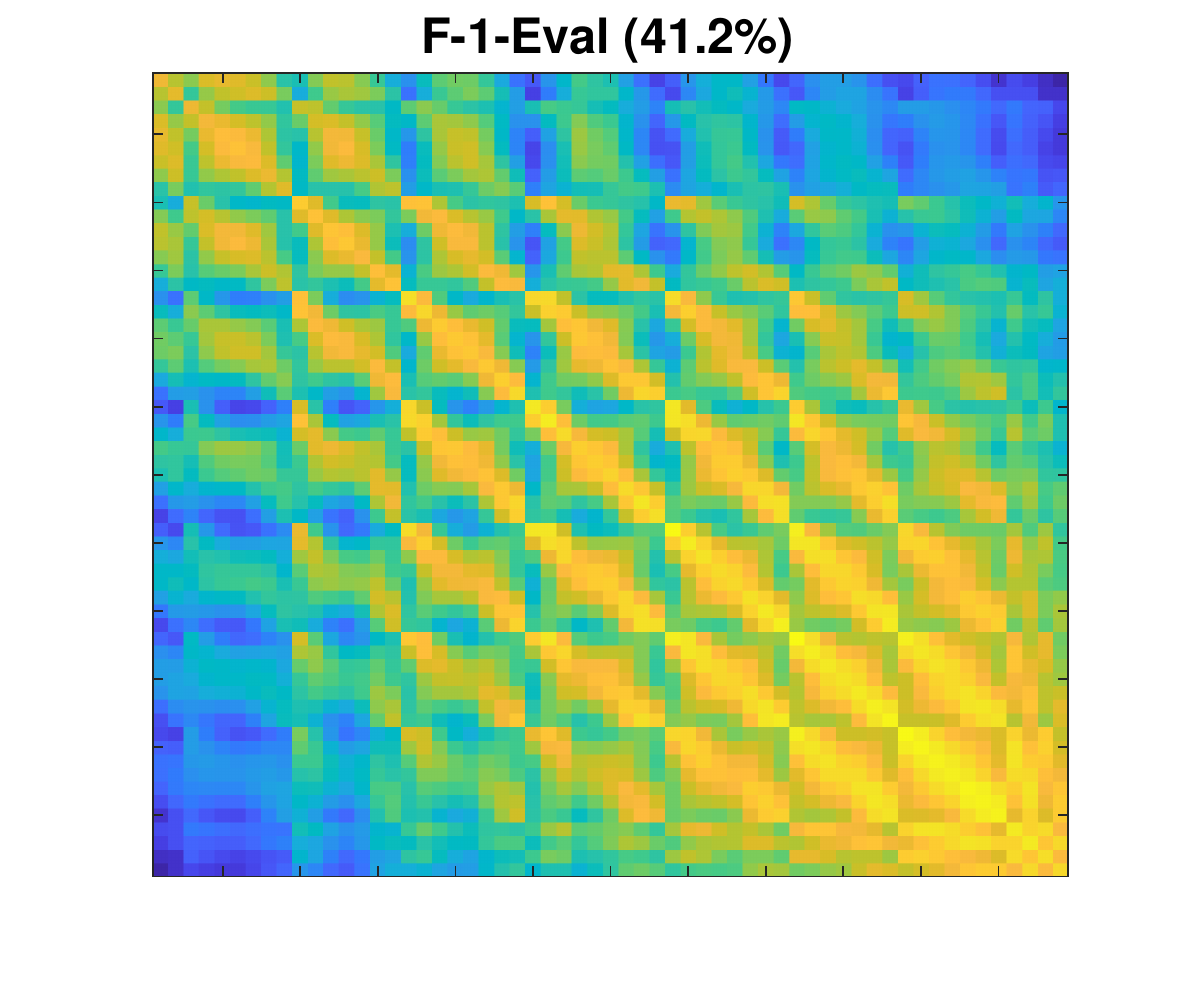}  &
\includegraphics[width = .27 \textwidth, keepaspectratio]{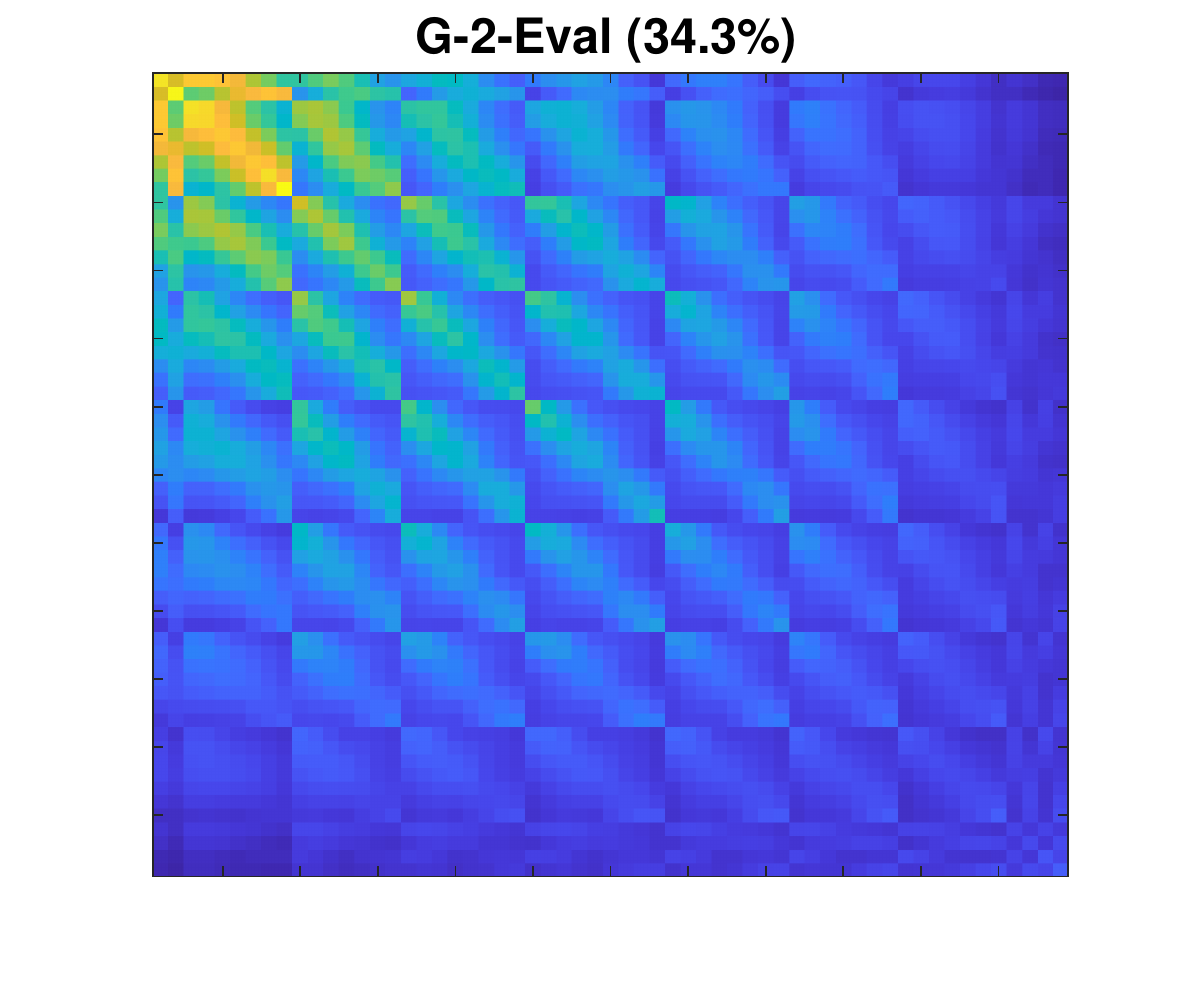}   & 
\includegraphics[width = .27 \textwidth, keepaspectratio]{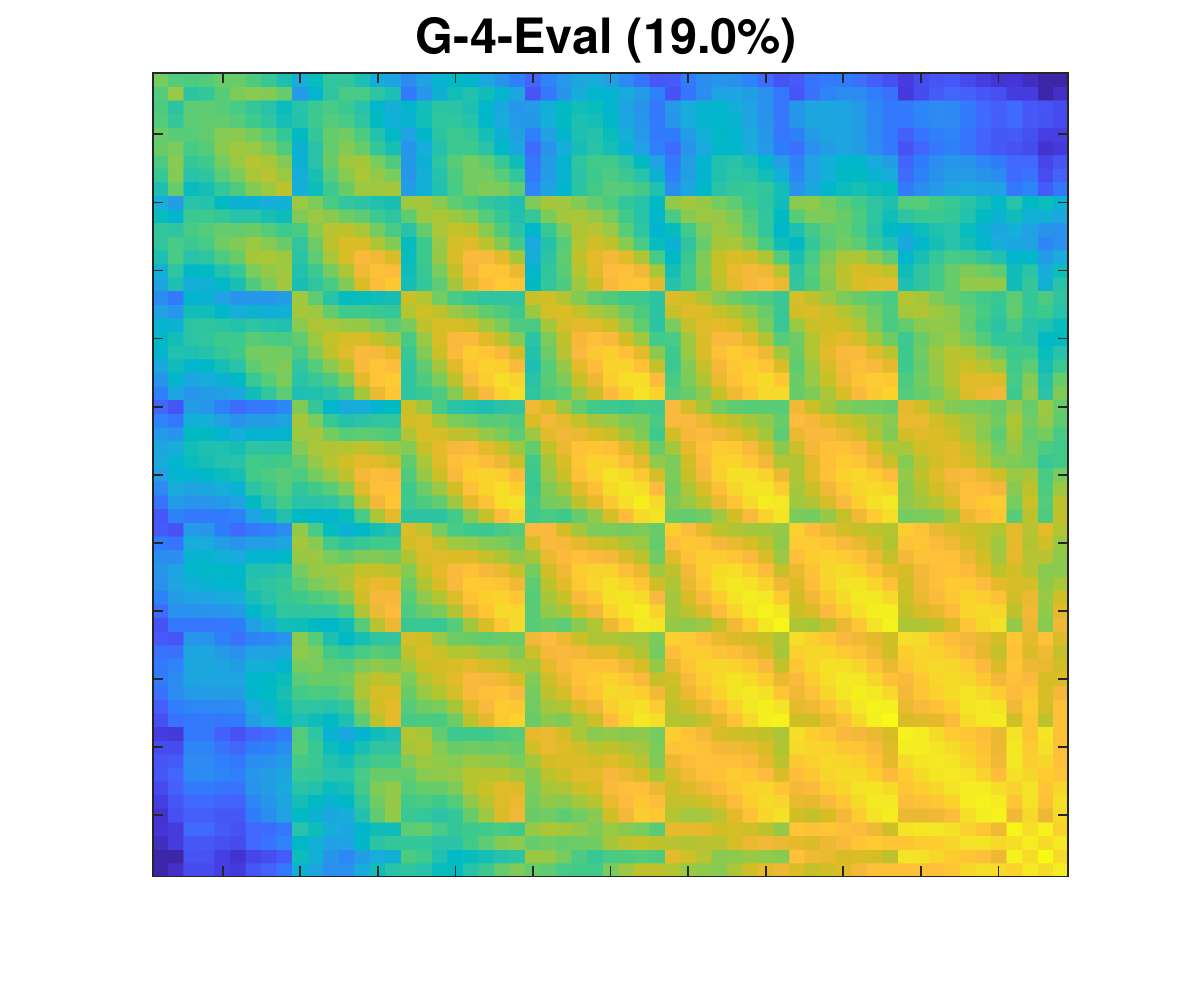} \\
\hline
\end{tabular}

\caption{Clustering of FC measures for the BCI data. Hierarchical clustering was applied to the covariance matrices $\widehat{\bSigma}_j^y$ derived from the switching SSM \eqref{switch-dyn} with $M=5,p=3,r=5$. 
All clusters of size 3 or more are displayed, separated by solid lines.  The  subject, regime, 
session, and dwell time in that regime are indicated on top of each plot. 
}
\label{fig: bci dfc clusters}
\end{center}
\end{figure}

\paragraph{Clustering FC measures.} 
The  above switching SSM was fitted to each dataset and the covariance matrices $\widehat{\bSigma}_j^y$ ($j=1,\ldots ,5$) were  derived as in Section \ref{sec: simulations}, resulting in  $8\times 5 = 40$ matrix features. To avoid redundant information, the lower triangular parts of these symmetric matrices were extracted and vectorized. 
  Treating the overall variance level as a confounder, we scaled the resulting vectors to unit (Euclidean) norm. 
Hierarchical agglomerative clustering was then applied to the 40 feature vectors using the Euclidean metric and the complete linkage criterion. (Reminder: given two sets $A,B$ and a metric $d$, the complete linkage criterion $\max_{a\in A, b\in B}d(a,b)$ 
ensures that all elements in a cluster are close to one another.) 
The associated dendrogram was cut at a height such that all pairs of vectors in a cluster form an angle of at most 8 degrees. This small angle was necessary to produce homogeneous clusters in sufficient number and size. Specifically, $k=23$ clusters were produced, of which 14 were singletons. 
Figure \ref{fig: bci dfc clusters} shows all clusters of size 3 or more. In these clusters, all or nearly all covariance matrices come from the same subject. Besides, the clusters span all subjects, which suggests that variations in FC tend to be much smaller within subjects (even when pooling sessions) than between subjects.  
While there are no obvious common patterns to all 20 covariance matrices, the associated correlation matrices (not shown here) have two common features. First, the average correlation level between all pairs of EEG channels is fairly high (0.5 and above); second, correlation decreases with distance and the decrease is faster in the left-right direction than in the anterior-posterior direction. Concerning spatial patterns in variance levels, we note that each cluster shows one privileged type of variation in the left/right direction: either smooth increase or decrease, uniform (horizontal line), or with high values at the extremes (U-shaped). As for the anterior/posterior direction, clusters mostly show monotonic variations, typically slow but occasionally fast as for the cluster associated with subject G.

\paragraph{Variations in FC between subjects and sessions.} 
Clustering analysis reveals groups of similar FC features without regard to subject or session. 
Here, in contrast, we seek to answer the question: given two pairs of subject/session, 
how well-matched are the associated sets of FC features? 
For this purpose  we consider the weighted average distance 
\begin{equation}\label{eq: weighted matched distance}
  \frac{1}{2} \sum_{i=1}^{M} w_{i}^{A} \min_{1\le j \le M} \| \mathbf{a}_i - \mathbf{b}_j \|_2^2 + \frac{1}{2}
 \sum_{j=1}^{M} w_j^{B} \min_{1\le i \le M} \|  \mathbf{a}_j - \mathbf{b}_i \|_2^2  
\end{equation}
where $(\mathbf{a}_1 ,\ldots, \mathbf{a}_M )$ and $ (\mathbf{b}_1 ,\ldots, \mathbf{b}_M ) $ are FC feature vectors for two pairs of subject/session which we denote by  $A$ and $B$. 
We observe that the implicit matching between feature vectors in \eqref{eq: weighted matched distance} may not be one-to-one. This is no concern as we seek to compare FC features globally. Recent perspectives on one-to-one matching across collections of feature vectors can be found in  \cite{Degras2021}.
The FC features considered here are the covariance matrices $\widehat{\bSigma}^y$, more precisely  their lower triangular parts suitably vectorized and scaled. 
The quantity $w_i^A = \#\{t: \hat{S}_{t}^{A} = i\} / T_A$  is the dwell time in regime $i$ for the data associated with $A$, with  $T_A$ the time series length and $(\hat{S}_{t}^{A})_{1\le t \le T_A}$ the estimated regimes; $w_j^B$ is defined in the same way.  It holds that $\sum_{i=1}^M w_i^A =  \sum_{j=1}^M w_j^B = 1$. 
To provide a meaningful basis for comparison, we also calculate the weighted variance for each pair $A$ 
of subject/session:
\begin{equation}\label{eq: weighted variance}
  \frac{1}{1 - \sum_{i=1}^{M} (w_i^A)^2 } \sum_{i=1}^M \sum_{j=1}^{M} w_{i}^{A} w_{j}^A \| \mathbf{a}_i - \mathbf{a}_j \|_2^2 \,.
\end{equation}

Table \ref{table: BCI matched} shows all weighted distances and variances based on \eqref{eq: weighted matched distance}-\eqref{eq: weighted variance}.
The diagonal of this table shows the within-session variance \eqref{eq: weighted variance}, which is highest for subjects A  and G  while subject F comes in third position. Subject B has by far the lowest within-session variance for both sessions. The diagonal also shows that the within-session variance is larger in the calibration session than in the evaluation session, especially for A and G.  
The fourth superdiagonal (.05, .02, .12, .06) shows the weighted distance between a subject and themself across sessions. 
Comparing its entries to the corresponding rows and columns, we see that each of subject A and B are better matched to themselves across sessions than they are to any other subject. This is also very nearly the case for subject G.  Looking at average weighted distances between a subject and the three others, subject G is the most distant from the pack: 0.1230 versus 0.1054,
    0.1073, and 0.1031 for subjects A, B, C, respectively.

\begin{table}[ht]
\begin{center}
\setlength{\tabcolsep}{4pt}
\begin{tabular}{  cc |   *{4}{c}  *{4}{c} | }
\hline
\multicolumn{2}{|c}{Session}& \multicolumn{4}{|c|}{Calibration} &   \multicolumn{4}{c|}{Evaluation} \\
\multicolumn{2}{|c|}{Subject} & A & B & F & \multicolumn{1}{c|}{G} & A & B & F & G \\
\hline
\multicolumn{1}{|c}{\multirow{4}{*}{\rotatebox{90}{\footnotesize{Calibration}\hspace*{0.5mm} } }} 
& A&  0.43 &0.12 &   0.12 &   0.13 &   0.05 &   0.14 &   0.11&    0.12\\
\multicolumn{1}{|c}{} &   B&           &   0.05  &   0.08  &  0.18 &   0.10 &   0.02 &   0.08 &   0.09\\
\multicolumn{1}{|c}{} &       F&      &      &     0.17   &    0.16  &  0.07 &   0.11&    0.12 &   0.10\\
\multicolumn{1}{|c}{} &      G&           &       &     & 0.55        & 0.15 &   0.19    &0.17  &  0.06\\
      \cline{1-2}
      \multicolumn{1}{|c}{\multirow{4}{*}{\rotatebox{90}{\footnotesize{Evaluation} \hspace*{-0.4mm}  } }} 
   &      A&       &        &       &          &   0.14    & 0.06    &0.08   & 0.05\\
  \multicolumn{1}{|c}{} &          B&      &        &         &        &         &   0.04     &0.08   & 0.06\\
   \multicolumn{1}{|c}{} &           F&     &     &          &       &             &     &   0.14    & 0.08\\
  \multicolumn{1}{|c}{} &          G&      &    &             &    &                &  &             &  0.23   \\
                \hline
\end{tabular}
\caption{Weighted average distances \eqref{eq: weighted matched distance} between matched FC measures (off-diagonal) and weighted variance \eqref{eq: weighted variance} (on-diagonal) for the BCI data. For each subject and session, the FC measures consist in $M=5$ covariance matrices (one per FC regime).}
\label{table: BCI matched}
\end{center}
\end{table}%

\paragraph{FC regimes and experimental tasks.} 
Although our switching SSM methodology is primarily designed to investigate dFC in EEG data
and not to predict external outcomes such as epileptic activity or motor imagery, 
it is natural to interrogate the relationship between the derived FC regimes $(\widehat{S}_t)$ 
and the sequence of experimental conditions, say $(C_t)$. 

Table \ref{tab: FC regimes and tasks} shows the distribution of the FC regimes $(\widehat{S}_t)$ across experimental tasks for the BCI calibration data. Similar results hold with the evaluation data. For each dataset, the regime estimates $(\widehat{S}_t)$ were obtained as $\widehat{S}_t = \argmax_{1\le j \le M} P_{\widehat{\btheta}}(S_t = j | \by_{1:T})$ after fitting model \eqref{switch-dyn} with $M=5,p=3, r=5$. For the experimental condition sequence $(C_t)$, task 1, rest, and task 2 were respectively 
coded as $1,2,3$. During the calibration session, about 21\% of the session was spent in task 1, 21\% in task 2, and the remaining 58\% in resting condition. As can be seen in the table, the $M=5$ estimated FC regimes are distributed across the experimental tasks more or less in the same 21/21/58 proportions. This result, which holds for the selected model and also for other choices of $(M,p,r)$, suggests that the FC regimes are largely independent of the experimental tasks. In other words, motor imagery only plays a small part in the changes in FC as measured by the regime indicators $(S_t)$.

\begin{table}[ht]
\begin{center}
\begin{tabular}{c l *{5}{c}}
&& \multicolumn{5}{c}{Regime} \\
\cline{3-7}
Subject & Task & 1 & 2 & 3 & 4 & 5 \\
\hline
 \multirow{3}{*}{A} &  left &  20.9&   21.0   &   22.4 &    23.0&     18.5\\
&    foot &     22.6& 22.5  &  20.0&   21.7&    18.9\\
& rest &     56.5&  56.5 &    57.6 &    55.3&   62.6    \\

\hline
 \multirow{3}{*}{B} &  left &22.9  &  18.1&22.8 &20.9 &14.4  \\
&    right &  21.5& 22.7 &18.2 & 19.1& 19.1 \\
& rest & 55.6 & 59.2 &59.0 &59.9 &66.5  \\
\hline
 \multirow{3}{*}{F} &  left & 20.9 &  19.2&21.1 &21.1 & 21.7 \\
&    foot & 21.0 & 21.3 & 21.3&21.1 & 20.8 \\
& rest & 58.1 &  59.5& 57.6& 57.8& 57.5 \\
\hline
 \multirow{3}{*}{G} &  left & 19.9 & 20.5 &16.7 & 22.8& 23.7 \\
& right & 19.4 & 22.3 & 25.4& 21.8& 17.8 \\
&    rest & 60.7 & 57.2 & 57.8& 55.4& 58.6 \\
\hline
\end{tabular}
\caption{Distribution of the estimated FC regimes $(\widehat{S}_t)$ across tasks (dwell time \%) for the BCI calibration data. 
''Left" and ``Right" refer to hand motor imagery. For each subject, each task (two among ``Left", ``Right", and ``Foot") lasted $\sim$21\% of the session while ``Rest" lasted $\sim$58\% of the the session.}
\label{tab: FC regimes and tasks}
\end{center}
\end{table}%

In light of the quasi-independence between estimated FC regimes and experimental tasks, one may ask: would the association between the two be stronger if the switching SSM was fitted to the data in a supervised way? More precisely, such supervised approach would consist in: (i) estimating the parameters $\btheta$ of model \eqref{switch-dyn} with $M=3$ (the number of tasks) while fixing the regimes $(S_t)$ to the task indicators ($C_t$), and (ii) estimating  the regimes $(S_t)$ while fixing  $\btheta$ to the estimates $\widehat{\btheta}$ of the first step. Interestingly, the answer to this question is no: the ``supervised" regimes $(\widehat{S}_t)$ remain largely independent of the experimental task indicators. Again, this observation holds  for a wide range of values $(p,r)$ beyond the  model retained for our analysis. By fitting a separate switching SSM 
for each experimental condition with a sufficient number of regimes ($M>1$), and by combining the three separate models into a bigger switching SSM, one may arrive at estimated regimes $(\widehat{S}_t)$ that match the tasks $(C_t)$ much more closely. 
This is  however accomplished at the price of overfitting, and the ensuing model does not generalize well to the evaluation data.


\section{Discussion}
\vspace*{-2mm}

\paragraph{Summary.} 

In this work we have developed new computational techniques, inferential methods, and software for Markov-switching state-space models in view of high-dimensional applications, in particular to the analysis of dynamic functional connectivity (dFC) in neuroimaging data. Our approach utilizes maximum likelihood estimation and its EM implementation.

We have considered three Markov-switching SSMs general enough to apply to various scientific fields. At the same time we have  motivated and interpreted these models in the specific context of dFC. The first model enables dimension reduction as well as different dynamics for each regime (``switching dynamics"). The switching VAR model, a special case of the first, specifies dynamics directly at the observation level without dimension reduction. 
This model enjoys a low memory footprint and fast computations for time series of relatively low dimension ($N$ in the tens). It is however not suitable for high-dimensional time series (at least not without additional rank or sparsity constraints) due to its number of model parameters $O(MpN^2)$. The third model (``switching observations") posits multiple state processes along with a Markov regime sequence indicating which of the processes generates the observations at a given time. The multiplicity of state processes and the latent Markov chain represent the uncertainty about the data-generating mechanism. 

Building on existing EM methodology, we have contributed initialization methods for the EM, numerical tools for handling optimization constraints, and a simple acceleration scheme that fill important gaps in the literature. The utility and effectiveness of these methods has been demonstrated in the simulations and data analyses. 
Good initialization methods are essential for the EM, especially in high-dimensional models, given the tendency of this algorithm to get stuck in local maxima. The proposed optimization methods make it possible to flexibly specify a Markov-switching SSM with fixed coefficient constraints, scaling constraints, equality constraints across regimes, and eigenvalue constraints on the state transition matrices. 
Fixed coefficient constraints, for example,  are useful to impose diagonality or more general sparsity structures on high-dimensional parameters. 
They can also serve to compare supervised model fits (with known regimes) to unsupervised fits like in the BCI data analysis. 
Eigenvalue constraints guarantee that the modeled processes are stable and asymptotically stationary. To our knowledge, the proposed algebraic method is the only one to enforce stability in the context of maximum likelihood estimation for VAR models. The proposed EM acceleration scheme is conceptually simple and converges in a reasonable time with large datasets, which usual EM acceleration techniques fail to do. 

One of the most important contributions of this work is a new parametric bootstrap  for Markov-switching SSMs. 
This bootstrap can be harnessed to infer model parameters and functions thereof such as regime-specific stationary covariance, correlation and autocorrelation structures which are paramount in dynamic functional connectivity analysis. For high-dimensional time series, given the intractability of likelihood-based inference and the inaccuracy of nonparametric bootstrap, our methodology is the only viable route to rigorous frequentist inference.

A simulation study has shown good to excellent performance of the  EM/maximum likelihood approach  in regime and parameter estimation  in the switching dynamics model for all data dimensions $(N,T)$.  The approach fared well in the switching VAR model for small $N$ but not for larger values, suggesting that satisfactory performance is attainable when $T \gg N$. Numerical instability and poor statistical accuracy were observed in the switching observations model, which indicates that finer approximations to the likelihood function are required there. 
In most simulation setups, the EM-based approach has substantially higher accuracy than its least squares initialization method and the sliding windows method. Confidence intervals based on the parametric bootstrap have shown excellent performance (near-nominal pointwise coverage) in the switching dynamics model for all $(N,T)$ and in the switching VAR model for $N$ sufficiently small compared to $T$.

In applications of Markov-switching SSMs to EEG data, we obtained new insights on brain function during  epileptic states and motor imagery. These significant findings do not appear in previous research on the same data 
and are made possible by the methodological and computational developments of this paper. 
 In the study of epileptic activity, after model selection, the retained switching VAR model correctly identified the three clinical phases (pre-ictal, ictal, and  late-ictal) of the seizure event. It  also revealed a finer-grained segmentation of the EEG in five regimes of functional connectivity (FC) that mostly but not fully align with the clinical phases. One of these  regimes was predominant in the pre-ictal phase, then disappeared in the ictal phase and its immediate aftermath, and returned at the very end of the recording. This observation is useful in assessing the time necessary for brain function to go back to normal after a seizure. 
In comparison to previous studies that decompose the same data in two regimes (ictal/nonictal or pre-ictal/ictal), 
our fine-grained segmentation revealed remarkable dynamics in the ictal phase (damped oscillations) but also stark contrasts between the pre- and  late-ictal phases (e.g.,   much larger spatial heterogeneity in autocorrelation patterns in the  late-ictal phase). 
We used our parametric bootstrap methodology to conduct simultaneous inference on FC graphs based on partial correlation. By considering all $N=16$ available EEG channels, as opposed to previous works that only considered specific channels associated with the seizure event, we were able to document not only fluctuations in FC graphs induced by the seizure event but also FC subgraphs that were preserved throughout the seizure.  
The second data analysis of the paper concerned motor imagery in brain computer interfaces (BCI). We used a switching dynamics model to reduce the dimension of the EEG channels ($N=59$) to a much smaller number 
of state variables ($r=5$). After fitting the model to EEG data of subjects recorded over multiple sessions, 
we extracted regime-specific stationary covariance matrices as FC measures and applied hierarchical clustering to  them. Common patterns were difficult to discern in covariances but the associated correlation and variance structures displayed consistent patterns of variation in the left/right and the anterior/posterior directions. In complement to this unsupervised approach, we quantified the variations in FC between subjects and sessions through an average distance between sets of FC measures that accounts  for the dwell times of FC regimes. This analysis showed that differences in FC  are strong between subjects but also that individual FC patterns are repeatable across sessions. We also sought to compare the time segmentations produced by the switching SSMs, i.e., the FC regime sequences, to the experimental task sequences. The comparison of unsupervised model fits to supervised fits with  regimes fixed to the task sequence showed that the derived FC regimes were essentially independent of the experimental tasks. This confirms that changes in FC induced by motor imagery are very small compared to natural FC variations.

Last but not least, we have developed a highly optimized MATLAB toolbox that implements Markov-switching SSMs (\url{https://github.com/ddegras/switch-ssm}). In addition to EM implementations of maximum likelihood estimation for the three models studied in this paper, the toolbox also contains functions for initializing the EM, 
extracting regime-specific stationary measures (covariance, correlation, coherence, autocorrelation, cross-covariance...),  performing parametric bootstrap serially or in parallel, calculating confidence intervals, and simulating the models. Example code is given in  Appendix \ref{sec: matlab}.

\paragraph{Extensions.}

The present work can be fruitfully extended in a number of directions. For example, the Markovian dynamics of the regime sequence $(S_t)$ may be too restrictive in some applications. It would be beneficial to extend the EM algorithm to semi-Markovian dynamics  for greater modeling flexibility \citep[e.g.,][]{Alaa2016}, or to pursue this extension from a Bayesian perspective with the Gibbs sampler or sequential Monte Carlo methods. 
A simpler alternative could be the method of  ``state aggregates" \citep{Langrock2011}
which can approximate arbitrary well the distribution of semi-Markov chains using only Markov chains.
In relation to this, it was observed that when fitting a Markov-switching SSM to data with a large number of regimes, some regimes may be duplicated, i.e., have nearly identical model parameters except for their transition probabilities. As a result, any meta-regime associated with duplicated regimes has a dwell time with negative binomial distribution that generalizes the geometric distribution.

For simplicity  intercept vectors were not included in the state equations 
of models \eqref{switch-ssm}--\eqref{switch-obs} nor were covariates included in the observation equation, 
but both quantities can easily be handled with the EM. 
More ambitious extensions of the general switching SSM \eqref{switch-ssm} would comprise state equations with 
richer dynamics such as structural vector autoregressive and VARMA models.

Regarding statistical inference, 
the  parametric bootstrap proposed in this paper
 has been used in the construction of confidence intervals (CI), 
 and its accuracy was measured in terms of average pointwise coverage.
In future work it would be good to examine the simultaneous coverage properties 
of the proposed percentile, basic, and normal bootstrap CIs, 
and maybe to develop more sophisticated CI methods.

\bibliographystyle{elsarticle-harv} 
\bibliography{EEG,fMRI,stats}

\newpage

\appendix

\section{Derivation of stationary covariance matrices}
\label{sec: stationary covariance} 

This appendix details how to obtain the stationary covariance matrices $\boldsymbol{\Sigma}_j^{x}$ in \eqref{eq: stationary covariance}, an important calculation which we have not found elsewhere in the literature.
Define the expanded state vector $\tilde{\bx}_{t} = \mathrm{vec}(\bx_t ,\ldots, \bx_{t-p+1} )$ 
(in models \eqref{switch-dyn}-\eqref{switch-var}, replace $\bx_{t }$ by $\bx_{t j}$ in model \eqref{switch-obs}). Also define the companion matrix $\widetilde{\bA}_j$ to the state transition matrices $\bA_{1j},\ldots , \bA_{pj}$, 
and the expanded state noise covariance matrix  $\widetilde{\bQ}_j$ by 
$$ 
\widetilde{\bA}_{j} = 
\left( \begin{array}{c c c c} 
 \bA_{1 j} &  \bA_{2 j} & \cdots &  \bA_{p j} \\
\bI_r & \mathbf{0}_{r\times r} & \cdots & \mathbf{0}_{r\times r}  \\
\mathbf{0}_{r\times r} & \ddots & \ddots& \vdots \\
\mathbf{0}_{r\times r}&\mathbf{0}_{r\times r}&\bI_r & \mathbf{0}_{r\times r} 
\end{array} \right) , \qquad 
\widetilde{\bQ}_j = \left( 
\begin{array}{cc} 
\bQ_j &  \bzero_{r \times (p-1)r}  \\ 
 \bzero_{(p-1)r \times r}  & \bzero_{(p-1)r \times (p-1)r} 
\end{array}
\right).
$$ 
We assume here that the eigenvalues of $ \widetilde{\bA}_{j}$ are less than 1 in modulus, 
which guarantees that the state process $(\tilde{\bx}_t)$ is asymptotically stationary. 
The  stationarity equation for $\boldsymbol{\Sigma}_{j}^{\tilde{x}} = \lim_{t \to \infty} V(\tilde{\bx}_t | S_1 = \cdots = S_t = j )$ expresses as  
\begin{equation}
\label{stationarity}
\boldsymbol{\Sigma}_{j}^{\tilde{x}} = \widetilde{\bA}_{j}  \boldsymbol{\Sigma}_{j}^{\tilde{x}} \widetilde{\bA}_{j}' +  \widetilde{\bQ}_j 
\end{equation}
and $\boldsymbol{\Sigma}_j^{x}$ equals any of the $p$ 
diagonal blocks of $\boldsymbol{\Sigma}_{j}^{\tilde{x}}$ of size $r\times r$. 

If $ \widetilde{\bA}_{j}$ is of full rank,  
this equation can be transformed into the Sylvester equation 
\begin{equation}
 \widetilde{\bA}_{j}^{-1}\boldsymbol{\Sigma}_{j}^{\tilde{x}}   - \boldsymbol{\Sigma}_{j}^{\tilde{x}} \widetilde{\bA}_{j}' = \widetilde{\bA}_{j}^{-1}  \widetilde{\bQ}_j 
\end{equation}
 and solved with standard methods \citep[e.g.,][]{Golub1979}.

On the other hand, if $ \widetilde{\bA}_{j}$ is not of full rank, 
\eqref{stationarity} can be vectorized as 
\begin{equation*}
\bv_j = ( \widetilde{\bA}_{j} \otimes  \widetilde{\bA}_{j} ) \bv_j + \mathbf{q}_j
\end{equation*}
where $\bv_j = \mathrm{vec}(\boldsymbol{\Sigma}_{j}^{\tilde{x}}  )$, $\mathbf{q}_j = \mathrm{vec}(\bQ_j )$, 
and $\otimes $ denotes Kronecker product. Noting that $\bv_j $ contains only $r(r+1)/2 + (p-1)r^2$ unique elements, 
there exists a matrix $\mathbf{M}$ filled with zeros except for a single value one in each row such that 
  $\bv_j = \mathbf{M} \tilde{\bv}_j $. Problem \eqref{stationarity} can then be solved as the linear system
\begin{equation}
 (\mathbf{I}_{(pr)^2} - \widetilde{\bA}_{j} \otimes  \widetilde{\bA}_{j} ) \mathbf{M} \tilde{\bv}_j = \mathbf{q}_j . 
\end{equation}

\section{Example code for the toolbox \texttt{switch-ssm}}
\label{sec: matlab}

The statistical methods of this paper are implemented in a MATLAB toolbox %
available at \url{https://github.com/ddegras/switch-ssm}.  
Here we demonstrate the toolbox  \texttt{switch-ssm}  with a small example
using continuous EEG data recorded from an open/closed eyes experiment. 
The data and their description can be found at \url{https://archive.ics.uci.edu/ml/datasets/EEG+Eye+State}.  
The data are also included in the toolbox after artifact removal and  standardization. 
The code below is a  simplified edited version of the file \emph{eeg\_example.mat} in the toolbox. 

{\footnotesize    
\begin{verbatim}
% Load the EEG eye state data 
load('eeg_eye_state.mat')

% Visualize the EEG signals
visualize_mts(eeg,channel)

% Fit a standard (non-switching) VAR model
M = 1; p = 6; S = ones(1,T) % set all regimes S(t) = 1
pars0 = fast_var(eeg,M,p,S);

% Subtract the common stationary component from the data
e = eeg; e(:,1:p) = e(:,1:p) - pars1.mu;
for l = 1:p
    e(:,p+1:T) = e(:,p+1:T) - pars0.A(:,:,l) * eeg(:,p+1-l:T-l);
end

% Fit a switching VAR to the residuals
M = 6; % number of regimes
[~,Ms,~,Shat,pars,LL] = switch_var(e,M,p); 

% Extract regime-specific stationary quantities
[ACF,COH,COV,VAR] = get_covariance(pars);

% Calculate 100 bootstrap replicates of the MLE
B = 100; parallel = true;
[parsboot,LLboot] = bootstrap_var(pars,T,B,[],[],[],[],[],parallel);

 % Match bootstrap estimates to MLE based on matrix A 
parsboot = bootstrap_match(parsboot,pars,'A');

% Build bootstrap confidence intervals for the model parameters
ci = bootstrap_ci(parsboot,0.95,'normal');
\end{verbatim}
 }

 In this example, rather than directly fitting a switching SSM to the data,  we first estimate  a single VAR model with the function \verb@fast_var@ and then fit a switching SSM to the residuals with the function \verb@switch_var@.  The reason for this approach is that if a switching SSM is directly  fitted to the data, the parameter estimates are all nearly equal across regimes and identical  to the common SSM component. In contrast, removing the common SSM component first reveals interesting and diverse FC regimes in the residuals. The model orders $M=6$ and $p=6$ are reasonable choices obtained after fitting and examining a range of models. The regime-specific stationary covariance, variances, and autocorrelation and coherence structures are computed via \verb@get_covariance@. After that, $B=100$ bootstraps of the MLE are calculated in parallel with \verb@bootstrap_var@ and matched to the MLE by regime based on the matrices $\widehat{\bA}_1, \ldots, \widehat{\bA}_M$. 
 This step is accomplished with  \verb@bootstrap_match@. (The matrices $\widehat{\bA}_j$ are used here for matching  and not, say, the noise variance matrices $\widehat{\bQ}_1, \ldots, \widehat{\bQ}_M$ which  are nearly equal to each other. The matching could also be based on stationary covariance matrices as in Section \ref{sec: simulation setup}.)
 Finally, normal bootstrap confidence intervals are built for all model parameters and stationary quantities 
 at the level 95\% with \verb@bootstrap_ci@.

\end{document}